\documentclass[]{spie}  

\usepackage{amsmath,amsfonts,amssymb}
\usepackage{graphicx}
\usepackage[colorlinks=true, allcolors=blue]{hyperref}

\usepackage{caption}
\usepackage{subcaption}

\usepackage{appendix}

\usepackage{wrapfig}
\usepackage{sidecap}

\title{Polarization Modeling and Predictions for DKIST Part 3: Focal Ratio and Thermal Dependencies of Spectral Polarization Fringes and Optic Retardance}

\author[a]{David M. Harrington}
\author[b]{Stacey R. Sueoka}
\affil[a]{National Solar Observatory, 8 Kiopa'a Street, Ste 201 Pukalani, HI 96768, USA}
\affil[b]{National Solar Observatory, 3665 Discovery Drive, Boulder, CO, 80303, USA}

\authorinfo{Further author information: (Send correspondence to D.M.H) \\
D.M.H.: E-mail: dharrington@nso.edu, Telephone: 1 808 572 6888}

\begin{document} 
\maketitle

\begin{abstract}

Data products from high spectral resolution astronomical polarimeters are often limited by fringes. Fringes can skew derived magnetic field properties from spectropolarimetric data. Fringe removal algorithms can also corrupt the data if the fringes and object signals are too similar. For some narrow-band imaging polarimeters, fringes change the calibration retarder properties, and dominate the calibration errors. Systems-level engineering tools for polarimetric instrumentation require accurate predictions of fringe amplitudes, periods for transmission, diattenuation and retardance. The relevant instabilities caused by environmental, thermal and optical properties can be modeled and mitigation tools developed. We create spectral polarization fringe amplitude and temporal instability predictions by applying the Berreman calculus and simple interferrometric calculations to optics in beams of varying F/ number. We then apply the formalism to super-achromatic six crystal retarders in converging beams under beam thermal loading in outdoor environmental conditions for two of the worlds largest observatories: the 10m Keck telescope and the Daniel K. Inouye Solar Telescope (DKIST). DKIST will produce a 300 Watt optical beam which has imposed stringent requirements on the large diameter six-crystal retarders, dichroic beamsplitters and internal optics. DKIST retarders are used in a converging beams with F/ ratios between 8 and 62. The fringe spectral periods, amplitudes and thermal models of retarder behavior assisted DKIST optical designs and calibration plans with future application to many astronomical spectropolarimeters. The Low Resolution Imaging Spectrograph with polarimetry (LRISp) instrument at Keck also uses six-crystal retarders in a converging F/ 13 beam in a Cassegrain focus exposed to summit environmental conditions providing observational verification of our predictions.

\end{abstract}


\section{MOTIVATION: DKIST THERMAL ISSUES AND F/ NUMBERS}
\label{sec:motivation}  

In many astronomical spectropolarimeters, spectral fringes in intensity and polarization are the dominant source of error. These errors can either involve corrupting of the measured signals or a skewing of the calibrations. Fringe amplitudes can be over 10\% with strong changes in fringe characteristics over time, field angle, wavelength and optical configuration. These fringes often have similar characteristics to the solar polarimetric signals. This similarity complicates the data analysis as fringe removal techniques can corrupt the measurement and skew the properties of the object derived from those measurements. Accurate tools to estimate fringe amplitudes and polarization characteristics are critical for assessing optical designs, evaluating the trade-offs in retarder location and preparing techniques for fringe removal in post-facto processing of instrument data products. Fringes must be estimated in converging or diverging beams along with dependence on optical design properties such as cover windows, oil layers and anti-reflection coatings. This must be coupled to thermal behavior as environmental and optical heat load control is critical for the instrument design and fabrication process. Particular challenges arise in modern solar instrumentation where beams are steeply converging and heat loads can be severe. 

The Daniel K. Inouye Solar Telescope (DKIST) on Haleakal\={a}, Maui, Hawai'i is under construction and planning on science operations beginning in 2020. The off-axis altitude azimuth telescope has a 4.2m diameter F/ 2 primary mirror (4.0m illuminated). The secondary mirror creates an F/ 13 Gregorian focus. Five more mirrors then relay this beam to a suite of polarimetric instrumentation in the coud\'{e} laboratory \cite{2014SPIE.9145E..25M, Keil:2011wj, Rimmele:2004ew}. Modulating retarders are used in each of these instruments with beams with focal ratios varying from F/ 8 to F/ 62.  Many of the proposed science cases rely on high spectral resolution polarimetry. We recently adopted the Berreman calculus to model many-crystal retarders along with anti-reflection coatings, oils and bonding materials and we refer to this work as H17 here \cite{Harrington:2017jh}. We use the Berreman calculus along with interferrometric calculations and thermal modeling to create fringe amplitude and Mueller matrix predictions for the DKIST instruments. We show how to predict fringe properties as well as to anticipate their amplitude in converging and diverging beams during the instrument design process. With our thermal modeling, we also can assess impacts from design choices on retarder performance and temporal instabilities limiting calibrations.

DKIST uses seven mirrors to feed the beam to the rotating coud\'{e} platform \cite{Marino:2016ks, McMullin:2016hm,Johnson:2016he,2014SPIE.9147E..0FE, 2014SPIE.9147E..07E, 2014SPIE.9145E..25M}. Operations involve four polarimetric instruments spanning the 380 nm to 5000 nm wavelength range. At present design, three different retarders are in fabrication for use in calibration near the Gregorian focus \cite{2014SPIE.9147E..0FE,Sueoka:2014cm,Sueoka:2016vo}. These calibration retarders see a beam with 300 Watts of optical power, a focal ratio F/ 13 with an extremely large clear aperture of 105 mm. A train of dichroic beam splitters in the collimated coud\'{e} path allows for rapid changing of instrument configurations. Different wavelengths can be observed simultaneously by three polarimetric instruments covering 380 nm to 1800 nm all using the adaptive optics system \cite{2014SPIE.9147E..0FE, 2014SPIE.9147E..07E, 2014SPIE.9147E..0ES, SocasNavarro:2005bq}. Another instrument (CryoNIRSP) can receive all wavelengths using an all-reflective beam to 5000 nm wavelength but without adaptive optics.

Complex polarization modulation and calibration strategies are required for such a mulit-instrument system \cite{2014SPIE.9147E..0FE,2014SPIE.9147E..07E, Sueoka:2014cm, 2015SPIE.9369E..0NS, deWijn:2012dd, 2010SPIE.7735E..4AD}. The planned 4m European Solar Telescope (EST), though on-axis, will also require similar calibration considerations \cite{SanchezCapuchino:2010gy, Bettonvil:2011wj,Bettonvil:2010cj,Collados:2010bh}.  Many solar and night-time telescopes have performed polarization calibration of complex optical pathways \cite{DeJuanOvelar:2014bq, Joos:2008dg, Keller:2009vj,Keller:2010ig, Keller:2003bo, Rodenhuis:2012du, Roelfsema:2010ca, 1994A&A...292..713S, 1992A&A...260..543S,  1991SoPh..134....1A, Schmidt:2003tz, Snik:2012jw,  Snik:2008fh, Snik:2006iw, SocasNavarro:2011gn, SocasNavarro:2005jl, SocasNavarro:2005gv, Spano:2004ge, Strassmeier:2008ho, Strassmeier:2003gt, Tinbergen:2007fd, 2005A&A...443.1047B, 2005A&A...437.1159B}. We refer the reader to recent papers outlining the various capabilities of the DKIST first-light instruments \cite{McMullin:2016hm, 2014SPIE.9147E..07E, 2014SPIE.9145E..25M, 2014SPIE.9147E..0FE, Rimmele:2004ew}. 

Berreman (1972) \cite{1972JOSA...62..502B} formulated a 4x4-matrix method that describes electromagnetic wave propagation in birefringent media. The interference of forward and backward propagating electromagnetic waves inside arbitrarily oriented stacks of biaxial material is included in this very general theory. This Berreman calculus can be used to describe wave interference in multiple birefringent layers, crystals, chiral coatings and other complex optical configurations with many birefringent layers of arbitrary optical axis orientation. A recent textbook by McCall, Hodgkinson and Wu (MHW) has further developed and applied the Berreman calculus \cite{2014btfp.book.....M}.  In this work we assume basic familiarity with the MHW textbook \cite{2014btfp.book.....M} and the basic thin film calculations by Abeles and Heavens matrices \cite{Heavens:1965uq}. This formalism is in common use in coating modeling software such as TFCalc or Zemax coating reports. 

We adapted the Berreman formalism to the six-crystal achromatic retarders used in DKIST along with many-layer anti-reflection coatings, oil layers and cover windows \cite{Harrington:2017jh}. In this paper, we use the Berreman calculus and add interference effects from converging and diverging beam variation across the aperture. We then show thermal models for our retarders under absorptive loads in the 300 Watt Gregorian beam. With associated spectral measurements of parts-per-million level absorption caused by anti-reflection coatings and crystal bulk material, we can accurately assess the spectral absorption though these retarder optics and predict thermal performance. The appendix details the thermal modeling. The fringe temporal instability caused by thermal loading is also measured in simple laboratory experiments to verify sensitivity. We predict fringe amplitudes and thermal timescales for DKIST retarders with application to typical solar telescope heat loads on similar calibration optics.

\begin{wrapfigure}{r}{0.37\textwidth}
\vspace{-6mm}
\begin{equation}
{\bf M}_{ij} =
 \left ( \begin{array}{rrrr}
 II   	& QI		& UI		& VI		\\
 IQ 	& QQ	& UQ	& VQ	\\
 IU 	& QU	& UU	& VU		\\
 IV 	& QV	& UV		& VV		\\ 
 \end{array} \right ) 
\label{eqn:MM}
\end{equation}
\vspace{-6mm}
\end{wrapfigure}

The Berreman calculus contains all polarization phenomena and is very general \cite{2014btfp.book.....M}. We can compute non-normal incidence interference effects through multiple birefringent layers or thick crystals as required for converging beams. The main limitation of the Berreman formalism is in the assumption of complete beam overlap using plane waves of infinite spatial extent. In the Berreman formalism for a finite sized beam at non-normal incidence, the multiple reflections inside a thick plate will, in practice not overlap with the incoming beam. In the limit of no beam overlap, the Jones formalism is recovered. Berreman always assumes infinite coherence lengths, and that all multiple reflections stay within the optical path. For most astronomical applications, this beam overlap assumption is reasonably valid as the crystals are thin compared to the beam diameter and the back-reflected footprint is within a few percent of the diameter of the incoming beam. As we show in this paper, most optical systems with beams slower than F/ 5 and retarders placed not exactly in focal planes will have amplitudes and fringe characteristics well-estimated by the Berreman formalism. 

In this work, we follow standard notation for propagation of polarization through an optical system. The Stokes vector is denoted as {\bf S} = $[I,Q,U,V]^T$. The Mueller matrix is the 4x4 matrix that transfers Stokes vectors \cite{1992plfa.book.....C, Chipman:2014ta, Chipman:2010tn}. Each element of the Mueller matrix is denoted as the transfer coefficient \cite{Chipman:2010tn, 2013pss2.book..175S}. For instance the coefficient [0,1] in the first row transfers $Q$ to $I$ and is denoted $QI$. The first row terms are denoted $II$, $QI$, $UI$, $VI$. The first column of the Mueller matrix elements are $II$, $IQ$, $IU$, $IV$. In this paper we will use the notation in Equation \ref{eqn:MM}

\begin{wrapfigure}{l}{0.43\textwidth}
\vspace{-6mm}
\begin{equation}
 \left ( \begin{array}{rrrr}
 II   		& QI/II	& UI/II	& VI/II	\\
 IQ/II 	& QQ/II	& UQ/II	& VQ/II	\\
 IU/II 	& QU/II	& UU/II	& VU	/II	\\
 IV/II	 	& QV/II	& UV	/II	& VV	/II	\\ 
\end{array} \right ) 
\label{eqn:MM_IntensNorm}
\end{equation}
\vspace{-6mm}
\end{wrapfigure}

We also will adopt a standard astronomical convention for displaying Mueller matrices.  We normalize every element by the $II$ element to remove the influence of transmission on the other matrix elements as seen in Equation \ref{eqn:MM_IntensNorm}.  Thus subsequent Figures will display a matrix that is not formally a Mueller matrix but is convenient for displaying the separate effects of transmission, retardance and diattenuation in simple forms.

\section{Equal Inclination Fringes: Fringe dependence on AOI \& F/}

Retarders are often used in converging and diverging beams. A range of incidence angles are present across the beam footprint for these optics. We compute the expected fringe amplitudes under some simple assumptions to compare with laboratory data. 

\begin{wrapfigure}{r}{0.63\textwidth}
\centering
\vspace{-4mm}
\begin{tabular}{c} 
\hbox{
\hspace{-0.4em}
\includegraphics[height=9.2cm, angle=0]{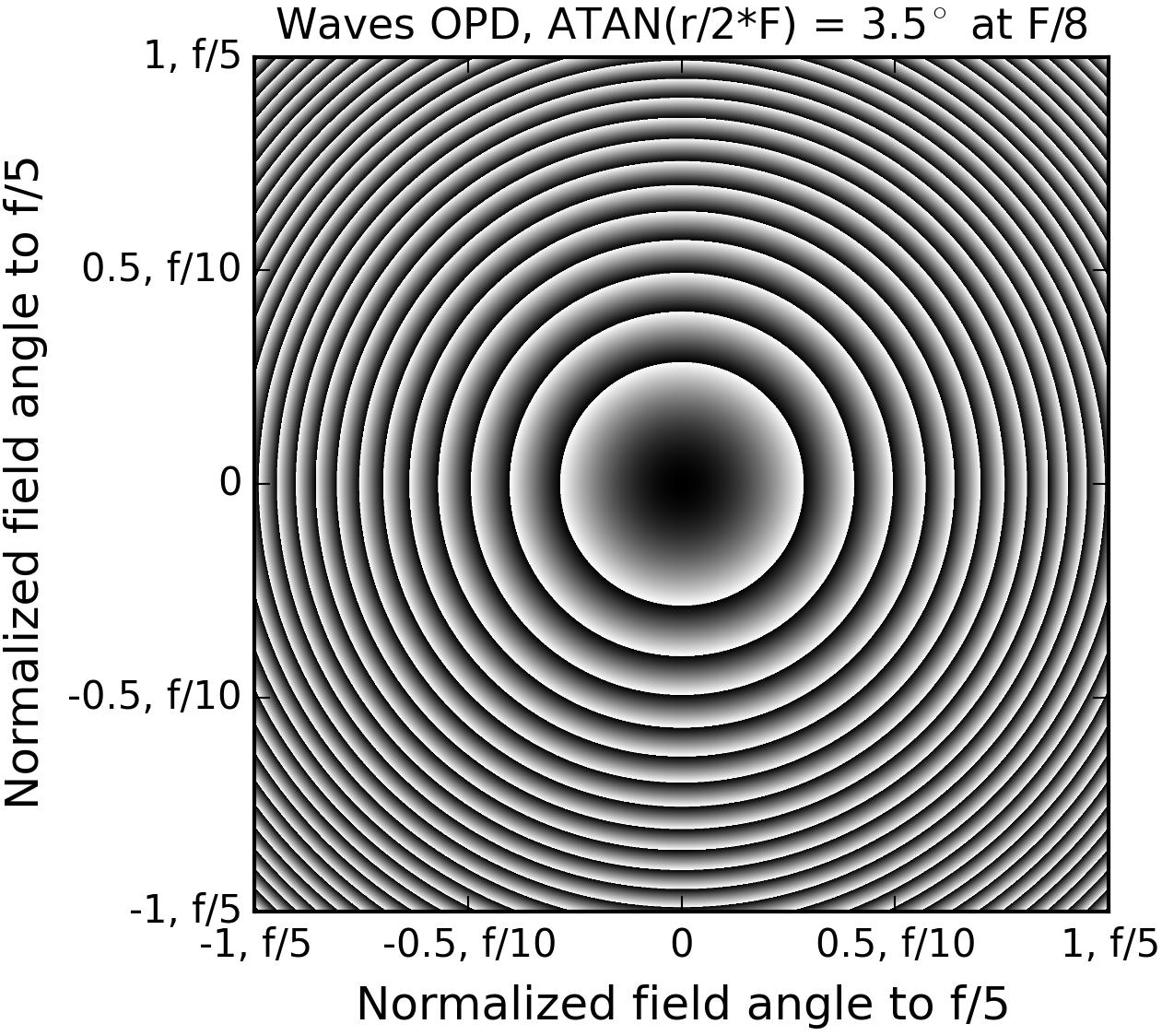}
}
\end{tabular}
\caption[Waves OPD] 
{\label{fig:waves_OPD} Waves OPD across a rectangular footprint for a beam at F/ 5. Field edges at r=1 correspond to an internal beam propagating at an angle of 3.91$^\circ$ refracted into an index n=1.46 medium. For a 1.1 mm thick fused silica window, the back-reflected chief ray traverses 5253.683 waves optical path. The back-reflected F/ 5 marginal ray sees 11.6 waves of additional path in addition to double the incidence angle.}
\vspace{-4mm}
\end{wrapfigure}

We consider the limiting case of a thin window where we can neglect the incomplete overlap between the back-reflected beam and the incoming beam. In this situation, we recover a simple division of amplitude type interferrometer for {\it fringes of equal inclination} sometimes called Haidingers fringes. Detailed descriptions are in several optical textbooks including Born \& Wolf Chapter 7 \cite{Born:2012un} and Hariharan Chapter 2 \cite{Hariharan:2007uk}.  By tracing both the first-surface reflected ray and the ray that reflects off the back surface, a trigonometric relation between the two parallel but displaced reflected rays can be created. The optical thickness of the window is computed as o = 2dn/$\lambda$. The phase difference between front-surface-reflected and back-surface-reflected rays is 2$\pi$o$\cos\theta$ where $\theta$ is the propagation angle in the medium. For small incidence angles, we can use the approximation that $\theta$ = $\theta_{air}$/n. We get bright fringes for constructive interference when 2dn$\cos\theta$ plus the half-wave of phase upon reflection gives integer waves of path.  We get destructive interference at half-wave integer multiples.  

For a beam of a given F/ number in air, the marginal ray represents the highest incidence angle in the beam at $\theta = \tan^{-1}(1/2F)$. The fastest beam seen by the DKIST and the Meadowlark high resolution spectrograph we use here has an F/ 8 beam which sees a maximum incidence angle of 3.67$^\circ$. The DKIST Gregorian focus at F/ 13 would see a 2.20$^\circ$ incidence angle for the marginal ray. For the calculation of fringes, we must divide by the material refractive index to get the propagation angle in the medium.

We compute a simple example of the interference pattern across the clear aperture of a fused silica window. We use the Meadowlark Optics provided Heraeus Infrasil 302 sample, as measured in H17\cite{Harrington:2017jh}. The thickness is measured to be 1.1335 mm with the Heidenhain metrology system and we compute a refractive index of n=1.46 at a measurement wavelength around 630 nm using the vendor provided equations. The optical thickness is 5253.7 waves for the on-axis beam. For a beam traveling through the part with a marginal ray incident at 3.67$^\circ$ for the Meadowlark Spex spectrograph F/ 8 beam, the refracted ray travels at 2.45$^\circ$ incidence inside the optic ($\theta$/n). The thickness for an inclined beam is 2dn/$\lambda \cos\theta$. The marginal ray traverses a part thickness of 5258.5 waves. The difference is about 4.8 waves path between on-axis chief ray and the marginal ray for the F/ 8 beam. When computing the interference path difference we use the {\it equal inclination fringe} equation 2$\pi$o$\cos\theta$ and we get the same 4.8 waves of path length between rays. The optical path difference (OPD) between chief and marginal rays can be computed as the factor $(1/\cos\theta - 1)$. Once the optical path is known, the interference amplitudes can be calculated across the footprint as the ray incidence angle changes.  

Figure \ref{fig:waves_OPD} shows the OPD in waves across a rectangular aperture for this Infrasil sample. The beam is at F/ 5 on the extreme diagonal corners of the rectangle. We choose a nominal wavelength of 630 nm and the metrologized thickness to compute nearly destructive interference at the center of the optic oscillating over many waves of optical path across the rectangular aperture.  We encode waves of OPD as the gray-scale color where white is integer multiples of 1 wave of path difference. Black is integer multiples of zero waves of path difference. The field angle was normalized from 0 to 1 along the X and Y axes of the image.  The inner part of the beam footprint representing an F/ 20 or slower beam is within less than 1 wave of interference variation across the aperture.  For a beam of F/ 10 illuminating more of the part, a few waves of interference would be seen.

\begin{wrapfigure}{r}{0.60\textwidth}
\centering
\vspace{-2mm}
\begin{tabular}{c} 
\hbox{
\hspace{-1.2em}
\includegraphics[height=6.95cm, angle=0]{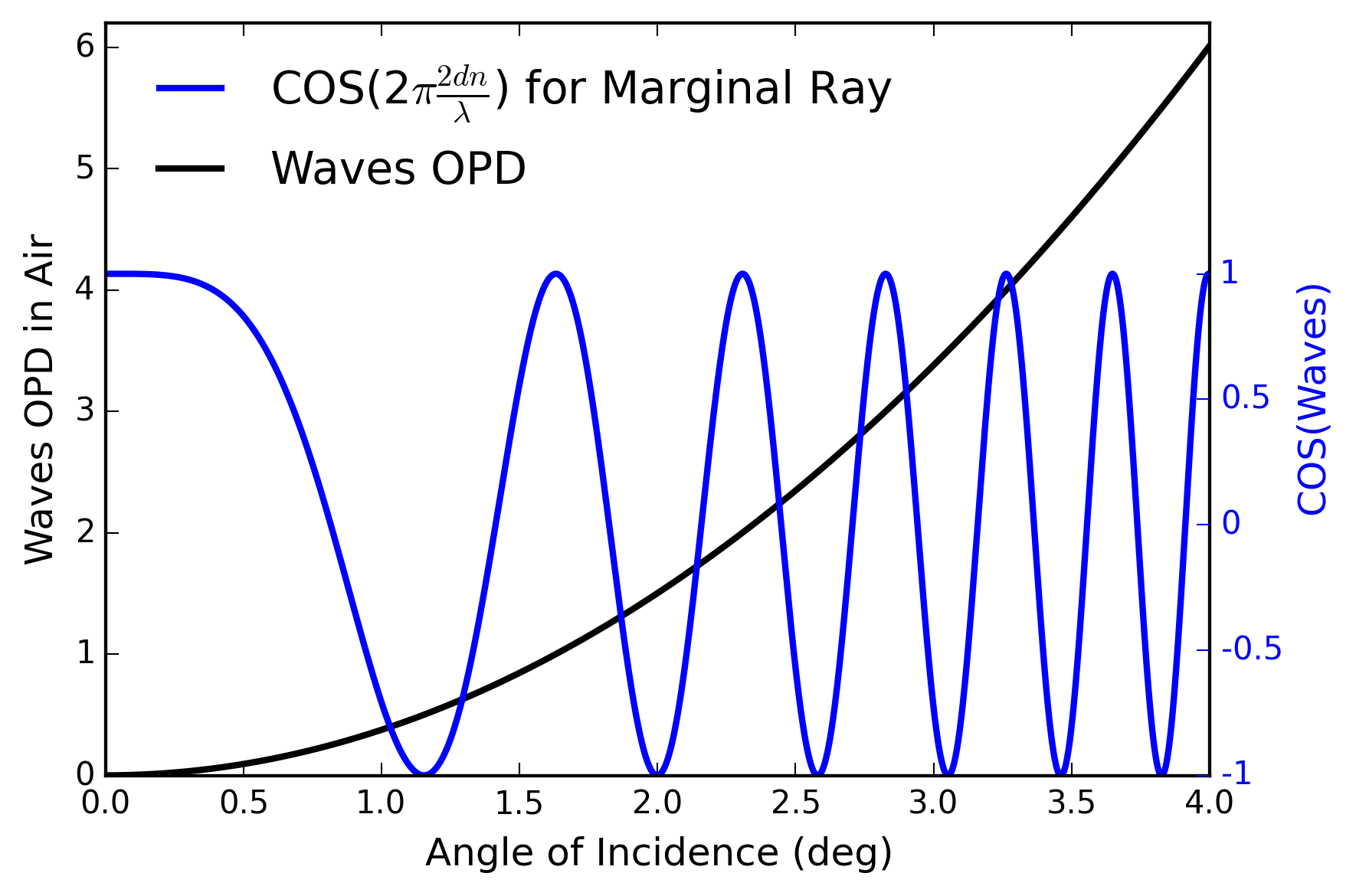}
}
\end{tabular}
\caption[Waves OPD] 
{ \label{fig:transmission_waves_solve} The waves optical path difference between chief and marginal rays across the center of the rectangular aperture from Figure \ref{fig:waves_OPD} is shown using the left hand Y axis. The cosine of this term is shown in blue using the right hand Y axis. This determines the magnitude of interference and multiplies the square root field amplitudes.}
\vspace{-4mm}
\end{wrapfigure}

The standard equation for summing interfering waves of the same frequency is $A_1^2 + A_2^2 + 2 A_1 A_2\cos{\phi}$ where the wave amplitudes are denoted $A$ and the relative phase between waves is $\phi$. Figure \ref{fig:transmission_waves_solve} shows the waves of phase path difference between chief and marginal rays from the center of the clear aperture. The in-air incidence angle runs from 0$^\circ$ at the center of the optic to 4$^\circ$, near an F/ 7 beam following Figure \ref{fig:waves_OPD}. As this phase represents the coherent interference term, the cosine of this optical path difference becomes $\phi$ in the interference equation multiplying the two root-amplitude coefficients. We can see the blue curve of Figure \ref{fig:transmission_waves_solve} sees seven peaks with constructive interference and six peaks with destructive interference as the optical path difference changes from zero to over six waves for a beam from collimated to F/ 7.

We next translate the interference pattern across the clear aperture to a transmitted intensity at each incidence angle. We do this using the simple interference equation where the fringe amplitude is 4 $\sqrt{I_{front}}\sqrt{I_{back}}$. For Infrasil at 630 nm wavelength, the single-surface uncoated reflection is nominally 3.5\%. The back-reflection from the internal Infrasil-to-air interface would have an intensity of 96.5\% of 3.5\% which is 3.4\%. As electric fields add coherently, we take the square root of the intensity and add the fields. If the phase is 180$^\circ$, destructive interference reduces the transmission of the optic to 86.25\%. If the phase is 0$^\circ$ then coherent interference increases the transmission of the optic to 99.9\%. As the effective angle of the incident ray is increased, the optic will have the thickness vary by several waves giving multiple constructive and destructive interference peaks.

\begin{figure}[htbp]
\begin{center}
\vspace{-3mm}
\hbox{
\hspace{-1.0em}
\includegraphics[height=7.4cm, angle=0]{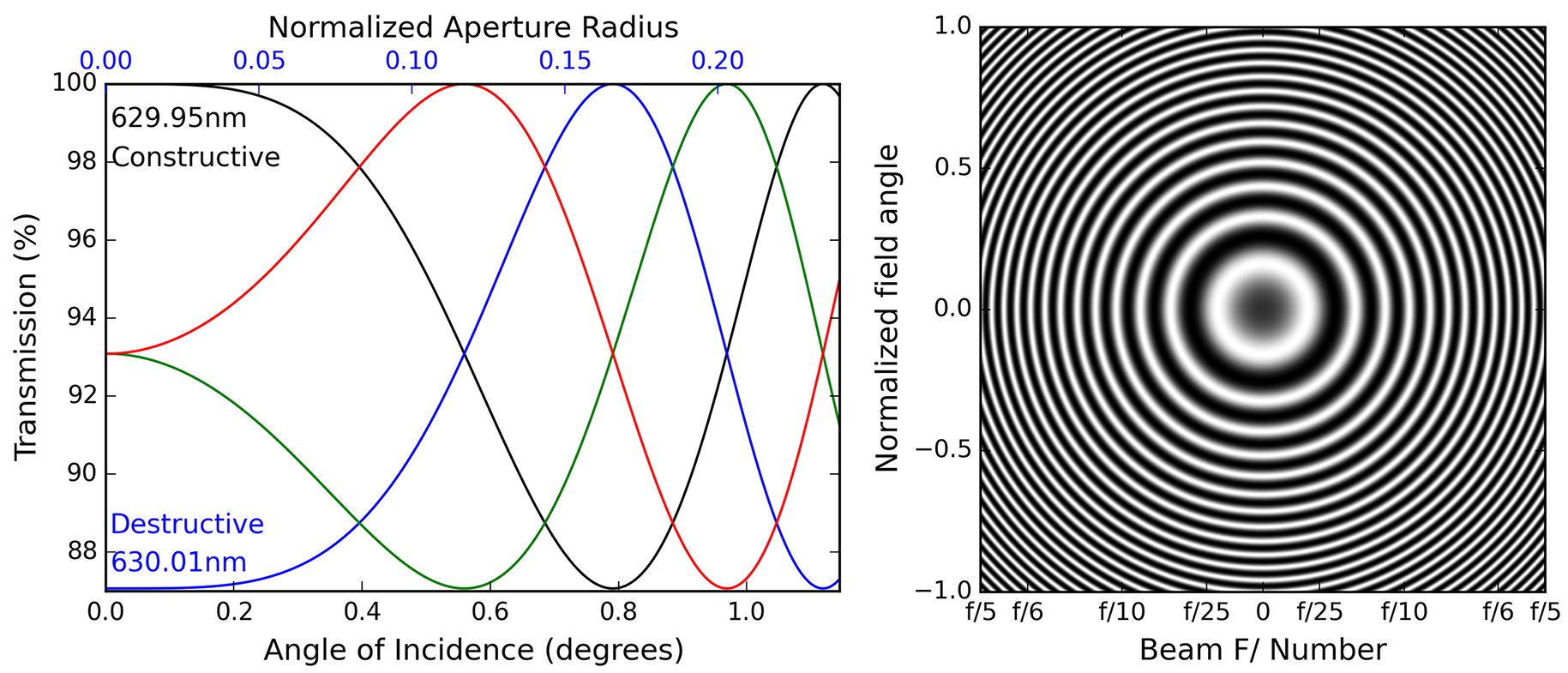}
}
\caption[] 
{The right plot shows a grey-scale image of transmission of the Infrasil using a wavelength of 630 nm computed by coherently summing electric fields. The grey scale goes linearly from 86.25\% to 100\% transmission for a fringe amplitude of 13.75\% peak-to-peak. The optical path is 5253.683 waves for the back-reflected chief ray at this wavelength. The F/ 5 reflected marginal ray sees an incidence angle of 3.91$^\circ$ in the index=1.46 medium and 5.71$^\circ$ in air. This gives an optical path increase of 11.6 waves when reflected back to the first surface in the medium. The left graph shows transmission functions for wavelengths near 630 nm chosen to be exactly 0, 0.25, 0.5 and 0.75 waves optical interference path thickness for the chief ray.  At part center, these rays see transmission of 99.9\%, 93.1\%, 86.3\% and 93.1\% transmission respectively. A wavelength of 629.95 nm gives perfect constructive interference and corresponds to the black curve.  A wavelength of 630.01 nm is exactly half a fringe period later and gives perfect destructive interference at the center of the optic. The green and red curves show optical paths at 0.25 and 0.75 waves interference path which provide average transmission of 93.1\% at the center of the optic.  As the incidence angle increases towards the edge of the beam footprint, the ray sees increasing path length and oscillations of constructive and destructive interference. Note that a 1$^\circ$ incidence angle in air corresponds to F/ 28.5. \label{fig:interference_vs_aperture}  }
\vspace{-6mm}
\end{center}
\end{figure}

Figure \ref{fig:interference_vs_aperture} shows an example of the electric field interference calculation across a simulated rectangular aperture for this Infrasil sample. The left hand graphic shows four separate wavelengths solved to have 5253 waves plus 0.5, 0.75, 0 and 0.25 waves of path for the back reflected chief ray. Using these wavelengths corresponding to integer multiples of quarter-wave optical thickness, we can show the transmission for rays as functions of incidence angle across the footprint of the beam on the optic. The integer-wave multiple wavelength sees complete constructive interference hence 99.9\% transmission for the chief ray at zero incidence angle. As the incidence angle is increased, we see the first minimum transmission of 86.25\% occur around 0.8$^\circ$ incidence angle corresponding to F/ 36 or only the inner 0.16 of the aperture radius. As the incidence angle increases and the ray encounters increasingly larger path lengths, we see oscillations between maximum and minimum transmission. 

For the two curves of Figure \ref{fig:interference_vs_aperture} at multiples of quarter-wave thickness, the chief ray sees a transmission of 93.1\% which is the non-interferometric transmission computed by independently considering the 3.5\% loss from the first surface and 3.4\% loss from the back surface. The first minimum and / or maximum occurs at incidence angles around 0.6$^\circ$ corresponding to a normalized radius of 0.12 or a beam of F/ 47.

The right hand image in Figure \ref{fig:interference_vs_aperture} shows the calculation of interference fringes at the single 630 nm wavelength across the full rectangular clear aperture out to the extreme edges of the F/ 5 beam. Given these fringes, an Infrasil window at 1.13 mm thickness and 630 nm wavelength would be expected to show high fringe amplitudes only for circular beams slower than roughly F/ 40 where the part is less than half-wave interference across the beam footprint. For beams faster than F/ 20, we are spatially averaging more than a full wave of optical path interference across the converging beam footprint. We should note that our Berreman calculus scripts were used to compute the curves in Figure \ref{fig:interference_vs_aperture}. To assess the fringe amplitude as a function of beam focal ratio, we can easily compute the average transmission over a footprint of a given F/ number by doing an intensity weighted aperture average.

We compute the dependence on beam F/ number and wavelength by running a large simulation over a full fringe spectral period.  We selected 100 F/ numbers between F/ 6 and F/ 120.  For each of these F/ numbers, we choose 100 wavelengths to cover at least a full fringe period. For the Infrasil window, we selected 0.15 nm of spectral bandpass to more than fully cover the 0.12 nm spectral fringe period.

\begin{figure}[htbp]
\begin{center}
\vspace{-3mm}
\hbox{
\hspace{-2.0em}
\includegraphics[height=5.8cm, angle=0]{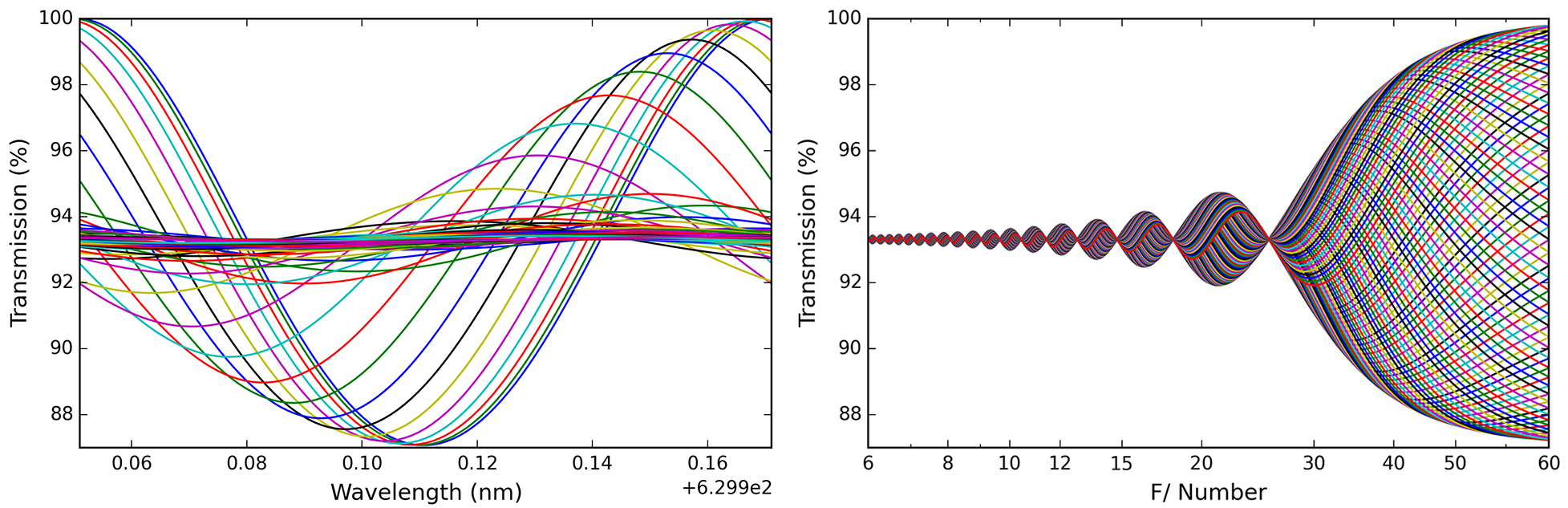}
}
\caption[] 
{The left panel shows the transmission spectrum at each simulated F/ number from F/ 120 to F/ 16.  The fringes have the expected range from 86.25\% to 99.99\% when the beam is nearly collimated.  As the F/ number approaches F/ 20, the fringe amplitude approaches nearly zero and the transmission is spectrally constant around 93\%. The right panel shows the transmission as a function of beam F/ number. In this panel, each wavelength is a different curve showing how the influence of F/ number creates alternating patterns of constructive and destructive interference at any individual wavelength that oscillates about the mean as the F/ number increases. \label{fig:infrasil_model_transmission_vs_fnum_and_wavelength}  }
\vspace{-6mm}
\end{center}
\end{figure}

For each of these simulations, we compute the spatial interference pattern across the aperture for transmission through the part as in Figure \ref{fig:interference_vs_aperture}. For each of these apertures, we can select transmission within a restricted F/ number to create the transmission function averaged over that aperture.  We repeat this aperture average for all 100 F/ numbers and all 100 wavelengths.  In Figure \ref{fig:infrasil_model_transmission_vs_fnum_and_wavelength}, we show the typical transmission spectra for F/ numbers from 16 to 120 in the left panel.  We show all 100 wavelengths in the right panel as a function of F/ number.

When averaging over the aperture, this simple geometric model predicts the fringe amplitude will go to zero at specific beam F/ numbers. This effect has a simple intuitive geometric explanation. When the marginal ray sees an additional half-wave of optical path difference from the chief ray, we will be averaging over a spatial pattern that has equal spatial areas of constructive and destructive interference. As we're averaging the beam spatially over an aperture, this would bring the transmission to a common average value.  As these parts are typically several thousand waves thickness, all wavelengths in Figure \ref{fig:infrasil_model_transmission_vs_fnum_and_wavelength} share common null points.

Each ray sees an optical thickness of $(1-\cos(\theta/n)) 2dn/\lambda$ waves. We can simply solve for integer multiples as $\theta = \cos^{-1}(1-m2dn/\lambda)$ where m is an integer (0,1,2....). The F/ number for this fringe null is then computed as f = 1/2$\tan\theta$. As an example, the 1.13 mm thick Infrasil window at 630 nm wavelength sees half-wave multiples for beam F/ numbers of (17.6, 12.4, 10.1, 8.8, 7.8, 7.1). To compute fringe maxima, simply calculate using half-wave multiples. For a 12 mm thick quartz optic, the F/ numbers of the null points are F/ 55, 39, 32, 28, etc.  An immediate conclusion is that the F/ 13 beam beam near DKIST Gregorian focus should be sufficiently fast that fringe amplitudes in 12mm thick crystal retarders will be averaged over many waves of optical path difference. Fringe amplitudes for the faster fringe periods will be reduced by factors of few to $>$20 for the DKIST super achromatic calibration retarders (SARs) and the polychromatic modulator (PCM) optics provided the beam F/ number is sufficiently fast. Given the DKIST retarders range from F/13 to F/ 62 and cover close to four octaves of wavelength, case-by-case consideration will be required.

\begin{wrapfigure}{r}{0.50\textwidth}
\vspace{-6mm}
\begin{equation}
\int_{0}^{r} \sqrt{ R_f + R_b + 2\sqrt{R_f} \sqrt{R_b} \cos(2\pi \frac{2dn}{\lambda \cos\frac{\theta}{n}}) } r dr
\label{eqn:area_integral_fnumber}
\end{equation}
\vspace{-6mm}
\end{wrapfigure}

The amplitude of the fringes decreases with F/ number as the aperture average drives the transmission towards the nominal average value. For the reflected beam, the equation is somewhat simpler to write in terms of the coherently summed electric field values.  We note that the reflectivity ($R$) is the square of the E field and use the standard equation for summing two waves of the same frequency but different phase offset. Equation \ref{eqn:area_integral_fnumber} shows the circular area integral over the clear aperture. This equation considers an optic of normalized aperture radius r where the incidence angle relates to the F/ number through $\tan\theta$=r/2F as r goes from 0 to 1. We can easily imagine integrating the area in a circular aperture weighted by the transmission functions of Figure \ref{fig:interference_vs_aperture}.

\begin{wrapfigure}{r}{0.63\textwidth}
\centering
\vspace{-3mm}
\begin{tabular}{c} 
\hbox{
\hspace{-1.0em}
\includegraphics[height=6.95cm, angle=0]{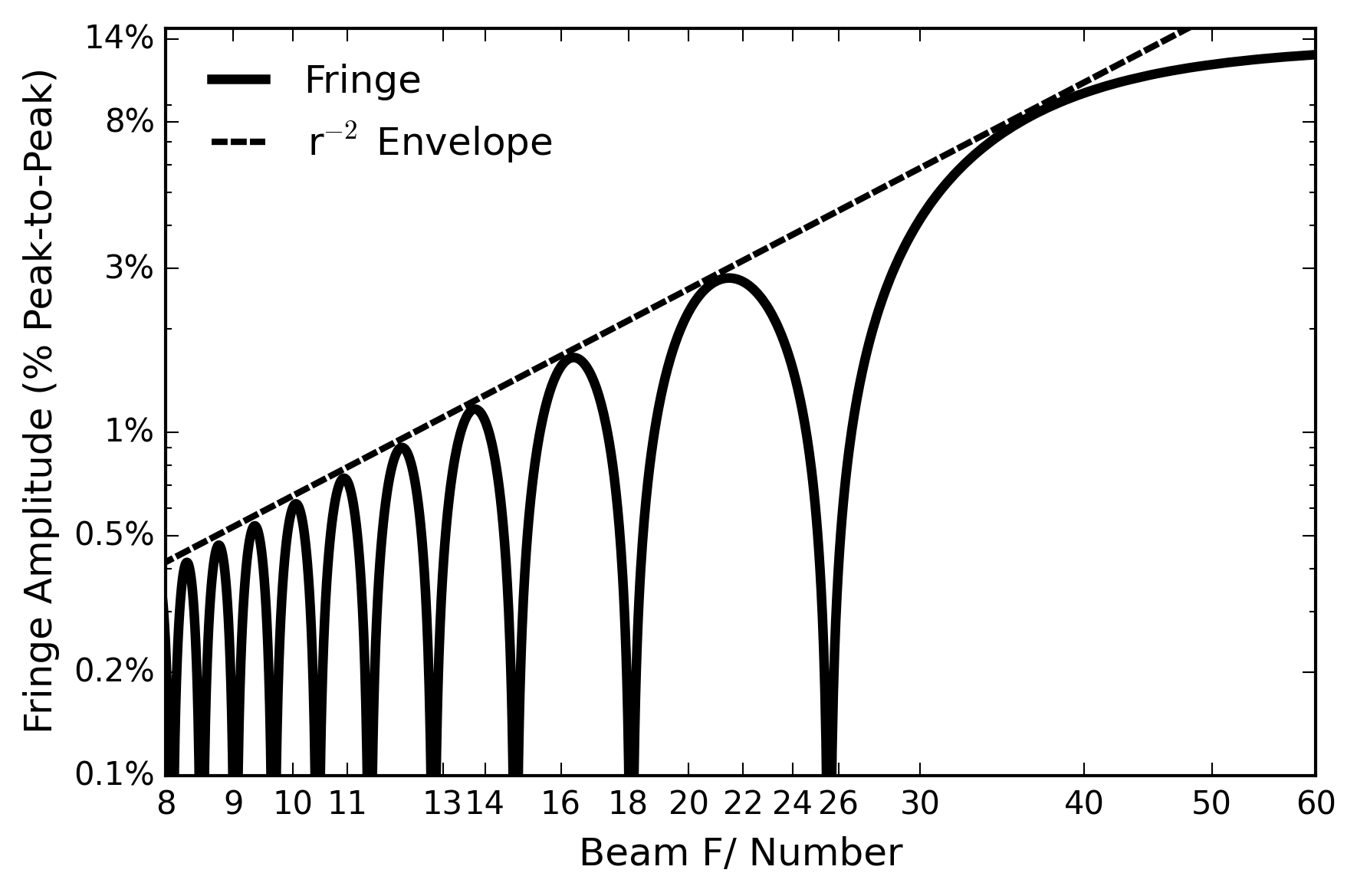}
}
\end{tabular}
\caption[Waves OPD] 
{ \label{fig:transmission_maxmin} The peak-to-peak fringe amplitude about the  93.1\% average transmission as a function of F/ number for the 1.13mm Infrasil window at 630 nm wavelength.}
\vspace{-4mm}
\end{wrapfigure}

Figure \ref{fig:transmission_maxmin} shows the deviation from the nominal average transmission of 93.1\% as the beam F/ number is changed. The peak to peak fringe amplitude is roughly 14\%. The Infrasil window at 630 nm wavelength shows a 2.8\% fringe peak to peak at the first maximum near F/ 21. This maximum amplitude would occur when integrating over the aperture from the center out to an integer multiple of quarter-wave interference path corresponding to constructive interference on the outer annulus of the aperture. This 2.8\% fringe is reduced by a factor of five from the collimated beam 14\% fringe amplitude. The second maximum is near F/ 16 with 1.6\% fringe, corresponding to an amplitude reduction factor of nine when averaging over more than two waves of aperture interference. The reduction factor is roughly 22 for five waves of aperture interference. 

The dashed black line of Figure \ref{fig:transmission_maxmin} shows the $r^{-2}$ envelope expected for fringes as the integrated area increases with beam F/ number and fringes successively average over multiple fringe cycles. Given the relatively simple dependence of these {\it equal inclination fringes} on optic thickness, beam F/ number and wavelength, we can construct amplitude reduction factors for the various spectral fringe components in the DKIST retarders for the wide range of operating wavelengths.

\subsection{Summary of Fringe Amplitude Reduction Estimates in a Converging Beam}

We have adapted a simple analytical theory for {\it equal inclination fringes} to show how we can scale fringe amplitudes in a single plane parallel window by a $r^{-2}$ envelope depending on wavelength, F/ number and material thickness. This simple $r^{-2}$ envelope will be used in later sections to estimate fringe amplitude reduction for many-crystal retarders in converging beams like DKIST.

\clearpage
\section{Laboratory Measurements: Fringes with Beam F/ Number}
\label{sec:labdata}  

Laboratory measurements are easily done with well characterized samples and controlled environments. We use windows and crystal retarders of known thickness, low beam deflection and small wavefront error in beams of controlled shape to verify the fringe behavior. In the Meadowlark facility, they have a SPEX 1401 double grating 0.85-meter Czerny-Turner spectrometer. The light source is an Energetiq broad band fiber coupled plasma source using a 200 $\mu$m diameter core fiber. The fiber output is nominally collimated to a $\sim$10 mm diameter beam by a Thor labs 90$^\circ$ fold angle silver-coated off-axis parabola mirror with an effective focal length of 15 mm. The fiber light source and the OAP collimating mirror will produce some polarization expected to be at amplitudes less than a few percent at visible wavelengths. For this mounted OAP, the beam diameter is set by the exit of the housing after the mirror at an 11 mm diameter.  

The mirror is oversized and mounted before this aperture, giving rise to a small field dependence and some spatial dependence on sampling the fiber exit illumination. Some mild non-uniformity is seen across the beam. The system is set up to have a 10 mm diameter collimated beam that is focused on the spectrograph entrance slit via a 50 mm focal-length singlet. The fiber core is magnified by the 15 mm to 50 mm ratio and fills a 0.67 mm tall SPEX slit. Given the 200 $\mu$m diameter fiber and the 50/15 magnification, the range of angles across the field is $\pm$0.38$^\circ$. At visible wavelengths, the measured resolving power is $\frac{\lambda}{\delta\lambda}$ in the 30,000 to 45,000 range. The slit is 35 $\mu$m in width and over 1 mm high to pass the full magnified 200 $\mu$m fiber core image.  The F/ 5 beam entering the spectrograph is stopped to F/ 8 beam by a rectangular aperture on the collimating mirror inside the spectrograph. 
 
\begin{wrapfigure}{r}{0.61\textwidth}
\centering
\vspace{-5mm}
\begin{tabular}{c} 
\hbox{
\hspace{-1.0em}
\includegraphics[height=6.85cm, angle=0]{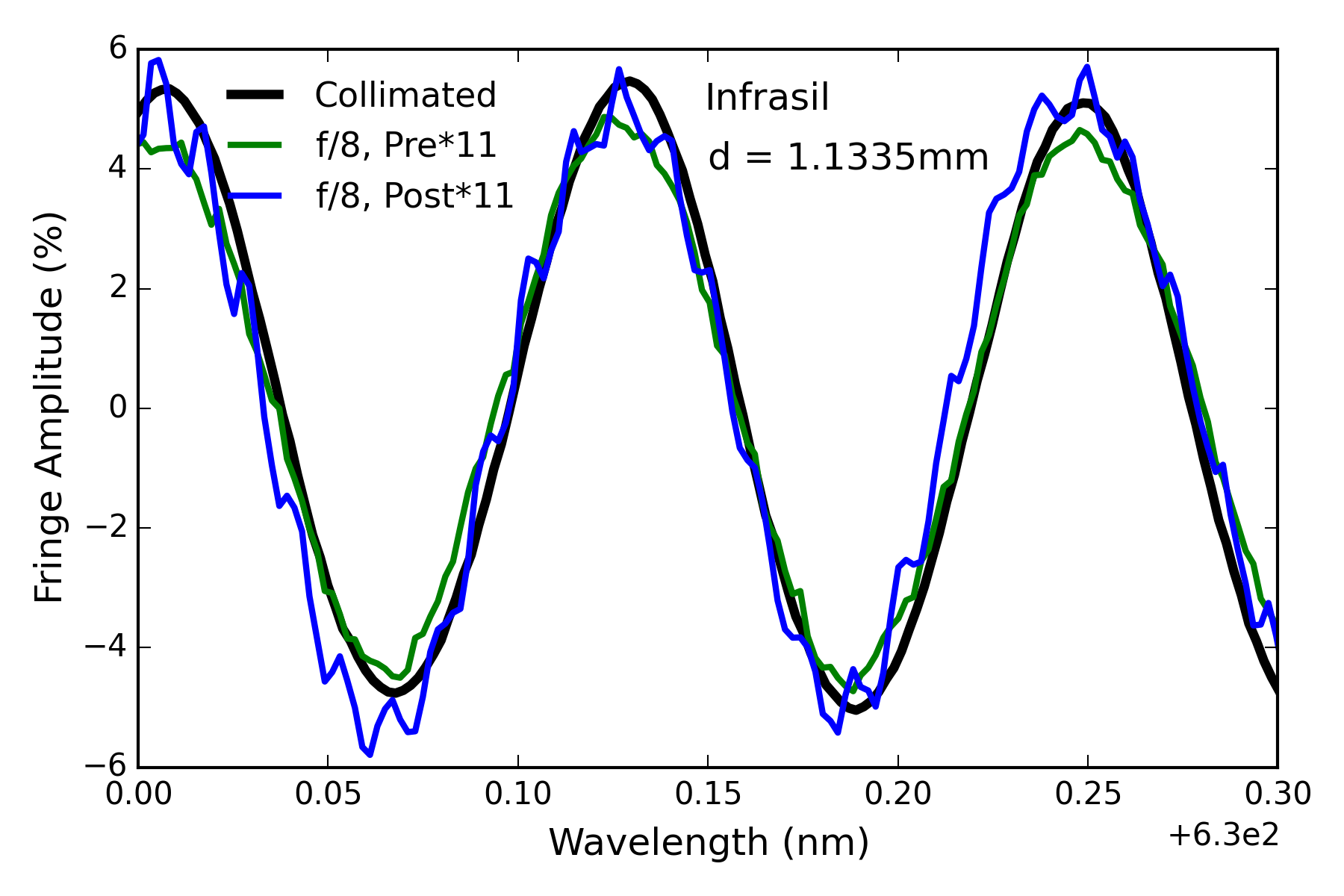}
}
\end{tabular}
\caption[Intensity Transmission Infrasil] 
{ \label{fig:transmission_fringes_Infrasil} The transmission fringes for the Infrasil 302 window. The black curve shows fringes in the collimated beam with $\pm$5\% amplitude. The blue and green curves show the measured fringes with the window in a converging or diverging F/ 8 beam multiplied by a factor of 11 to match amplitudes. This factor shows the amplitude decreases strongly without a fringe period change in the diverging beam. The green curve shows the Infrasil mounted before the slit while blue shows the Infrail mounted after the slit. }
\vspace{-4mm}
\end{wrapfigure}

The system uses photo-multiplier tubes (PMT) to cover a range of wavelengths from the UV to NIR. For our nominal integration times, the standard PMT delivers a measured statistical signal to noise ratio around 1000. The system noise level is dominated by systematic errors for integration times longer than 0.1 seconds through drifts in the baseline count levels. The baseline count rate was measured to vary by roughly 10\% in 200 minutes with a mostly linear trend, however some erratic behavior of the bias offset was observed.  We typically complete a measurement in a few minutes with a baseline scan measured before and after.

\subsection{Infrasil Window Fringes in Collimated \& F/ 8 Beams}

The first sample tested is a window of 1.1335 mm thickness of Heraeus Infrasil 302. The physical thickness was measured by a Heidenheim MT 60M metrology system with $\sim$0.5 $\mu$m thickness accuracy.  Meadowlark measured the transmitted wavefront error (TWE) at 632.8 nm wavelength to be 0.021 waves peak-to-valley (P-V) over an aperture of 12 mm diameter. The beam deviation through the Infrasil was measured to be 0.26 arc-seconds.  

This window was illuminated 10 mm beam footprint when mounted in the collimated beam ahead of the 50 mm focal length lens. As reported in H17\cite{Harrington:2017jh}, the nominal data sets recover the predicted fringe period at moderate spectral sampling of 0.080 nm step per measurement. In data sets presented here, we increased the spectral sampling to cover smaller bandpass at spectral steps of 0.002 nm giving an effective sampling at $\frac{\lambda}{\delta\lambda}$ of about 315,000.  We thus sample the 16 pm full width half maximum (FWHM) instrument profile with about 8 points giving us hundreds to thousands of measurements over a few fringe cycles. 

We tested this Infrasil 302 window mounted in a converging beam before the slit, in the diverging beam after the slit and also in the collimated beam before the focusing lens.  Figure \ref{fig:transmission_fringes_Infrasil} shows the collimated beam fringe amplitude is 11 times larger than the fringes detected in the converging and diverging beams. The Infrasil data sheet from the manufacturer (Heraeus) gives a refractive index of 1.457 at 632.8 nm wavelength. We compute the period as $\frac{\lambda^2}{2dn}$ = 0.120 nm  for both collimated and F/ 8 beams. A Fourier analysis of the fringe data shows that the fringe period does not significantly change, as expected. Given that the slit is a spatial filter at the focal plane and the beam is F/ 5 before hitting the internal spectrograph collimating mirror stop, comparing these measurements allows us to rule out significant impact of the slit spatial filtering. When mounted before the slit, the footprint was about 8 mm diameter as the optic was 10 mm down-stream of the focusing lens. When mounted behind the slit, the footprint was of similar size.

The theoretical calculation gives minimum and maximum transmissions of 86.25\% and 99.99\% for a fringe amplitude of about 13.75\% at infinite spectral resolving power.  In this data set, we achieve amplitudes of roughly 10\%. We used our Berreman calculus Python code to compute fringes at a spectral sampling of $\frac{\lambda}{\delta\lambda}$ of 500,000. We then convolved the resulting Berreman fringes with Gaussian profiles of the appropriate full width half maximum (FWHM) to simulate reduced spectral resolving power. At reduced resolving power of R=40,000, we see a reduction of the fringe amplitude to 10\% peak to peak, matching our measurements. For this window, the optical thickness seen by the nominal back-reflected interfering wave is 2dn/$\lambda \sim$5250 waves of path.  For an F/ 8 beam, the incoming ray in air sees an incidence angle of $\tan^{-1}(1/2F)$ = 3.7$^\circ$. refracted marginal ray propagates at an angle of 2.5$^\circ$ in the medium and would see an optical thickness difference of roughly 5.0 waves.

\subsection{Quartz Crystal Retarder: Measured Fringes in Collimated \& F/ 8 Beams}

\begin{wrapfigure}{r}{0.61\textwidth}
\centering
\vspace{-5mm}
\begin{tabular}{c} 
\hbox{
\hspace{-1.0em}
\includegraphics[height=6.9cm, angle=0]{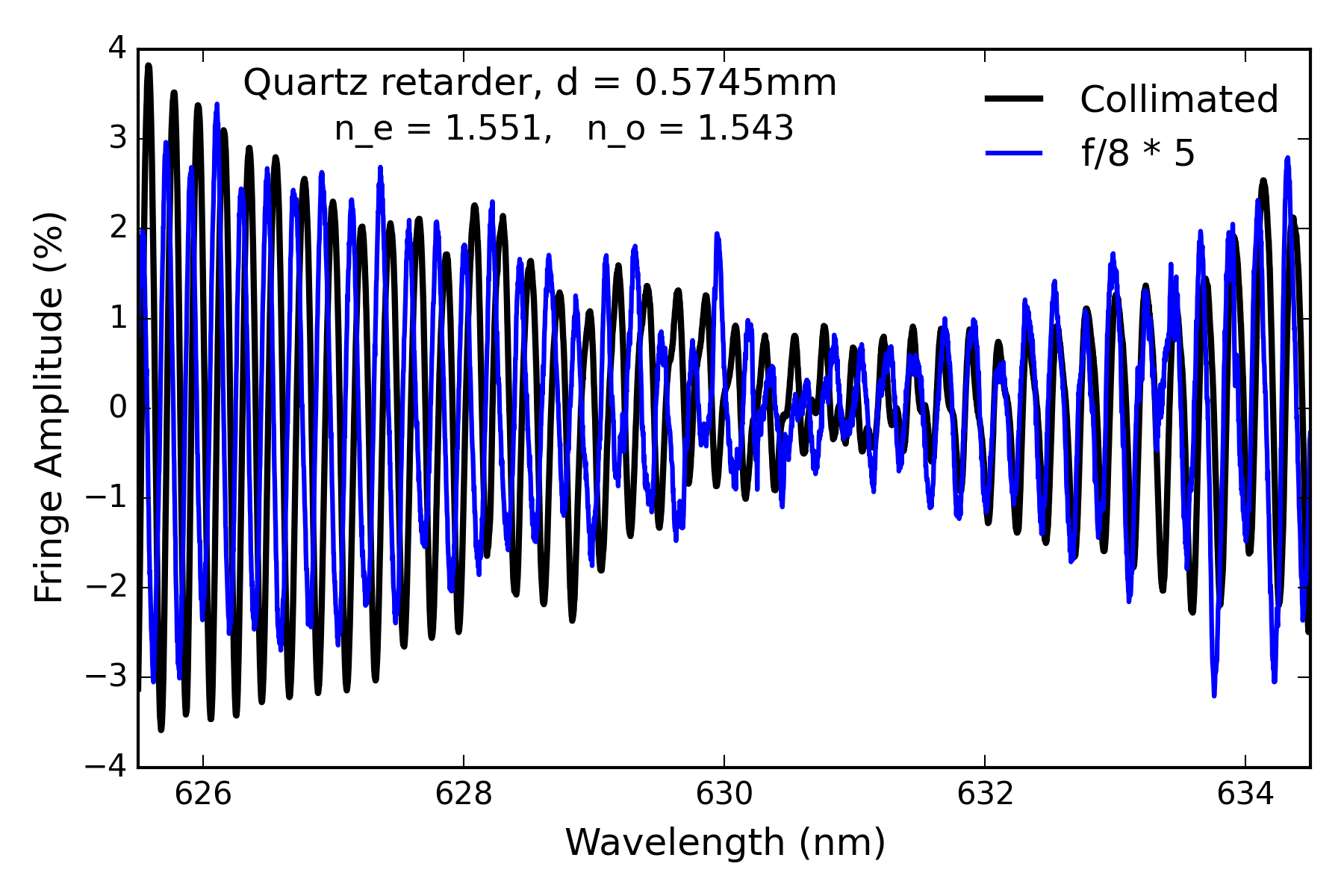}
}
\end{tabular}
\caption[] 
{ \label{fig:quartz_fringes} The fringes measurd in the Quartz retarder. Black shows measurements in the collimated beam. Blue shows measurements the F/ 8 beam multiplied by 5 to roughly match the collimated beam fringe amplitude. See text for details.   }
\vspace{-3mm}
 \end{wrapfigure}

A quartz crystal retarder sample was measured to have 575.4 $\mu$m physical thickness $\pm$0.5 $\mu$m. The retarder was oriented with fast and slow axes at 45$^\circ$ orientation to the grating rulings and mirror fold orientations. The retarder was mounted in a collimated beam as well as in the F/ 8 beam mounted ahead of the slit. The baseline scans without the sample in the beam were also recorded. The TWE for the crystal is 0.034 waves peak to peak at 632.8 nm wavelength over a clear aperture of 12 mm diameter. Beam deviation was measured to be 0.21 arc-seconds. 

A Fourier analysis of the collimated and F/ 8 data set give nearly identical periods of 0.22 nm.  The power spectrum is dominated by a single somewhat broad peak at 0.22 nm without any other significant features in the 0.05 nm to 0.5 nm period range.  The theoretical period of $\lambda^2$/2dn is 0.2226 nm for the extraordinary beam of refractive index 1.551 and 0.2239 nm for the ordinary beam at a refractive index of 1.543. We created fringe predictions with our Berreman code similar to H17\cite{Harrington:2017jh}. The fringes were derived at spectral sampling of $\delta\lambda/\lambda$ = 500,000.  The theoretical fringe spectral period for this optic was 1.8x longer than the Infrasil sample. As such, the resolving power of the spectrograph has less influence on the detected fringe amplitude. We use this quartz crystal retarder for the analysis of the F/ number dependence. 

The higher refractive index Quartz crystal has transmission ranging from 99.99\% to 81.8\% for a fringe amplitude around 18\%. When convolving this theoretical curve with a Gaussian profile at resolving power of R=40,000 the amplitude only decreases to about 15\%. In addition, the interference between the extraordinary and ordinary beams gives rise to a much slower amplitude modulation at a period of roughly 35 nm at 630 nm wavelength.  The measurements show the minimum fringe amplitude clearly around 631 nm wavelength in Figure \ref{fig:quartz_fringes} with fringe amplitudes rising quickly to shorter and longer wavelengths. This is very similar to our Berreman calculation.

\begin{wrapfigure}{l}{0.60\textwidth}
\centering
\vspace{-3mm}
\begin{tabular}{c} 
\hbox{
\hspace{-1.0em}
\includegraphics[height=6.6cm, angle=0]{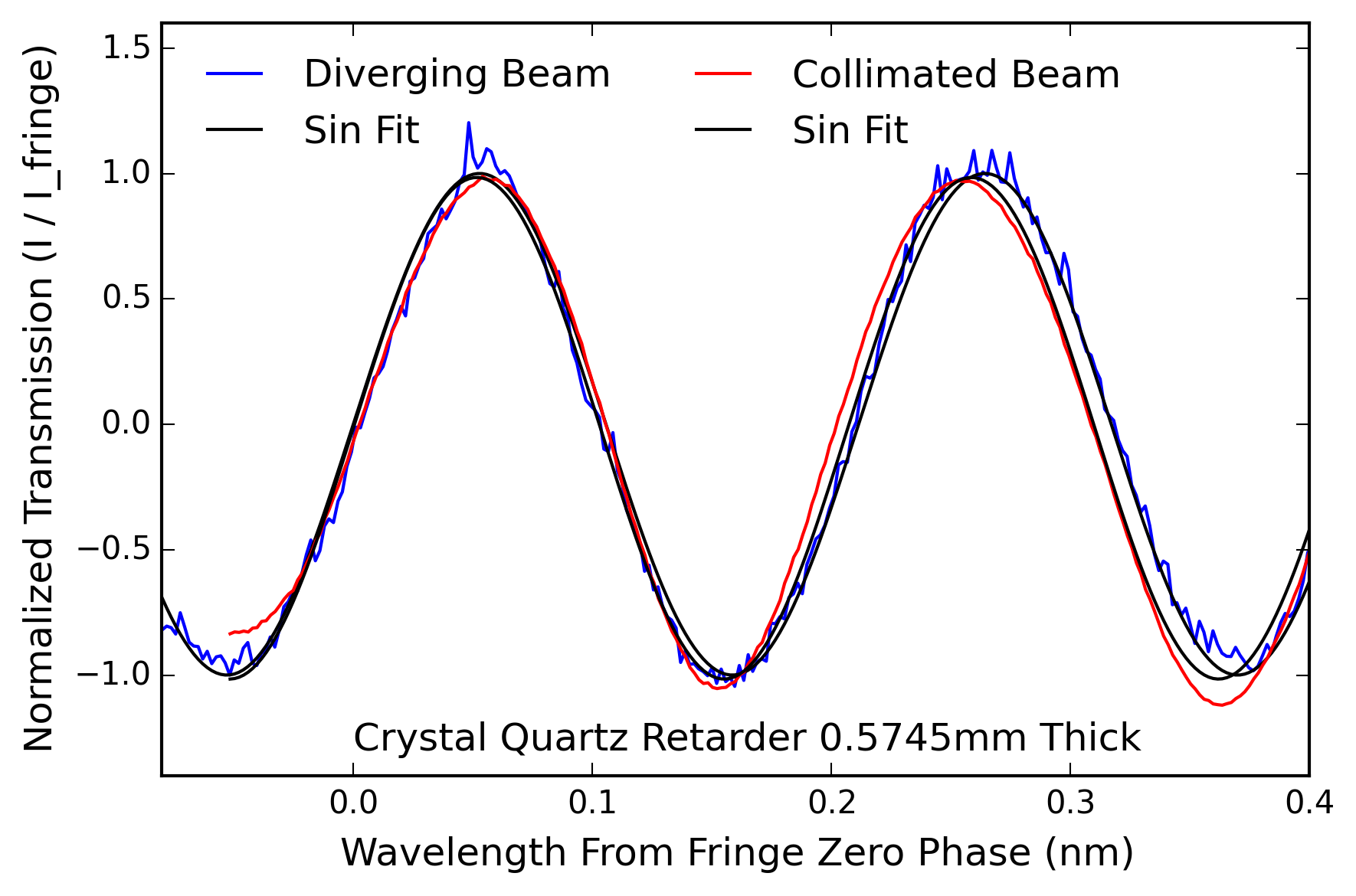}
}
\end{tabular}
\caption[] 
{ \label{fig:quartz_fringes_fit_sinfunc} The fringes measured in the Quartz crystal retarder normalized and adjusted with sinusoidal fit parameters. Black shows the sine function fits. Blue shows the data with the crystal in the diverging F/ 8 beam after the slit.  Red shows the data with the crystal in the collimated beam. As all curves overlap and are difficult to distinguish, we consider the sin function fits successful. See text for details.   }
\vspace{-5mm}
\end{wrapfigure}

For this crystal data set, we measured with a 0.002 nm spectral step size from 614.0 nm to 614.5 nm to cover 2 fringes but keep the measurement time to less than 2 minutes. This is significantly faster measurement time than the data sets on the Infrasil window in Figure \ref{fig:transmission_fringes_Infrasil}. The exact value of the baseline scan was somewhat more erratic for this data set even though the measurement time was significantly shorter. Baseline values changed at amplitudes up to several percent even for immediately repeated measurements without any perturbation of the system. The spectral shape of the baseline was much more stable. Given this uncertainty, we determined fringe amplitudes and phases by fitting sinuosoidal functions allowing for a constant offset.  

An example of a data set comparing collimated to diverging fringes on our crystal quartz sample is seen in Figure \ref{fig:quartz_fringes_fit_sinfunc}. The curves have been fit by sinuosoidal functions then normalized and centered using the fit parameters. The red curve shows the data with the crystal quartz retarder in the collimated beam.  The fit fringe amplitude was 16.1\% peak-to-peak. Blue shows the data with the crystal retarder in the diverging beam with a 1.9\% peak-to-peak fringe amplitude. The ratio of fringe amplitudes computed using the sin-fit parameters is 8.4. Significantly more statistical noise is seen in the blue curve of Figure \ref{fig:quartz_fringes_fit_sinfunc} as the curves are normalized by the fit fringe amplitude. The periods are essentially identical again showing that converging beams do not impact the period calculation.  We see excellent agreement in both curves in Figure \ref{fig:quartz_fringes_fit_sinfunc}. We can conclude the SPEX setup is sufficient to measure fringe period, amplitude and to assess impact of diverging beams on fringe properties. Some slight differences are seen between the Infrasil window fringe data set of Figure \ref{fig:transmission_fringes_Infrasil} and the thinner quartz crystal data set of Figure \ref{fig:quartz_fringes_fit_sinfunc} after normalization by a sin-fit to data with a larger fringe amplitude. Differences arise from the increased spectral sampling, decreased measurement time, higher cadence of baseline scans and the 1.8x slower spectral fringe period of the crystal sample.

\begin{wrapfigure}{r}{0.65\textwidth}
\centering
\vspace{-1mm}
\begin{tabular}{c} 
\hbox{
\hspace{-1.0em}
\includegraphics[height=7.15cm, angle=0]{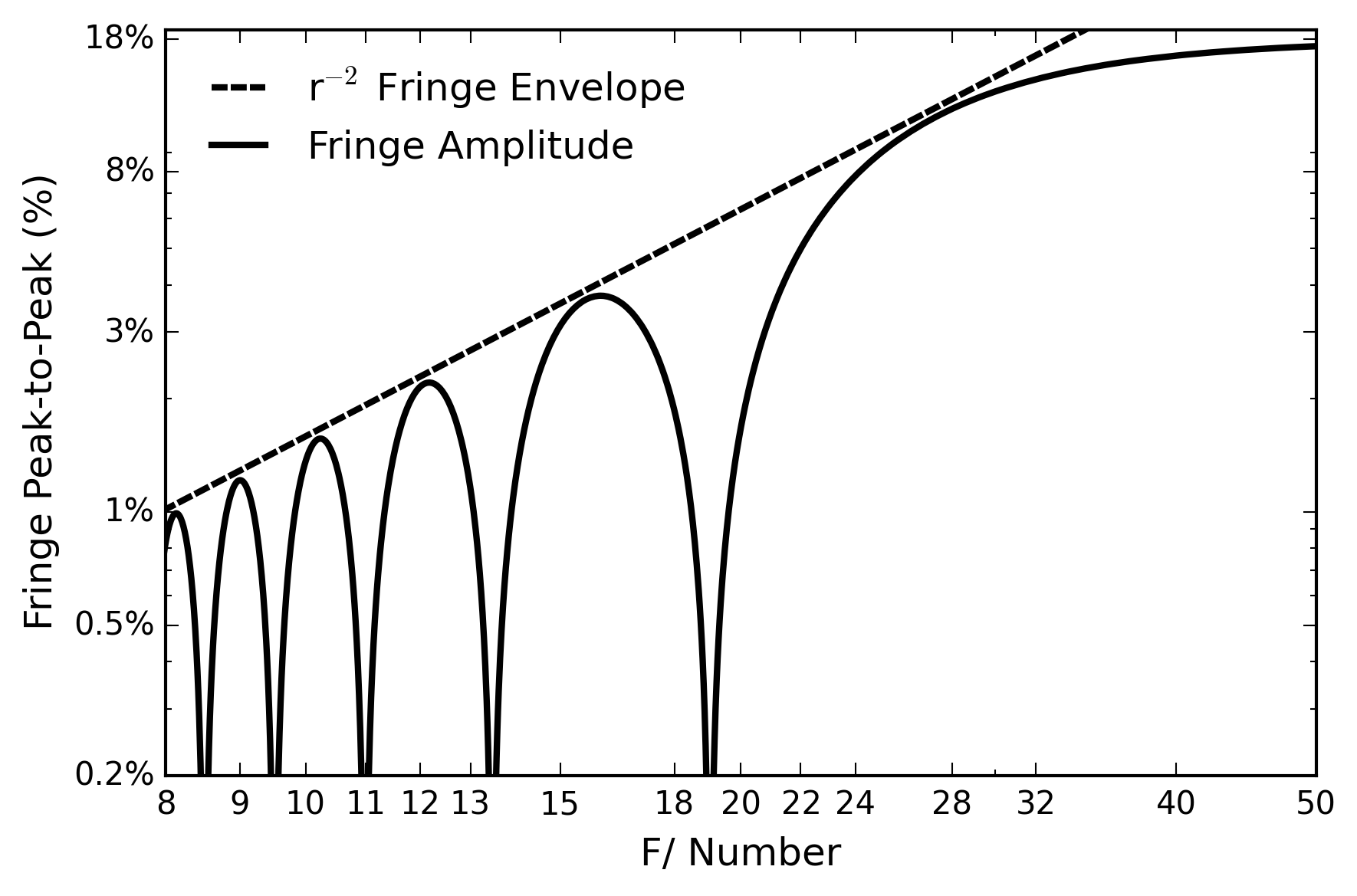}
}
\end{tabular}
\caption[Waves OPD] 
{ \label{fig:transmission_maxmin_quartz} The peak-to-peak fringe amplitude about the average transmission as a function of F/ number for the 0.5745 mm crystal quartz retarder at 614 nm wavelength. The $r^{-2}$ amplitude envelope begins around half peak fringe amplitude.}
\vspace{-4mm}
\end{wrapfigure}

Figure \ref{fig:transmission_maxmin_quartz} shows the deviation from the nominal average transmission as the beam F/ number is changed. The theoretical peak to peak fringe amplitude is roughly 18\%.  The quartz retarder at 614 nm wavelength shows a $\pm$1.7\% fringe at the first maximum near F/ 16. This peak would occur when integrating over the aperture from the center out to an integer multiple of quarter-wave interference path corresponding to constructive interference on the outer annulus of the aperture. This 3.5\% fringe is reduced by a factor of five from the collimated beam 18\% fringe amplitude. The second maximum is near F/ 12 with $\pm$1\% fringe, corresponding to an amplitude reduction factor of nine when averaging over more than two waves of aperture interference. 

We were suspicious that the residual angular divergence in the input light source may have been reducing spatial fringe amplitudes.  As a further test of the SPEX system, we changed the fiber and feed optics to ensure that any small angular divergence from our light source did not impact measured fringe amplitudes or periods.  We changed from a 200 $\mu$m diameter fiber to a 50 $\mu$m diameter fiber.

With the 50mm focal length collimator and four times smaller fiber diameter, the field divergence in the beam decreased by a factor of 13.3. The field divergence in the beam is now $\pm$ 0.03$^\circ$ corresponding to the outer diameter of the fiber. When this 50 $\mu$m core fiber is used, the crystal quartz measurement has a 16.5\% fringe amplitude peak to peak. This represents a slight increase from the 16.1\% amplitude found with the 200 $\mu$m core fiber and shows that the impact of light source field divergence. In the diverging beam, the fringe amplitude significantly increased to 3.7\% peak to peak. Changing the light source roughly doubled the amplitude of the fringes detected compared to using the 200 $\mu$m core diameter fiber.

We performed a simple experiment to manually change the system F/ number by closing the iris in the collimated beam ahead of the lens focusing the beam on the spectrograph entrance slit. The nominal setting with the iris open gives a 100 mm round beam on the spectrograph collimating mirror.  The full collimating mirror aperture is a 110 mm square that is fully illuminated without the iris. We would sequentially close the iris and manually measure the beam diameter on the collimating mirror mask.  

Fringe measurements were performed with the crystal retarder mounted both in the collimated beam and again in the diverging beam.  Occasional repeated measurements were performed with slight adjustments to the optical alignment to verify that the detected fringe amplitude was not sensitive to optical alignment or system stability. 

\begin{wrapfigure}{r}{0.64\textwidth}
\centering
\vspace{-2mm}
\begin{tabular}{c} 
\hbox{
\hspace{-0.3em}
\includegraphics[height=7.15cm, angle=0]{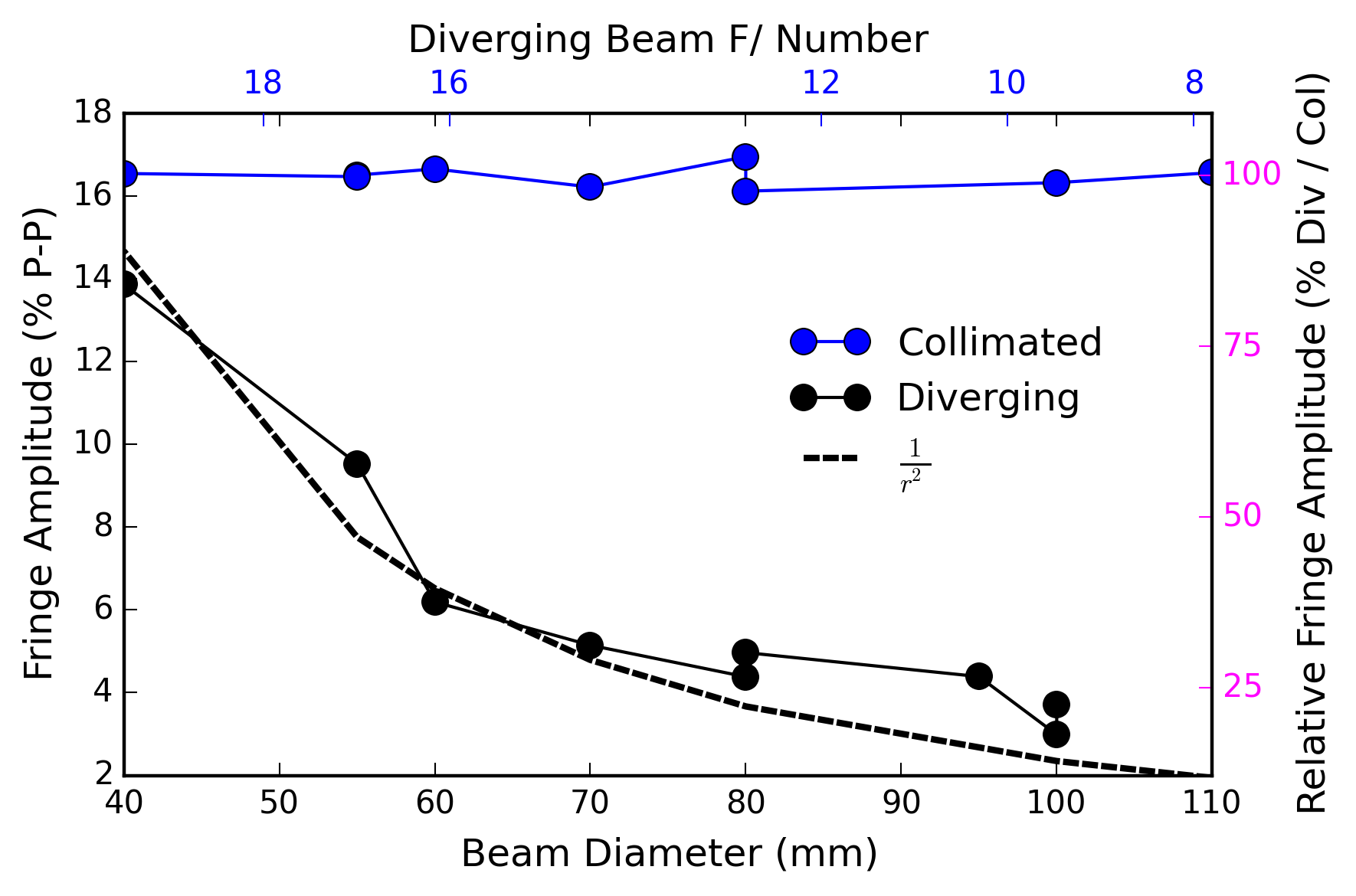}
}
\end{tabular}
\caption[] 
{ \label{fig:quartz_fringes_fnumber} The fringe amplitudes measured in the Quartz crystal retarder as a function of F/ number. Black shows the fringe amplitudes with the crystal quartz retarder mounted in the diverging beam. Dashed black shows a r$^{-2}$ law scaled to 16.25\% fringe at 38 mm beam diameter.  Blue shows the fringe amplitude in the same optical setup but with the crystal quartz retarder moved to the collimated beam before the slit. Blue shows the change in beam diameter did not degrade fringe amplitude in the collimated beam. The left hand Y-axis shows the detected fringe amplitude in percent. The fringe amplitude reduction between blue and black curves as seen in the right hand Y-axis. The bottom X-axis shows the manually measured beam diameter on the collimating mirror.  The top X-axis shows a conversion of that beam diameter to F/ number within the spectrograph. See text for details.   }
\vspace{-1mm}
\end{wrapfigure}

Figure \ref{fig:quartz_fringes_fnumber} shows the compiled results of this data set. Blue shows the fringe amplitudes detected with the crystal retarder mounted in the collimated beam.  Fringe amplitudes remained near the nominal 16.5\% fringe amplitude to within a small fraction of a percent as the iris was closed and the beam diameter reduced from 100 mm to 40 mm. The 16.5\% fringe amplitude is detected in all cases, showing the system resolution and optical alignment is stable upon changing the beam diameter with the iris. The black curve shows the fringe amplitudes detected when the retarder was mounted in the diverging beam.  The bottom X- axis shows the manually measured beam diameter while the top X-axis shows the corresponding spectrograph beam F/ number.  Diameters ranged from 40 mm for the slowest beam of F/ 20 to 110 mm for the fastest beam of F/ 8.  The collimating mirror aperture is a square at 110 mm per side.  The iris was set to 100 mm at the widest, corresponding to roughly F/ 9.

The right hand Y-axis of Figure \ref{fig:quartz_fringes_fnumber} shows how the fringes were reduced in amplitude at the corresponding F/ number and beam diameter.  For instance, with the beam at 100 mm diameter and the system at F/ 9, the fringes were roughly 3.5\% amplitude which is only 20\% of the nominal 16.5\% fringe amplitude collimated. For the smallest beam diameter of 40 mm near F/ 20, the fringes detected with the crystal in both collimated and diverging beams are nearly the same with an amplitude of 80\% that of the collimated case.  We note that the F/ 9 limit here derived a fringe reduction factor of roughly 5 while the experiments above without an iris in the older setup derived a factor of 8.4.  Given the potential issues with spectrograph alignment and manual optical positioning, this magnitude of uncertainty may be expected in our simple experiments.

At 614 nm wavelength, this 574.5 $\mu$m thick quartz crystal has a refractive index of n=1.552 and the back-reflected chief ray sees 2905 waves of optical path.  At F/ 10, the marginal ray is traveling at an incidence angle of 2.9$^\circ$ and will see 1.5 waves of additional optical path compared to the chief ray. At F/ 17.4, the marginal ray would see exactly half a wave of path difference and destructive interference. This also corresponds to the F/ number where the measured fringe amplitude drops significantly in Figure \ref{fig:quartz_fringes_fnumber}. Our system uses a multi-mode fiber-coupled plasma light source with imperfect coherence and mild filed divergence. We do not expect to match the fully coherent predictions for Haidingers fringes with multiple oscillations of constructive and destructive interference.  However, we do recover the significant reduction by a factor of four in detected fringe amplitude when more than half a wave of path variation is seen across a beam footprint. These results were consistent and repeatable with low sensitivity to manual optical alignment. From the simple $r^{-2}$ behavior of Figure \ref{fig:transmission_maxmin_quartz}, we do expect to see a break in the fringe amplitude curve around F/ 30. With our setup, this rapid reduction in fringe amplitude occurs closer to F/ 20.  Given the alignment and potential issues with the double-grating spectrograph, we consider this agreement within the uncertainty of our simple manual experiments. We show additional details of the Meadowlark SPEX system in Appendix \ref{sec:appendix_spex}.

\subsection{Summary of Measured Fringe Amplitude \& Period Dependence on F/ Number}

We have experimentally verified the $r^{-2}$ behavior of measured fringe amplitudes for windows and crystal retarders with high spectral resolving power data in the lab. The measured fringe amplitudes and periods matched the Berreman predictions in a collimated beam. Fringe periods did not change the a converging beam, but fringe amplitudes were reduced by factors of 5 to 10 for the 0.5 mm thick crystal quartz and 1.1 mm thick Infrasil window.  These reduction factors are in agreement with the $r^{-2}$ envelope. Crystal retarders show the expected interference of fringes between ordinary and extraordinary beams, but otherwise the behavior of fringe ampliutde reduction with F/ number matches the $r^{-2}$ envelope.  Next we will take these $r^{-2}$ envelopes and assess the six-crystal DKIST retarder designs over a range of wavelengths and for all fringe periods predicted in the Berreman calculus.

\clearpage

\section{DKIST Retarders: Amplitude Vs Beam F/ Number Predictions}

With this simple $r^{-2}$ envelope for predicting fringe amplitudes as a function of beam F/ number, we can make simple predictions for the fringe amplitudes caused by the various DKIST retarder optics. We note that the various beam splitters and dichroics part of the Adaptive Optics system (WFS-BS1) and the Facility Instrument Distribution Optics (FIDO) are both wedged.  As such, there will be thousands of fringe periods spatially averaged across the clear aperture. We focus this paper on the crystal retarders which are strictly plane parallel optics mounted in converging beams.

\begin{wraptable}{l}{0.67\textwidth}
\vspace{-2mm}
\caption{Beam Properties vs Crystal Thicknesses}
\label{table:fringe_beam_properties}
\centering
\begin{tabular}{l l l l l l l}
\hline\hline
$\lambda$		& 2.1mm		& 4.2mm	& 6.3mm	& 8.4mm	& 10.5mm & 12.6mm	\\
\hline
\hline
393nm		& 16760 		& 33520 	& 50280	& 67039	& 83799	& 100559		\\
\hline
F/ 13			& 5.0			& 10.1	& 15.1	& 20.1	& 25.2 	& 30.2			\\
F/ 26			& 1.3			& 2.5		& 3.8		& 5.0		& 6.3		& 7.6			\\
\hline
525nm		& 12452		& 24904	& 37356	& 49808	& 62260	& 74712			\\
\hline
F/ 13			& 3.8			& 7.6		& 11.4	& 15.2	& 19.0	& 22.8			\\
F/ 26			& 1.0			& 1.9		& 2.9		& 3.8		& 4.8		& 5.7			\\
F/ 62			& 0.2			& 0.3		& 0.5		& 0.7		& 0.8		& 1.0			\\
\hline
854nm		& 7604		& 15209	& 22813	& 30418	& 38022	& 45626			\\
\hline
F/ 13			& 2.4			& 4.7		& 7.1		& 9.4		& 11.7	& 14.1			\\
F/ 26			& 0.6			& 1.2		& 1.8		& 2.4		& 2.9		& 3.5			\\
F/ 62			& 0.1			& 0.2		& 0.3		& 0.4		& 0.5		& 0.6			\\
\hline
1083nm		& 5982		& 11964	& 17947	& 23929	& 29911	& 35893			\\
\hline
F/ 13			& 1.9			& 3.7		& 5.6		& 7.4		& 9.3		& 11.1			\\
F/ 26			& 0.5			& 0.9		& 1.4		& 1.9		& 2.3		& 2.9			\\
F/ 62			& 0.1			& 0.2		& 0.2		& 0.3		& 0.4		& 0.5			\\
\hline
1565nm		& 4121		& 8243	& 12364	& 16486	& 20607	& 24729			\\
\hline
F/ 8			& 3.4			& 6.8		& 10.2	& 13.6	& 17.0	& 20.4			\\
F/ 13			& 1.3			& 2.6		& 3.9		& 5.2		& 6.5		& 7.7			\\
F/ 26			& 0.3			& 0.6		& 1.0		& 1.3		& 1.6		& 1.9			\\			
F/ 62			& 0.1			& 0.1		& 0.2		& 0.2		& 0.3		& 0.3			\\
\hline
3934nm		& 1452		& 2903	& 4355	& 5806	& 7258	& 8709	 		\\		  
\hline
F/ 18			& 0.3			& 0.6		& 0.9		& 1.2		& 1.5		& 1.8			\\
\hline
\hline
\end{tabular}
Optical path variation in waves for the chief and marginal ray experiencing back-reflection when propagating through successive numbers of 2.1mm thick crystals. Each column corresponds to increasing numbers of crystals from one (2.1 mm total thickness) to six (12.6 mm total thickness). Rows listing a wavelength also show the chief ray OPD.  Rows listing an F/ number show the marginal ray path difference between chief and marginal rays. An example, at 525 nm wavelength, the back-reflected chief ray sees 12,452 waves of optical path when propagating through a single 2.1 mm thick crystal while the marginal ray for an F/ 13 beam sees an additional 3.8 waves of OPD.
\vspace{-3mm}
\end{wraptable}

In Table \ref{table:fringe_beam_properties}, we show the optical thickness of the various crystal retarder interferrence paths. We also compute the associated physical thickness difference in waves (2dn/$\lambda \cos(\theta/n)$) seen by the marginal ray.  For the ViSP instrument, the calibration retarder is crystal Quartz working in an F/ 13 beam while the modulator is at F/ 26.  The marginal ray is at an incidence angle of 2.20$^\circ$ in air for F/ 13 and at 1.10$^\circ$ for the F/ 26 beam.  For the DL-NIRSP at F/ 8 , the marginal ray is at a 3.58$^\circ$ incidence angle while at 0.46$^\circ$ for the F/ 62 beam.

The number of waves path difference for the marginal ray is a proxy for the number of interference cycles across the clear aperture of the illuminated optic. The wavelength and F/ number dominate the behavior with some slight dependence on refractive index variation with wavelength.  Using the standard formula from CVI, we get refractive indices of 1.568 at a wavelength of 393 nm falling to 1.546 at a wavelength of 854nm and 1.536 at 1565nm wavelength. For the MgF$_2$ retarder, we derive a refractive index of 1.3596 at wavelength 3934nm intended for Cryo-NIRSP observations of the Si IX spectral line.  The modulating retarder for Cryo-NIRSP is in an F/ 18 converging beam mounted upstream of the spectrograph entrance slit.  Table \ref{table:fringe_beam_properties} shows the shortest period spectral fringe will see roughly 2 waves of interference over the converging beam aperture. The longest period fringe corresponding to a single crystal however will only see a small fraction of a wave.  Thankfully, the individual crystals are at refractive index 1.35 and the oil layers between crystals have an index of 1.3, significantly reducing this spectral component of the fringe.

Simulations for the fringe amplitude at specific wavelengths can be simply computed using the Berreman scripts or the {\it equal inclination fringe} equations for isotropic materials. An aperture integral converts the predicted intensity for each incidence angle to the total transmission for a footprint. For this simple example, we do not compute the fully birefringent fringe spatially across a retarder crystal. However, this is straightforward in the Berreman formalism.

\begin{wrapfigure}{r}{0.63\textwidth}
\centering
\vspace{-5mm}
\begin{tabular}{c} 
\hbox{
\hspace{-0.5em}
\includegraphics[height=6.9cm, angle=0]{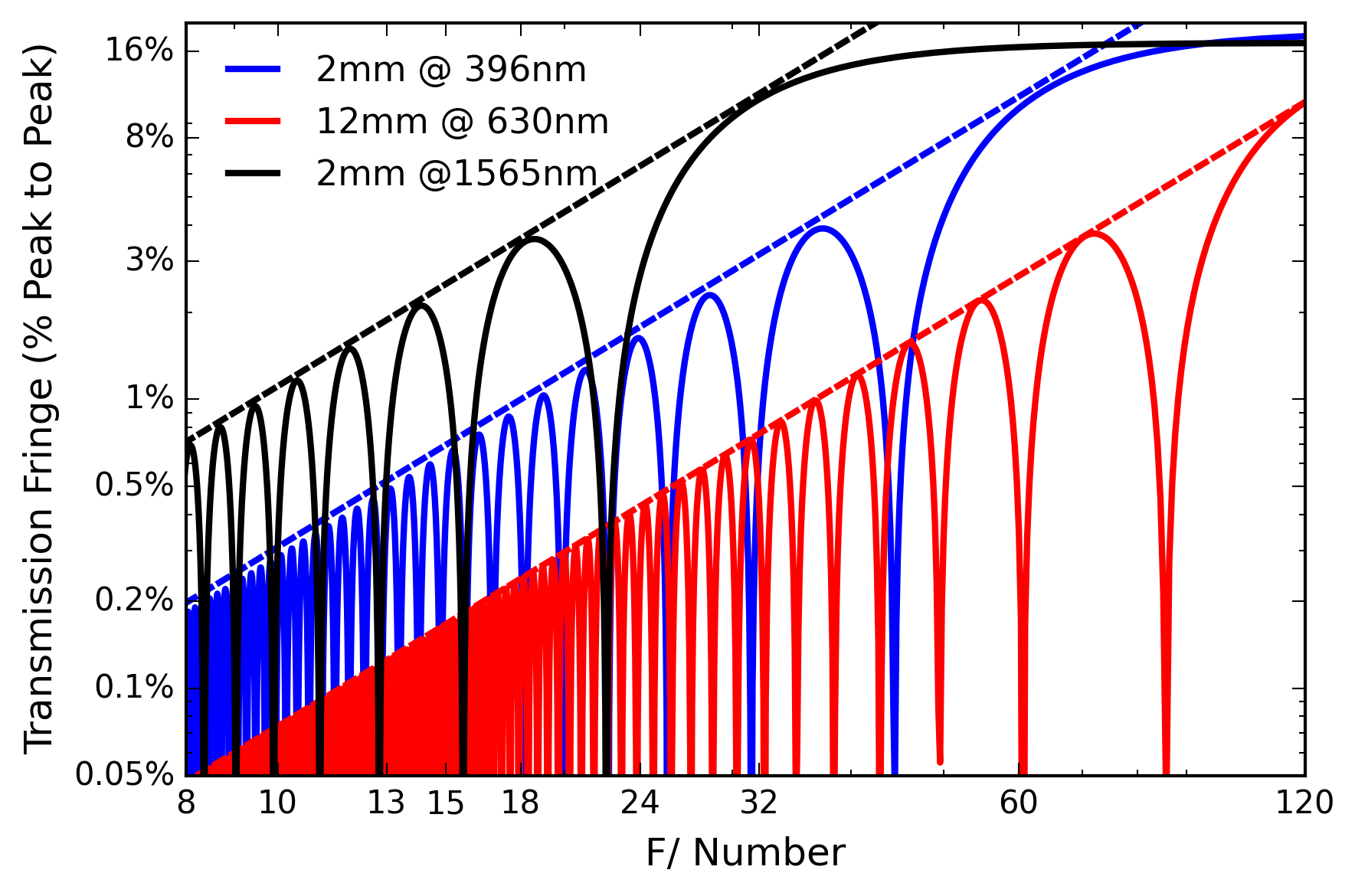}
}
\end{tabular}
\caption[] 
{ \label{fig:fringes_with_fnum_thickness_quartz_mgf2} The $\pm$transmission fringe amplitudes as functions of beam F/ number for uncoated crystals.  Blue shows 2 mm thickness of crystal quartz at a wavelength of 396 nm.  Black shows quartz at a wavelength of 1565 nm.  Red shows the full 12 mm thick six-crystal stack at 630 nm wavelength. The $r^{-2}$ envelopes are scaled for each curve.}
\vspace{-3mm}
\end{wrapfigure}

Figure \ref{fig:fringes_with_fnum_thickness_quartz_mgf2} shows examples of how the fringe amplitude depends on wavelength, F/ number and thickness. The deviation from the average un-fringed calculation is shown in $\pm$ transmission. Colors show different wavelengths and crystal thicknesses. Solid lines show the spectral fringe. Dashed lines show the r$^{-2}$ fringe amplitude decrease behavior. The blue curves show a 2 mm thickness of an isotropic material of index 1.55 corresponding to crystal quartz at 396 nm wavelength corresponding to the shortest wavelength for the ViSP instrument. The ViSP modulator sees a diverging F/ 26 beam and should see fringe amplitudes less than $\pm$1\% as compared to the over $\pm$8\% collimated fringe. The air-crystal interface through the six-crystal optic produces the largest amplitude fringes in the Berreman calculus, but the aperture average would reduce this spectral component to $\pm$0.15\% per the dashed blue line.

Black shows quartz but at a wavelength of 1565 nm appropriate for the science wavelength DL-NIRSP instrument camera arm. For the F/ 62 configuration, the beam is essentially collimated with minimal fringe amplitude reduction and peak amplitudes above $\pm$8\%.   The F/ 24 configuration would reduce the fringe magnitudes to below $\pm$3\%.

Red shows the full 12 mm thickness of the six-crystal stack at a wavelength of 630 nm as intended for ViSP, VTF and DL-NIRSP instruments.  The full crystal stack is six times thicker.  The fringes corresponding to this interface sees significant reduction of amplitude. The calibration and modulation retarders all would see $\pm$0.3\% instead of $\pm$8\%, a reduction of roughly 27 times. Anti-reflection coatings further reduce the fringe amplitude. 

The results show that simple scaling relations apply. From beams of F/ 28 to F/ 10 the peak fringe amplitudes decrease by roughly a factor of three.  As seen in Table \ref{table:fringe_beam_properties}, most optics in the converging beams see a few to several waves of optical path variation across the aperture.  This gives fringe amplitude reductions that follow the linear trends of Figure \ref{fig:fringes_with_fnum_thickness_quartz_mgf2}. In the six-crystal modulator however, this amplitude prediction is modified by the oil at refractive index 1.3 reducing the magnitude of the back reflection as in H17\cite{Harrington:2017jh}.

\begin{wraptable}{l}{0.18\textwidth}
\vspace{-3mm}
\caption{OPD}
\label{table:fringe_amplitude_reduction}
\centering
\begin{tabular}{c c}
\hline
\hline
OPD		& Fringe	\\
C-M		& Factor	\\
\hline
\hline
0.7		& 2		\\
1.5		& 4 		\\
2.5		& 8 		\\
5.5		& 16		\\
10.5		& 32		\\
20.5		& 64		\\
\end{tabular}
\vspace{-4mm}
\end{wraptable}

We compile a rough estimate of the fringe amplitude reduction using the $r^{-2}$ envelope in Table \ref{table:fringe_amplitude_reduction}. The left column shows the optical path difference between chief and marginal rays.  The right column shows the rough estimate of the fringe reduction factor for a single window or crystal. These rough estimates are simply rounded to factors of two where the circular clear aperture averages of Figure \ref{fig:fringes_with_fnum_thickness_quartz_mgf2} would have maxima. The table shows that a few waves of marginal ray optical path difference compared to the chief ray can reduce fringes by close to one order of magnitude. However, getting two orders of magnitude fringe reduction to levels below DKIST sensitivity, would require at least a few tens of waves and unrealistically thick optics. We note that this rough estimate is not rigorously applicable to many-crystal optics as the beam overlap, incidence angles and phase relationship between all the many internal reflections is not considered.

\begin{figure}[htbp]
\begin{center}
\vspace{-1mm}
\includegraphics[width=0.98\linewidth, angle=0]{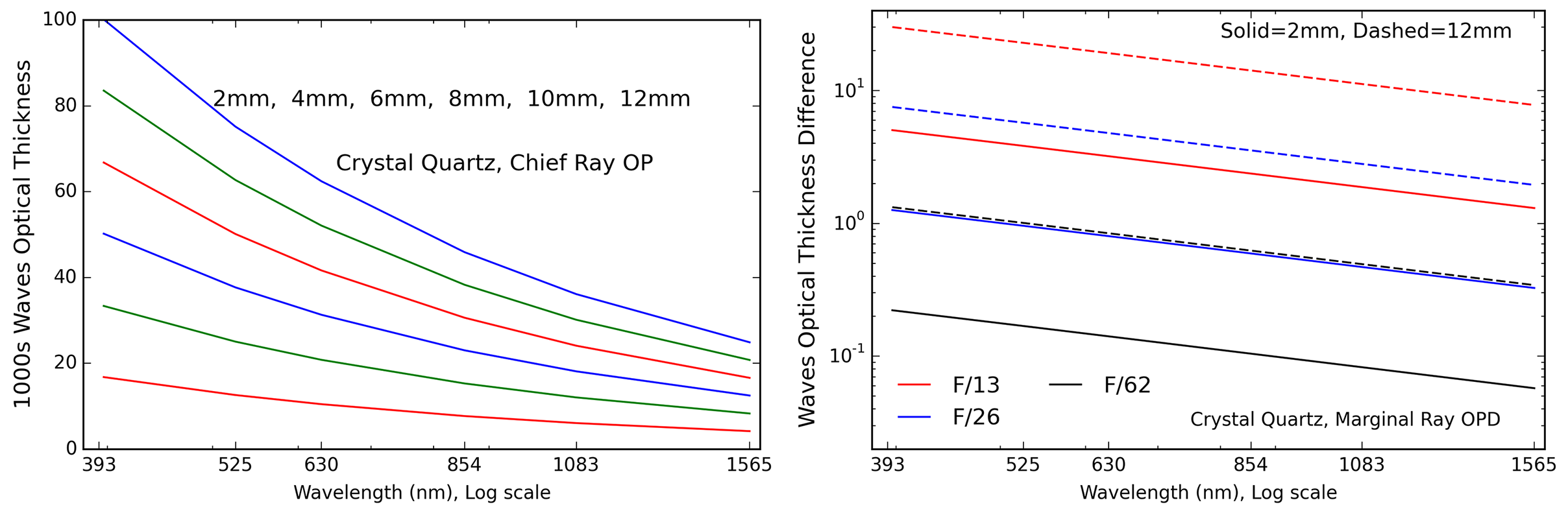}
\caption{The left panel shows the optical path seen by the interfering beam (2dn/$\lambda$) through the DKIST retarder crystals as a function of wavelength. The right panel shows the path difference between the chief and marginal rays when propagating through the DKIST retarders at varying F/ number. See text for details.}
\label{fig:fringe_optical_path}
\vspace{-2mm}
\end{center}
\end{figure}

To synthesize these results for easy application to retarder design and DKIST optical configurations, we compute fringe properties as functions of beam F/ number and wavelength. Figure \ref{fig:fringe_optical_path} shows the optical path as a function of wavelength for the DKIST retarders. The right plot of Figure \ref{fig:fringe_optical_path} shows the difference between chief and marginal ray optical paths for a few F/ numbers used for the DKIST retarders. For the six-crystal retarders at shorter DKIST wavelengths, the back-reflected optical path is close to 20,000 waves for the 2 mm crystal and over 100,000 waves for the entire crystal stack. However, when assessing amplitudes of which fringe spectral components at which F/ number and wavelength, we need many waves of interference over the aperture to significantly reduce the measured fringe amplitude.

The right hand plot of Figure \ref{fig:fringe_optical_path} shows that significant amplitude reduction is expected only the shortest wavelengths and shortest period spectral components for beams faster than F/ 30.  Thus for DKIST we expect to see a higher relative amplitude for the longer period spectral fringe components in steeper converinging beams.  This fringe period dependent amplitude reduction factor can now be coupled with the Berreman predictions from H17\cite{Harrington:2017jh} using anti-reflection coatings to assess what fringe components will be present for the various DKIST instruments at specific wavelengths with a calibration retarder at F/ 13 and modulating retarders from F/ 8 to F/ 62.

\subsection{Wedged Optics: DKIST Beam Splitters}

Similar calculations across the clear aperture can be made for wedged optics. The DKIST beam-splitter train includes a permanently mounted beam-splitter for the adaptive optics beam feed. All AO-assisted polarimetric instruments see this optic in transmission including ViSP, VTF and DL-NIRSP.  The Facility Instrument Distribution Optics (FIDO) dichroic beam splitters are interchangeable and have a variety of coatings to allow combinations of wavelengths to reach all instruments for simultaneous multi-wavelength use. All the beam-splitter designs include wedge of 0.5$^\circ$ in matched pairs. The collimated beam in coud\'{e} has a diameter of nearly 290 mm depending on the exact optical mounting station. As opposed to circular fringe patterns across the clear aperture, we will see fringes corresponding to tilted planes. The beam is collimated but the wedged optic introduces a tilt to the back-reflected beam. A 0.5$^\circ$ tilt over a 290 mm aperture corresponds to a 2.53 mm run-out over the clear aperture.  Computing the optical interferrence path 2dn/$\lambda$ gives a parametric equation as 7,500 waves scaled by the wavelength in microns. At 400 nm wavelength, we see 20,000 cycles of fringe variation across the clear aperture.  At 2000 nm wavelength, this drops to 3,800 waves.  Given the interference is roughly two orders of magnitude larger than for the crystal retarders, we can reasonably neglect fringe considerations from these optics from the polarization plans for DKIST.

\subsection{Fringe Amplitude Reduction Prediction for DKIST Calibration Use Cases}

Here we apply the rough amplitude estimates to the DKIST super achromatic calibration retarders use cases. The retarders are used in the converging F/ 13 beam near Gregorian focus to cover many wavelengths simultaneously. The quartz retarders are designed for wavelengths as long as 2500 nm while the MgF$_2$ based retarders cover wavelengths from 2500 nm to 5000 nm. 

\begin{wraptable}{l}{0.45\textwidth}
\vspace{-0mm}
\caption{F/ 13 Fringe Reduction w/ d \& $\lambda$}
\label{table:fringe_reduction_wavelength}
\centering
\begin{tabular}{l r r r r r r}
\hline\hline
$\lambda$		& 2.1			& 4.2		& 6.3		& 8.4		& 10.5	 & 12.6		\\
nm			& mm		& mm	& mm	& mm	& mm 	& mm	\\
\hline
\hline
393			& 16 			& 32		& $>$32	& 64		& $>$64	& $>$64		\\
525			& $>$8		& 16		& 32		& $>$32	& 64		& 64			\\
854			& 8			& 16		& $>$16	& 32		& 32		& $>$32		\\
1083			& 4			& $>$8	& 16		& $>$16	& 32		& 32			\\
1565			& 4			& 8		& $>$8	& 16		& 16		& $>$16		\\
\hline
3934			& 0			& 2		& 2		& 4		& 4		& 4	 		\\		  
\hline
\hline
\end{tabular}
\vspace{-3mm}
\end{wraptable}

In Table \ref{table:fringe_reduction_wavelength} we show the rough estimates of fringe amplitude reduction factor from the r$^{-2}$ envelope for the various crystal thicknesses ($d$) producing the dominant fringe periods in the calibration retarder for some common solar spectral observation wavelengths. The left column shows the wavelength of observation in nm. Subsequent columns take the marginal ray optical path difference from Table \ref{table:fringe_beam_properties} for an F/ 13 beam and roughly estimate the amplitude reduction of this fringe period component in the beam. We can see that for the shortest wavelengths where the Visible Spectropolarimeter (ViSP) instrument might calibrate the 396 nm solar spectral line, we would expect fringe magnitudes from the calibration retarder to be 16 to over 100 times smaller than the collimated Berreman prediction.  The marginal ray at this wavelength sees 30 waves of path difference compared to the chief ray for the fastest spectral fringe period produced by the full 12.6 mm crystal thickness. Thus the shortest period fringes are expected to be at quite low amplitudes. Conversely, the MgF$_2$ calibration retarder working with Cryo-NIRSP at 3934 nm wavelength would see fringes of magnitude quite similar to the collimated Berreman prediction with a mild factor of few reduction for the shortest period fringes.

\subsection{Summary of DKIST Fringe Amplitude Predictions}

We have shown in this section examples of how the optical properties of the DKIST calibration retarders relate to expected fringe amplitudes at F/ 13 Gregorian focus and for the modulating retarders located within in the DKIST instruments. For the modulators, the various instrument beams are F/ 18, F/ 24, F/ 32, F/ 62 posing a wide range of fringe magnitude possibilities. But with this simple $r^{-2}$ envelope, we can anticipate fringe amplitudes as functions of observing wavelength and fringe period. These simple analytic tools can be used to provide order-of-magnitude estimates for fringe properties when comparing designs for many-crystal achromatic retarders with alternate design strategies. We showed how the short-wavelength DKIST use cases at 396 nm can expect more than one magnitude amplitude reduction for the longest period fringes while expecting up to two orders of magnitude fringe suppression for the shortest period fringes during calibration. The longer wavelength DKIST instruments DL-NIRSP and Cryo-NIRSP are designed to do many types of observations at wavelengths of roughly 1000 nm to 5000 nm. The longest wavelength use cases do not get significant reduction of fringe amplitudes while the use cases around 1000 nm wavelength can expect some fringe reduction. In addition, the DL-NIRSP modulator when used at F/ 62 will see the full magnitude of fringes as the beam is essentially collimated.

Now we have a way to estimate fringe amplitudes in converging beams in addition to collimated beams using the Berreman calculus and this $r^{-2}$ envelope.  The major source of solar spectropolarimetric error is incomplete removal of polarized fringes during the calibration process.  To anticipate calibration errors and model our instruments accurately, we must also know how stable these fringes are with respect to temperature perturbations from the environment and from the heat loads imposed by the 300 Watt DKIST beam. By coupling these fringe amplitude predictions with thermal predictions of stability, we can begin to estimate the residual fringe calibration errors under a range of likely DKIST use cases.

\clearpage
\section{Fringe Thermal Stability: A Large Source of Error}

Temperature sensitivity is a major concern for the stability of a retarder. Temporal instability often is the ultimate calibration limiation and DKIST expects the 300 Watt beam to impose strong useage constraints. With the Berreman calculus and simple analytical calculations, we can show how fringes and retarder optical properties depend on thermal effects.  There are three main factors.  First is physical expansion (through the coefficient of thermal expansion, CTE, $\alpha$). Second is the change in refractive index with temperature through the thermo-optic coefficient (dn/dT, $TOC$). Third is the dependence of crystal birefringence on temperature (d(n1-n2)/dT).  All three parameters cause the fringes and retarder properties to change during typical system operation.

The polarized fringes from the calibration retarders are strongly temperature dependent. As we've seen in H17\cite{Harrington:2017jh}, fringes in these many-crystal retarders cause variation in all 3 retardance properties:  linear retardance magnitude, linear retardance fast axis orientation, circular retardance (ellipticity). To assess the calibration limitations, we need not only the amplitude predictions for all spectral components but the stability.

\begin{figure}[htbp]
\begin{center}
\vspace{-1mm}
\hbox{
\hspace{-1.0em}
\includegraphics[height=5.6cm, angle=0]{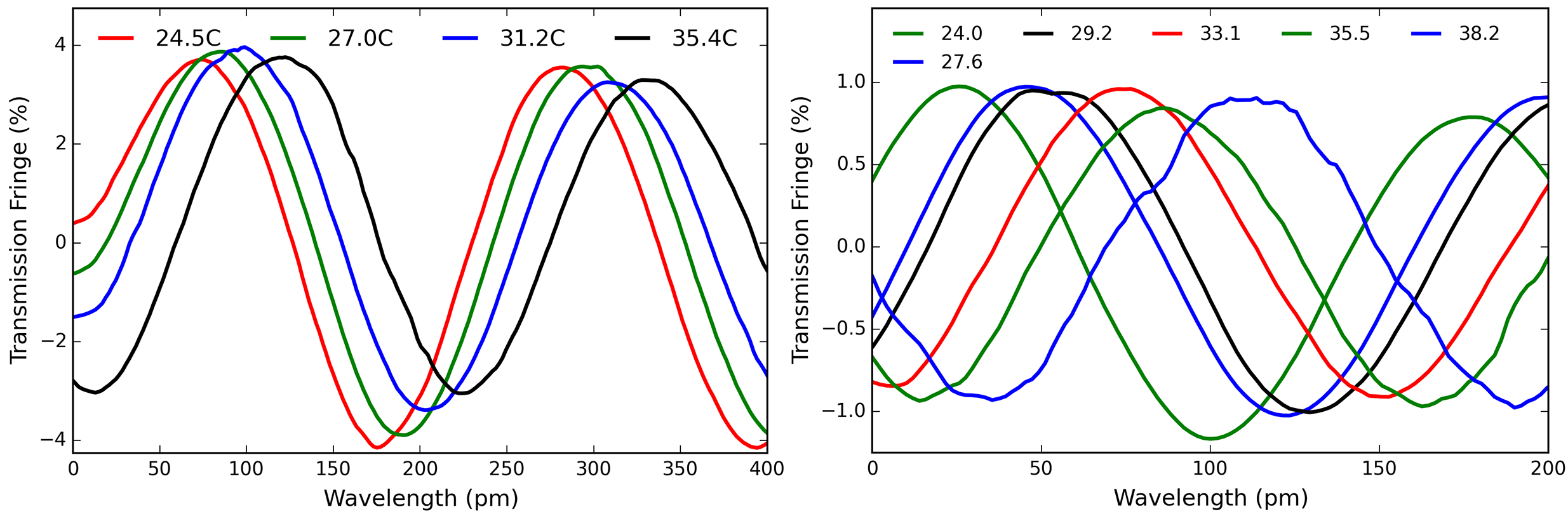}
}
\caption[] 
{The transmission fringes through the sample optic as the temperature is raised by over 10$^\circ$C.  The left panel shows the quartz retarder the right panel shows the MgF$_2$ retarder. \label{fig:fringes_vs_temp_MLO_spex}  }
\vspace{-6mm}
\end{center}
\end{figure}

As a demonstration, we collected laboratory measurements as functions of temperature for quartz and MgF$_2$ crystal as used in the DKIST retarders. In the Meadowlark Spex system, an enclosure with a heater was created for the crystal retarder sample.  This system was set to heat by roughly 10$^\circ$C over several hours.  Small entrance and exit ports were cut in the enclosure to allow the 10mm diameter beam to pass unobstructed through the enclosure. In addition to a heater, a temperature sensor was coupled to the optic mount as a direct reading of the optic temperature.  Given the slow heating rate and high crystal conductivity, we assume this temperature proxy is accurate enough for the purposes of demonstrating fringe drift with temperature.  

Figure \ref{fig:fringes_vs_temp_MLO_spex} shows the resulting high spectral resolution scans as functions of temperature for the crystals.  In both cases, the fringe pattern moved very roughly about half a wave period during the 10$^\circ$C to 12$^\circ$C of heating. Given that the samples are a fraction of a millimeter thick, the order of magnitude expected for the fringe drift is a fraction of a period per cm of optical path per $^\circ$C of heating.  In subsequent sections, we go through each physical effect.

We also note that we repeated this heating experiment for the Infrasil window sample.  Instead of scanning fringes in wavelength, we monitored a single wavelength at cadences faster than 1Hz. Similar behavior was recorded and no high-frequency errors were detected. Fringe drift temperature scales were similar. We focus on the crystal retarders here.

\subsection{Physical Expansion: Coefficient of Thermal Expansion $\alpha$}

Linear expansion coefficient,$\alpha$ = (1/L) dL/dT, is a normalized expansion coefficient with units of (1/C) which multiplies the optic physical thickness to compute the thermally perturbed thickness. In internal DKIST and vendor documentation, this coefficient in parts per million is $\alpha$= 13.5 for quartz, 9.2 for MgF$_2$ and 5.7 for sapphire. We note that birefringent crystals have different CTE values in the ordinary and extraordinary directions. As an example, the Crystran handbook shows 13.7 and 8.9 ppm/K for MgF$_2$ crystals. 

We compute a simple first-order estimate of the DKIST crystal temperature sensitivity to CTE by perturbing the fringe from a 10 mm thick piece of quartz. This substrate would have an optical thickness of roughly 30,000 waves path at visible wavelengths computed as 2dn/$\lambda$.  For quartz, we get a fringe drift of 0.8 waves per $^\circ$C of fringe sensitivity at 500 nm wavelength.  For the 13 mm physical thickness of MgF$_2$ in the CryoNIRSP retarder components, the sensitivity is a bit lower due to the smaller refractive index and lower CTE, but of the same order of magnitude. These values scale inversely with the wavelength and linearly with the thickness.  For the DKIST retarders, the many spectral components of the fringes do contribute and the fringe temporal stability decreases inversely as the fringe period increases. 

\subsection{Refractive Index Variation with Temperature: Thermo-Optic Coefficient dn/dT}
\label{subsub_refractive_index}

We now consider the refractive index variation with wavelength (dn/dT). For temperature dependence of the refractive index n, several articles show dn/dT in the 10's of parts per million range for various materials \cite{Tropf:2005uo,2008arXiv0805.0096L, 2006SPIE.6273E..2KL, 1999OptCo.163...95G, 1984ApOpt..23.1980D}. For most crystals, this term has opposite sign from the CTE ($\alpha$). As the crystal heats, the part becomes physically thicker but the refractive index drops. The two effects cancel each other to some degree. 

As above, we can show a first-order calculation by perturbing the fringes for a 10 mm thick piece of quartz using a 1.56 $\mu$m wavelength. We can separate the index perturbation from the nominal value and compute the sensitivity via 2d (dn/dT) / $\lambda$. We compute 0.1 waves of fringe change per C at a wavelength of 1560 nm.  For a coefficient of 5e-5 and and a shorter wavelength of 500 nm, the coefficient increases the value 5x and the wavelength increases the value 3x giving a value around 1 wave of fringe motion per C of at 500 nm wavelength. For the 13 mm of MgF$_2$ crystals with 30 ppm for dn/dT and visible wavelengths we get a similar sensitivity.  These factors must be included in the Berreman model to accurately predict fringe behavior.

\subsection{Birefringence Variation with Temperature}

\begin{wrapfigure}{r}{0.60\textwidth}
\centering
\vspace{-3mm}
\begin{tabular}{c} 
\hbox{
\hspace{-0.5em}
\includegraphics[height=6.6cm, angle=0]{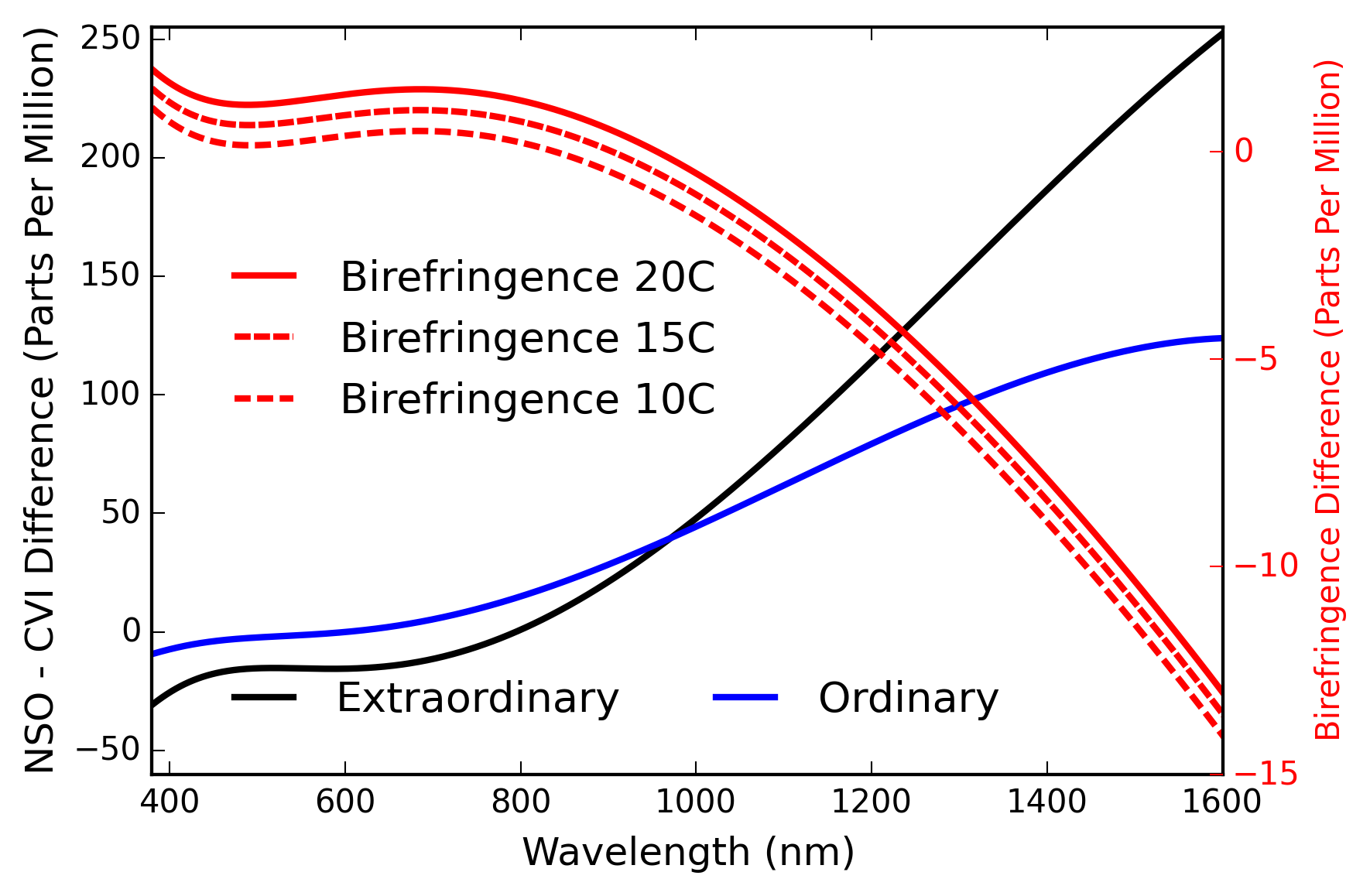}
}
\end{tabular}
\caption[] 
{ \label{fig:Quartz_refractive_Index} The refractive index difference between NSO and CVI catalog values for crystal quartz are shown in blue and black at amplitudes around 250 parts per million.  The three red curves show the difference in birefringence between CVI and the thermally perturbed NSO equations. Values are shown in red on the right hand Y axis and are differences at the level of 10 parts per million.}
\vspace{-2mm}
\end{wrapfigure}

The birefringence of a crystal optic is also a function of temperature \cite{Guimond:2004wv}. The extraordinary and ordinary rays do not see the same refractive index change with temperature, creating a differential effect. In DKIST designs, this effect was incorporated to address concern for the temperature sensitivity of birefringence impacting the basic retarder design \cite{2014SPIE.9147E..0FE, Guimond:2004wv}. Additionally, Sueoka found that the refractive index and birefringence models available in the literature did not adequately address the birefringence at longer wavelengths \cite{Sueoka:2016vo}.  The uncertainty was significant enough to require DKIST to perform an independent assessment and to adapt our designs accordingly.

The CVI Melles-Griot Materials Handbook entry for crystal quartz gives both ordinary and extra-ordinary refractive indices in terms of a Laurent series equation following Equation \ref{eqn:Laurent}. In Sueoka \cite{Sueoka:2016vo}, the Handbook of Optics from the Optical Society of America (OSA) provides a five-term Sellmeier equation following the style of Equation \ref{eqn:Sellmeier3}. Sueoka modified the ordinary refractive index following the equation plus the measured birefringence as a way to correct the equations for accurate birefringence predictions as required in the DKIST application \cite{Sueoka:2016vo}. In Table \ref{table:RefIndexQuartz} we show the coefficients for each equation.

In Figure \ref{fig:Quartz_refractive_Index} we show the difference between CVI Handbook and OSA Handbook refractive indices as blue and black curves. The two equations diverge at $\sim$100ppm amplitudes as well as diverge from each other by $\sim$100ppm at wavelenghts longer than 1000 nm. The modified birefringence difference is shown as the red curves using the red right-hand Y axis.  The birefringence only differs at levels of less than 15 parts per million, but this is enough to have impacted the modulation efficiency for retarders in the DKIST designs\cite{Sueoka:2016vo}. Included in the Sueoka \cite{Sueoka:2016vo} analysis is physical expansion and measurements of temperature perturbation from the thermo-optic coefficient. The birefringence is predicted to change at amplitudes of a few parts per million when temperature is changed by 10$^\circ$C as seen by the various red curves of Figure \ref{fig:Quartz_refractive_Index}.

\begin{wrapfigure}{r}{0.62\textwidth}
\vspace{-4mm}
\begin{equation}
n^2 = A_0 + A_1 \lambda^2  +  \frac{ A2} {\lambda^2 }  +  \frac{ A3} {\lambda^4 } +  \frac{ A4} {\lambda^6 } +  \frac{ A5} {\lambda^8 }
\label{eqn:Laurent}
\end{equation}
\begin{equation}
n^2 = 1 + \frac{ B_1 \lambda^2} {\lambda^2 - C_1} + \frac{ B_2 \lambda^2} {\lambda^2 - C_2} + \frac{ B_3 \lambda^2} {\lambda^2 - C_3} + \frac{ B_4 \lambda^2} {\lambda^2 - C_4} + \frac{ B_5 \lambda^2} {\lambda^2 - C_5}
\label{eqn:Sellmeier3}
\end{equation}
\vspace{-6mm}
\end{wrapfigure}

Several have reported on the temperature coefficients of quartz and crystal MgF$_2$ and athermalization of retarder designs. \cite{2011ApOpt..50..755M, 1988ApOpt..27.5146H, 1999OptCo.163...95G}  Bi-crystalline achromats can be constructed of positive and negative crystals to become thermally compensated at a single wavelength. By keeping crystal thickness ratios similar, athermal retarder designs can be created. \cite{2011ApOpt..50..755M, 2014SPIE.9147E..0FE}. The various Pancharatnam style designs \cite{Pancharatnam:1955iw} that have thin crystal components have lower thermal sensitivity than thicker many-order retarders. For multi-wavelength designs, typically you cannot exactly solve both for a retardance at multiple wavelengths as well as athermal peformance. However, you can balance thermal behavior against requirements on retardance, plate thickness, wavefront error, alignment tolerances, etc to decrease the sensitivity to various effects.

\begin{table}[htbp]
\vspace{-1mm}
\caption{Refractive Index Coefficients for CVI Laurent \& NSO Sellmeier}
\label{table:RefIndexQuartz}
\centering
\begin{tabular}{l l l l l l l l }
\hline
\hline
CVI Extraord.	&  	2.38490e+00	& -1.25900e-02		&  1.07900e-02 &  1.65180e-04 & -1.94741e-06 &  9.36476e-08 \\
\hline
CVI Ordinary   & 	2.35728e+00 	& -1.17000e-02 	&  1.05400e-02 &  1.34143e-04 & -4.45368e-07 &  5.92362e-08 \\
\hline
NSO Extraord. B	& 0.74700637 	& 0.45865921		& 0.17833250 	& 0.73225069 	& 8.7421747 & \\
NSO Extraord. C 	& 0.063458831 & 0.11266214		& 0.11288341 	& 9.1190338  	& 54.983117 & \\
\hline
NSO Ordinary B	& 0.663044	& 0.517852 		& 0.175912 	& 0.565380 	& 1.675299  & \\
NSO Ordinary C 	& 0.060		& 0.106	   		& 0.119    		& 8.844    		& 20.742    & \\
\hline
\hline
\end{tabular}
The coefficients for the CVI 6-term Laurent series of Equation \ref{eqn:Laurent} for both Extraordinary and Ordinary beams are in the first two rows.  The B and C coefficients of the 5-term Sellmeier equations (10 coefficients each) for the ordinary and extraordinary beams of the NSO modified fit are shown as the last four rows.
\vspace{-1mm}
\end{table}

The DKIST retarder designs used coefficients for birefringence changes near 10$^{-4}$ per $^\circ$C for quartz and 5*10$^{-5}$ per $^\circ$C for MgF$_2$ crystals. These numbers agree with the Handbook of Thermo-Optic Coefficients \cite{Ghosh:1998tl} and are similar to other athermal designs. \cite{2011ApOpt..50..755M, 1988ApOpt..27.5146H, 1999OptCo.163...95G}  Thus, the birefringence changes are the same order of magnitude as the refractive index changes.

\subsection{Fringe Thermal Sensitivity and Impact On Retarder Use Cases}

Translating the fringe temperature dependence into a specific quantifiable impact on the calibration or observation process depends on many estimates of materials properties, heat loads and calibration strategies.  The order of magnitude estimates presented above show that the fringes are expected to change for the DKIST quartz and MgF$_2$ retarders.  We apply a simple linear thermal perturbation analysis using our Berreman calculus in H17\cite{Harrington:2017jh} to show the expected magnitude and character of thermal perturbations for the DKIST retarders. 

We fit a sinuosoidal function to the fringes of Figure \ref{fig:fringes_vs_temp_MLO_spex}.  For the single crystal quartz at 0.5745 mm thickness and an observed wavelength of 625 nm, we found roughly a quarter wave of fringe drift in 10.9 per $^\circ$C of heating.  The fringe period is computed as $\lambda^2$ / 2dn which gives a spectral fringe period of 0.219 nm.  The fit periods were at 99.3\% of this value for the 24.5$^\circ$C data set and 98.7\% of the nominal period for the 35.4$^\circ$C data set.  The offset between fringes was computed at 83.7$^\circ$ phase or roughly 0.233 waves drift of the fringe. 

\begin{wrapfigure}{l}{0.40\textwidth}
\vspace{-6mm}
\begin{equation}
OP = \frac{ d (1+\alpha \Delta T) \hspace{2mm} n (1+TOC \Delta T) } {\lambda}
\label{eqn:OPD_perturbed}
\end{equation}
\vspace{-8mm}
\end{wrapfigure}

The Crystran Handbook of Optical Materials gives CTE values for crystal quartz as $\alpha$ = 7.1e-6 per $^\circ$C for the extraordinary beam and $\alpha$ = 13.2e-6 per $^\circ$C for the ordinary beam.The thermo-optic coefficients were much more similar in the Crystran Handbook with the TOC = -5.5e-6 per $^\circ$C for the extraordinary beam and -6.5e-6 per $^\circ$C for the ordinary beam. With these handbook values, we compute a simple perturbation of the thickness as d(1+$\alpha$) and the refractive index changes to n(1+TOC).  

The perturbed optical path is computed as dn/$\lambda$ using the perturbations in thickness and refractive index linearly in the temperature change $\Delta$T show in Equation \ref{eqn:OPD_perturbed}. In the Berreman calculus, we input the 3-dimensional refractive index data, crystal orientation and thickness.  Each value is perturbed separately in the Berreman formalism.  Note that for the 10$^\circ$C change in the Meadowlark lab, the 0.5 mm quartz sample only expanded by 76 nm and the refractive index changed by 65 parts per million. 

\begin{figure}[htbp]
\begin{center}
\vspace{-1mm}
\hbox{
\hspace{-0.5em}
\includegraphics[height=5.6cm, angle=0]{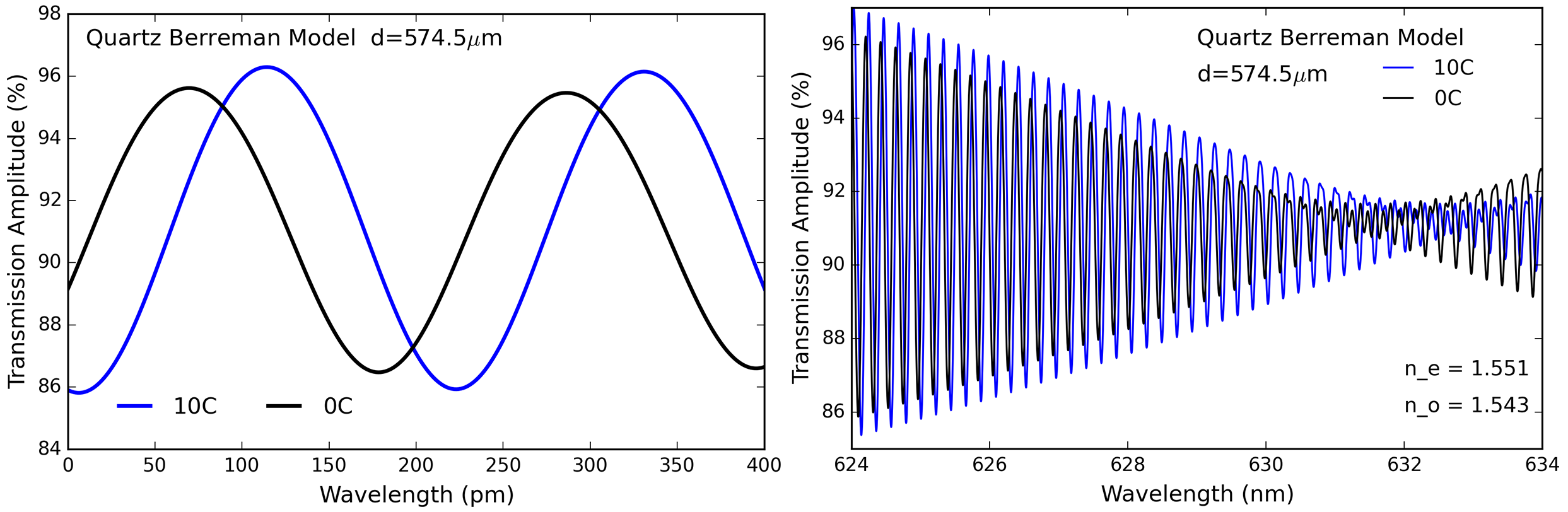}
}
\caption[] 
{Our Berreman model using a thermal perturbation of 10$^\circ$C following Equation \ref{eqn:OPD_perturbed}. The left plot shows a narrow wavelength range of 400 pm matching the Meadowlark Spex bandpass of Figure \ref{fig:fringes_vs_temp_MLO_spex} We perform a sin function fit and find a phase offset of 0.21 waves or 74.5$^\circ$ when using the Crystran Handook values. The right hand graphic shows a larger bandpass illustrating how the differential temperature sensitivity of extraordinary and ordinary beams causes a wavelength drift in the destructive interference. \label{fig:thermal_perturb_quartz_crystal_MLOspex}  }
\vspace{-6mm}
\end{center}
\end{figure}

We use this simple linear thermal perturbation in our Berreman calculus fringe models to verify our calculations match laboratory data for thermal drifts of the crystals.  Figure \ref{fig:thermal_perturb_quartz_crystal_MLOspex} shows such a calculation for this 0.57 mm thick crystal around the 625 nm wavelength used in the Meadowlark Optics test setup. 

Using this simple perturbation with Crystran Handbook values, our Berreman model gives a fringe phase shift of 74.5$^\circ$ or 0.207 waves of fringe thermal drift. This is very similar to the laboratory measured value of 83.7$^\circ$ and 0.233 waves drift using the Meadowlark Spex system. With this simple linear perturbation, we can easily compute the expected form and magnitude of thermal sensitivity in the DKIST and Keck LRISp six-crystal retarders.

\subsection{Summary of Fringe Instability: Thermo-Optic and Thermal Expansion Coefficient}

In this section we applied a simple thermal perturbation to Berreman models to fringes in SiO$_2$ and MgF$_2$ single crystal retarders as functions of temperature. We successfully compared these Berreman models to high spectral resolving power SPEX data sets where retarder crystals underwent $\sim$10$^\circ$C thermal change. We used linear perturbations of the refractive index differentially for the extraordinary and ordinary beams through the thermo-optic coefficient (TOC) for each crystal axis. This leads to changes both in refractive index and birefringence as functions of temperature. We also included simple models for physical thickness via the coefficient of thermal expansion (CTE, $\alpha$). The TOC for ordinary and extraordinary beams combined with physical expansion are required to assess the thermal stability of the DKIST six-crystal retarders in the summit environmental conditions of 0$^\circ$C to 40$^\circ$C as well as in response to heating caused by the 300 Watt DKIST beam. With experimental validation of our thermal perturbations in the Berreman calculus, we can now predict the fringes present in DKIST calibration optics and in modulated spectra measured by the DKIST instruments in response to the DKIST laboratory thermal environment ($\pm$1$^\circ$C) as well as the Gregorian focus summit environment ($\pm$20$^\circ$C) in response to the thermal loads imposed by the 300 Watt beam.

\clearpage

\section{Application to the DKIST Retarder in DL-NIRSP at F/ 62}

We apply both the Berreman fringe amplitude estimates as well as the thermal perturbation analysis to a DKIST instrument and the six-crystal modulator. In the Diffraction Limited Near Infrared Spectropolarimeter (DL-NIRSP) instrument, the DKIST project will install a six crystal polychromatic modulator (PCM) which includes anti-reflection coatings and oil layers between the crystals. The modulating retarder sees beams at either F/ 24 or F/ 62 depending on the configuration of the feed optics. We assess the worst-case F/ 62 beam for fringe amplitude and thermal stability. Our assessment above shows that the F/ 62 beam is essentially collimated and fringe amplitudes will not be reduced by the mild convergence of the beam. In addition, the DKIST coud\'{e} laboratory is only stabilized to $\pm$1$^\circ$C plus possible thermal instability caused by imperfect temperature control on the rotation stage motors driving the crystal modulator.

We use the linear perturbation of crystal thickness and refractive index in Equation \ref{eqn:OPD_perturbed} to modify the refractive indices and physical thickness for every layer in the six-crystal retarder design.  The DL-NIRSP instrument mounts the modulator (PCM) just ahead of the focal plane formed on the fiber-bundle integral field unit input to the spectrograph. The DL-NIRSP has two infra-red camera channels we will consider here. One channel is nominally used to observe two common solar lines at 1075 nm and 1083 nm wavelength. The second channel is optimized for two lines at 1430 nm and 1565 nm.  

\begin{figure}[htbp]
\begin{center}
\vspace{-3mm}
\hbox{
\hspace{-1.5em}
\includegraphics[height=11.3cm, angle=0]{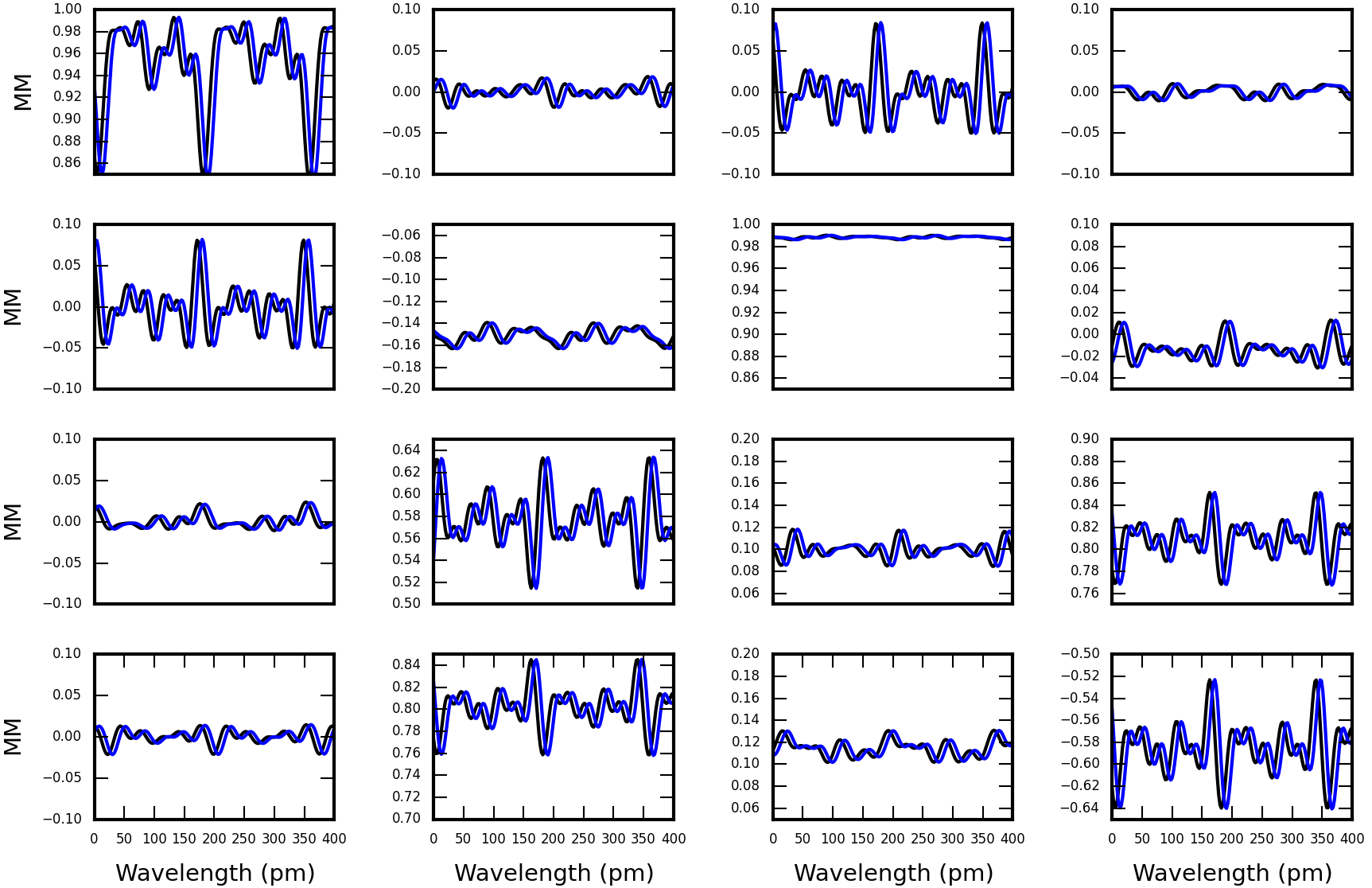}
\vspace{4mm}
}
\caption[] 
{The Berreman model for the DL-NIRSP modulator with a bandpass of 1083.0 nm to 1083.4 nm covering a wavelength range of 400 pm. Computations were done at spectral sampling of 500,000 and no degradation of resolving power.  We used a thermal perturbation of 1$^\circ$C following the temperature stability specification for the DKIST coud\'{e} laboratory and actively cooled rotation stages. Black shows the nominal model while blue shows an increase of 1$^\circ$C and perturbations to the optical path following Equation \ref{eqn:OPD_perturbed}.  \label{fig:thermal_perturb_DL_Modulator}  }
\vspace{-7mm}
\end{center}
\end{figure}

At these wavelengths, the nominal resolving power of the instrument is over 100,000.  This instrument plans to spectrally sample the beam with a variety of user-selected modes. For this discussion, we assume sampling is in the range of several pico-meters as designed to fully sample the instrument profile at these wavelengths. This modulator includes the refractive index $\sim$1.3 oil between crystal interfaces.  

\begin{wraptable}{l}{0.29\textwidth}
\vspace{-3mm}
\caption{DL PCM Retarder}
\label{table:DL_PCM_Design}
\centering
\begin{tabular}{l l l}
\hline
\hline
Material	& Thickness	& $\theta$		\\
		& $\mu$m		&  deg.		\\
\hline
\hline
AR 	& 0.2355		& -	\\
Qtz	& 2169.1		& 0.0	\\
AR 	& 0.2355		& -	\\
Oil	& 10.0		& -	\\
AR 	& 0.2355		& -	\\
Qtz	& 2099.1		& 90.0	\\
AR 	& 0.2355		& -	\\
Oil	& 10.0		& -	\\
AR 	& 0.2355		& -	\\
Qtz	& 2146.9		& 42.20	\\
AR 	& 0.2355		& -	\\
Oil	& 10.0		& -	\\
AR 	& 0.2355		& -	\\
Qtz	& 2099.1		& 132.20	\\
AR 	& 0.2355		& -	\\
Oil	& 10.0		& -	\\
AR 	& 0.2355		& -	\\
Qtz	& 2169.1		& 152.51	\\
AR 	& 0.2355		& -	\\
Oil	& 10.0		& -	\\
AR 	& 0.2355		& -	\\
Qtz	& 2099.1		& 241.52	\\
AR 	& 0.2355		& -	\\
\hline
\hline
\end{tabular}
\vspace{-7mm}
\end{wraptable}

As discussed in H17, we removed the 10 mm thick Infrasil cover windows from the optics. The design now only has six crystals, five oil layers and anti-reflection coatings on each crystal surface. Each coating is modeled as an isotropic MgF$_2$ layer designed as a quarter-wave of path at a central wavelength of 1300 nm. With a refractive index of about 1.38 at this wavelength, we compute 236 nm physical thickness for the coating. Table \ref{table:DL_PCM_Design} shows the Berreman stack of birefringent materials used in the model. As described later, the modulators are designed as elliptical retarders that deliver efficient modulation over wide wavelength bandpasses. For the DL-NIRSP, we achieve sufficient efficiency from 500 nm to 2500 nm when using six quartz crystals with the thicknesses and orientations specified in Table \ref{table:DL_PCM_Design}.

In Figure \ref{fig:thermal_perturb_DL_Modulator} we show the nominal Berreman model in black along with a 1$^\circ$C thermally perturbed Berreman model in blue. We used a wavelength grid for the model at a constant spectral sampling of $\lambda / \delta\lambda$ of 500,000. The spectral resolving power was infinite and no simulation of instrument profile resolution degradation was applied. We also applied a thermal perturbation to the oil layer with an assumed CTE value of $\alpha$ = 10$^{-4}$ consistent with other oils.  We do not have any data on the dn/dT value for the oil.

The dominant effect of the thermal perturbation is a shift of the entire pattern in wavelength by about 7.5 pm.  This is of roughly the spectral sampling for the DL-NIRSP. Given the temperature stability of the optic, it is possible that the fringe pattern could be stable at levels around a few resolution-elements. We can easily fit simple optical models to the thermally perturbed Mueller matrix.  In typical solar demodulation schemes, intensity fringes caused by transmission and / or diattenuation can be post-facto {\it filtered} in various ways. These techniques have various consequences for the fidelity of the derived solar signals when the fringes and real signals are similar.

We can use the Berreman calculus to highlight the impact of oil layers, bonding epoxies, coatings and other materials between the crystals. As an example of the oil layer impact, we ran a grid of models where the oil layer thickness was either 7, 10 or 13 microns for each layer computed against each other layer. Given the five oil layers and three possible thicknesses, we computed 243 separate Berreman models. Diattenuation is dominated by Stokes $U$ at this particular DL-NIRSP wavelength with amplitudes up to 15\% peak-to-peak. Figure \ref{fig:thermal_perturb_DL_Modulator} shows the $IQ$ and $IV$ terms are of order $\pm$1\%. The elliptical retardance fringes varies spectrally by over 6$^\circ$ peak-to-peak.

\begin{figure}[htbp]
\begin{center}
\vspace{-3mm}
\hbox{
\hspace{-1.5em}
\includegraphics[height=11.9cm, angle=0]{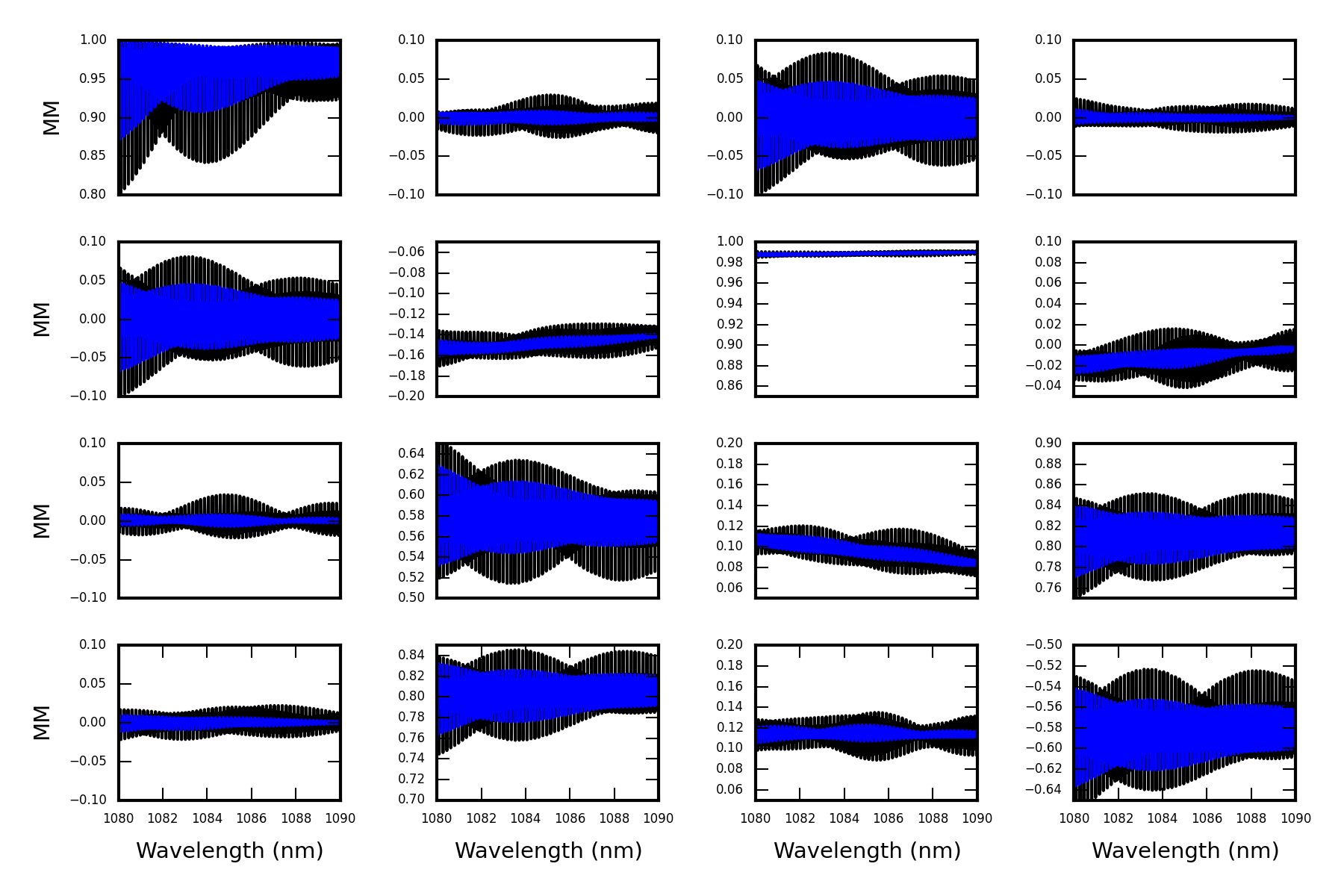}
}
\caption[] 
{The Mueller matrix following \ref{eqn:MM_IntensNorm} for the DL-NIRSP modulator around 1083nm wavelength.  The 1080 nm to 1090 nm bandpass at left, a narrow 1083.0 nm to 1083.5 nm bandpass at right.  \label{fig:DL_1083_MM_Examples_Oil}  }
\vspace{-7mm}
\end{center}
\end{figure}

The Fourier analysis shows that every layer gives rise to a fringe component at the appropriate spectral period.  However, the interplay between the relatively thin layers and the relatively thick crystals creates amplitude variation and also changes a much lower period amplitude envelope for the fringes. Simply changing the isotropic oil layer thickness by a few microns can strongly vary peak fringe amplitudes. As an example of the relatively slow spectral variation caused by the oil layers in the DKIST designs, we show two Mueller matrices in Figure \ref{fig:DL_1083_MM_Examples_Oil}.  The black curve shows the nominal 10 micron layer thickness while blue shows uniform 7 micron layer thickness.  The Fourier analysis of any narrow spectral bandpass used by a DKIST instrument would be nearly the same, but there is a spectrally slow amplitude modulation that changes strongly with varying oil layer thickness. 

Figure \ref{fig:DL_1083_MM_Examples_Oil} only covers a 10 nm wide spectral bandpass but the transmission fringes change by over 8\%, diattenuation can double and elliptical retardance change by degrees at specific narrow wavelengths used by solar spectropolarimeters.  We have done several spectrophotometric tests to determine oil layer thickness between various materials. Values range from 5 microns to over 20 microns. The detailed fringe spectra of each individual DKIST retarder will no doubt require testing at the highest spectral resolving powers for each specific wavelength planned.

\subsection{Thermal Fringe Behavior Summary: Temporal Stability Impacts Calibration}

The fringe stability and amplitude requirements imply a temperature stability requirement of the calibration and modulation retarders to be roughly a fraction of a degree Celsius per calibration cycle for the highest period fringes to be considered stationary. Otherwise, the fringes must be assumed to be variable and other mitigation strategies considered. In later sections of this paper and in Appendix \ref{sec:appendix_thermal}, we explore thermal behavior and outline mitigation strategies.

Some DKIST instruments such as the Visible Tunable Filter (VTF) are narrow-band Fabry-Perot type imagers and cannot apply spectral fringe filtering techniques. The VTF images are quasi-monochromatic as the etalons change the bandpass wavelength discretely in steps of roughly 6 pm (60 m{\AA}, 0.06 nm) with the pass-band at about as wide as the wavelength step size. Tunable imaging systems do not have simultaneous spectral measurements available to apply fringe filters and this requires that the calibration both include fringe properties and be temporally stable. If the calibration retarder varies in linear fast axis orientation, linear retardance magnitude {\bf or} circular retardance (QU frame rotation), there are direct and irrecoverable impacts on the ability to calibrate. The DKIST quartz calibration retarders will all have noticeably different retardance values for each wavelength step of the etalons in a scanning FP system. As we've shown in H17\cite{Harrington:2017jh}, a very approximate fringe amplitude for the DKIST retarders is a few degrees linear retardance variation, a few degrees fast axis rotation and a few degrees circular retardance (fully elliptical fringes). These fringes can vary the properties of the calibration retarder differently for every independent wavelength in the scan of the VTF imager.  The calibration retarder will also be highly time dependent through the fringe temperature sensitivity. These values obviously vary with wavelength, time and field of view but the Berreman calculus combined with the analytical tools here provides us a way to quantify and assess timescales for stability.  

As pointed out in H17\cite{Harrington:2017jh}, we have one calibration retarder made of crystal MgF$_2$. Thermal models suggest the heat loads are essentially negligible when this MgF$_2$ retarder is used in conjunction with a calibration polarizer upstream. The polarizer blocks all infrared wavelengths where the MgF$_2$ crystal absorbs and there are minimal other heating terms. We also have a calibration polarizer that includes a additional 25 mm thick Infrasil window that removes most of the heat load from the quartz calibration retarders. We explore thermal models in Appendix \ref{sec:appendix_thermal} of this paper. 

With a calibration optic unstable in time, any calibration process is either reduced in accuracy or the analysis must become more complex. For DKIST, this temperature sensitivity is likely the major limitation for VTF calibration using the nominal retarders. Spectral instruments will also face difficulty, but filtering and averaging techniques can somewhat mitigate. Other telescopes with many-crystal retarders will suffer as these types of instruments will not have a temporally stable calibration retarder. For many-crystal retarders, this should be a major design consideration when coupled with other system performance parameters (heat loads, bandpass and field of instruments available to use for fringe filtering, etc). 

We showed in this section that we can predict fringe thermal instabilities through single crystal and many crystal stack retarders.  A Berreman model using simple linear perturbation of thickness and refractive index was applied to Meadowlark data on single crystal retarders of known thickness and temperature, validating the Berreman models. Comparison of various literature values for refractive index, coefficient of thermal expansion and thermo-optic coefficients showed the perturbation analysis is not very sensitive to known uncertainties in refractive index. We then outlined a specific application to a six-crystal DKIST retarder including oil and anti-reflection coatings for the DKIST instrument DL-NIRSP at 1083 nm wavelength in the F/ 62 high spatial resolution mode, one of the commonly used solar spectral channels.  At F/ 62 we expect minimal reduction of fringe amplitude from the nearly collimated beam. Fringes are present at amplitudes over 12\% in transmission, 14\% in diattenuation and 7$^\circ$ elliptical retardance.  Thermal perturbations of 1$^\circ$C shift the fringes by 7.5 pm in wavelength, comparable to the resolving power of the instrument. We also showed how oil layer thickness variation changes broader spectral amplitude envelopes for the fringes but does not fundamentally change the underlying spectral periods.

\section{Summary: Predicting Retarder Fringe Amplitudes \& Temporal Stability In Converging Beams With Thermal Loads}

Polarization fringes are a major calibration limitation in astronomical spectropolarimeters. Designing systems with reduced fringe amplitudes and benign behavior is a challenge for modern large instrumentation. Calibration of DKIST instruments demands stringent temporal stability requirements as well as minimization of optical sensitivities to thermal changes. The temporal stability of optical components must be assured for DKIST in the presence of thermal loads from a 300 Watt beam and operations in the mountain summit environmental conditions. A systems-engineering level assessment of DKIST calibration processes requires these new tools for predicting polarization fringe amplitudes and their temporal behavior in converging and diverging beams. We showed simple calculations of Haidingers fringes (fringes of equal inclination) over a converging beam footprint to show fringe amplitude reduction dependence on beam F/ number. This combined with the Berreman formalism presents a tool to estimate full Mueller matrix and fringe behavior under design, thermal, and manufacturing perturbations. The fringe amplitude is subsequently reduced by the averaging over many waves of spatial fringes in converging or diverging beams but the underlying fringe spectral periods remain unchanged. We verified the $r^{-2}$ fringe amplitude scaling relation with laboratory data on crystal retarder and window samples. 

For the DKIST six-crystal retarders, the highest amplitude fringes from the air-crystal interfaces see the greatest reduction of amplitude in the converging beam as the marginal ray sees significantly more optical path upon back-reflection through the entire crystal stack. The amplitude of polarization fringes can be significantly reduced by placing the retarder in a steeply converging beam in addition to using anti-reflection coatings as was done for the DKIST calibration retarders and certain instrument modulator configurations. This fringe amplitude reduction benefit in converging beams must be traded against effects of spatial non-uniformity, depolarization (as outlined in Sueoka \cite{Sueoka:2014cm}) and exacerbated thermal issues. The temporal stability of the fringes was assessed for DKIST by including of physical expansion, the thermo-optic coefficient and the birefringence variation with temperature under heat load in Berreman models. These thermal sensitivities were also demonstrated for crystal retarders and windows in the lab with a high resolution spectrograph. 

These issues are common to any precision astronomical high resolution spectropolarimeter. We included in Appendix \ref{sec:appendix_keck} an on-sky demonstration. We showed fringe amplitude estimates and Berreman models for the six-crystal achromatic retarder used in the Keck 10 meter diameter telescope and LRISp spectropolarimeter on Maunakea. This retarder is an excellent comparison case for DKIST and other astronomical systems as both use F/ 13 beams and a six-crystal achromatic retarder design. The fringes in LRISp are detected at amplitudes of a small fraction of a percent with thermal evolution over a night in outdoor conditions as reported in H15. This small amplitude is consistent with the Berreman predictions presented here after accounting for F/ number and low spectral resolving power. Berreman predicts large fringe amplitudes for a collimated beam and substantial dependencies on cement layer thickness and refractive index. We predict and detect significant reduction when convolving with low resolution instrument profiles and averaging over the aperture in the F/ 13 converging beam. This on-sky demonstration of fringe properties validates the aperture-average in a converging beam as well as thermal perturbation when combined with the Berreman calculus. 

With the design tools presented here, the DKIST team was able to assess fringe behavior for optics in varying F/ number beams. This formalism was also used to re-assess the optical design with cover windows and to assess the temporal instabilities for the retardance as well as polarization fringe evolution. Considering the thermal protection provided by the calibration polarizer as well as an additional window mounted separately with the polarizer, predicted thermal loads are reduced by an order of magnitude and keep steady-state temperatures within 1$^\circ$C of ambient. We show detailed thermal analysis of our retarders under various beam configurations in Appendix \ref{sec:appendix_thermal} and \ref{sec:appendix_thermal_FEM}. 

This polarization fringe amplitude calculation was also used to predict the various fringe spectral component amplitudes for the DKIST modulating retarders, which work in beams from F/ 8 to F/ 62 and wavelengths from 380 nm to 5000 nm. We showed an example calculation for the DL-NIRSP instrument modulator in the F/ 62 configuration. The fringe amplitude $r^{-2}$ envelope calculation shows no significant fringe amplitude reduction for this configuration compared to a collimated beam. With spectral resolution up to 125,000, this DKIST instrument will see significant fringes at amplitudes over 10\% for transmission, 15\% for diattenuation and several degrees for elliptical retardance. A simple thermal perturbation analysis was performed to show the likely drift of this modulator Mueller matrix using the 1$^\circ$C temperature stability requirement for the DKIST coud\'{e} laboratory. This modeling tool should be useful for future solar and night time spectropolarimeters where fringes may be high amplitude, thermally unstable and possibly mitigated using a range of techniques. 

With this analysis we showed theoretical origins and laboratory verification of the $r^{-2}$ fringe amplitude envelope in converging beams. The Berreman calculus was used with thermal perturbations in refractive index through the thermo-optic coefficient and the physical thickness through the coefficient of thermal expansion.  These thermal perturbations were also experimentally verified in the laboratory. Predictions were made for DKIST instruments as well as for on-sky data from the Keck LRISp retarder in Appendix \ref{sec:appendix_keck}. By combining the Berreman calculus with thermal simulations and converging beam parameters, instrument designers now have tools to estimate likely fringe amplitudes for a wide variety of use cases and thermal conditions.

\section{Acknowledgements}

This work was supported by the DKIST project. The DKIST is managed by the National Solar Observatory (NSO), which is operated by the Association of Universities for Research in Astronomy, Inc. (AURA) under a cooperative agreement with the National Science Foundation (NSF). We thank David Elmore for his assistance, guidance and insight into the long history of work on the DKIST project. Some of the data presented herein were obtained at the W. M. Keck Observatory, which is operated as a scientific partnership among the California Institute of Technology, the University of California and the National Aeronautics and Space Administration. The Observatory was made possible by the generous financial support of the W. M. Keck Foundation. This research made use of Astropy, a community-developed core Python package for Astronomy (Astropy Collaboration, 2013). The authors wish to recognize and acknowledge the very significant cultural role and reverence that the summit of Maunakea has always had within the indigenous Hawaiian community.  We are most fortunate to have the opportunity to conduct observations from this mountain.

\appendix

\clearpage
\section{Thermal Impacts on Elliptical Retardance Errors}
\label{sec:appendix_thermal}

The fringe drift with temperature is only one of several thermal effects that limit polarization performance of the system. We show here how this simple perturbation analysis can be simplified to predict just elliptical retardance changes to the design caused by uniform and non-uniform temperature changes throughout the optic. We use the refractive index data from above to predict the theoretical retardance in the six individual crystal plates.  We then compute the Mueller matrix of the optic as the combined impact of the six theoretical matrices. The bulk temperature increase and established thermal gradients with depth effect the Mueller matrix elements. We've detailed the thermal performance models in the appendix and apply some of the depth gradients, radial gradients and temporal changes here. Temperature changes affects the birefringence and the apparent thickness of each of the crystal layers. These effects thus change the A-B-A bicrystalline achromat retardance, which in turn creates fully elliptical deviations from the design retardance.  The retarder will thus vary spatially and temporally in response to thermal perturbations.

Each retarder design is sensitive to bulk and depth temperature changes at different wavelengths in different ways. As a typical example, we consider an 8$^\circ$C bulk temperature rise and a 0.8$^\circ$C linear gradient with depth from a hot optic top to a cooler optic bottom. This case would be somewhat typical of the no-polarizer thermal gradient and $\sim$20 minutes of use to reach 8$^\circ$C above ambient consistent with some use cases shown in Appendix \ref{sec:appendix_thermal_FEM}. The thermal gradient effects on the ViSP and DL-NIRSP calibration retarders are slightly larger than the bulk temperature effects. In particular, at the shorter wavelength range the gradient effect is a factor of two larger for most Mueller matrix elements. For comparison, we also modeled the six-crystal modulator for the Cryogenic Near-Infrared Spectropolarimeter (Cryo-NIRSP).  This optic is made entirely of MgF$_2$ crystals and is optimized for the wavelength range 1000 nm to 5000 nm. The Cryo-NIRSP retarder models predict much smaller thermal gradients which reduces the thermal impact to the Mueller matrix elements. This retarder only experiences a small change due to bulk temperature rise with mild impact to the Mueller matrix at shorter wavelengths.

\begin{figure}[htbp]
\begin{center}
\vspace{-0mm}
\hbox{
\hspace{-0.3em}
\includegraphics[height=6.3cm, angle=0]{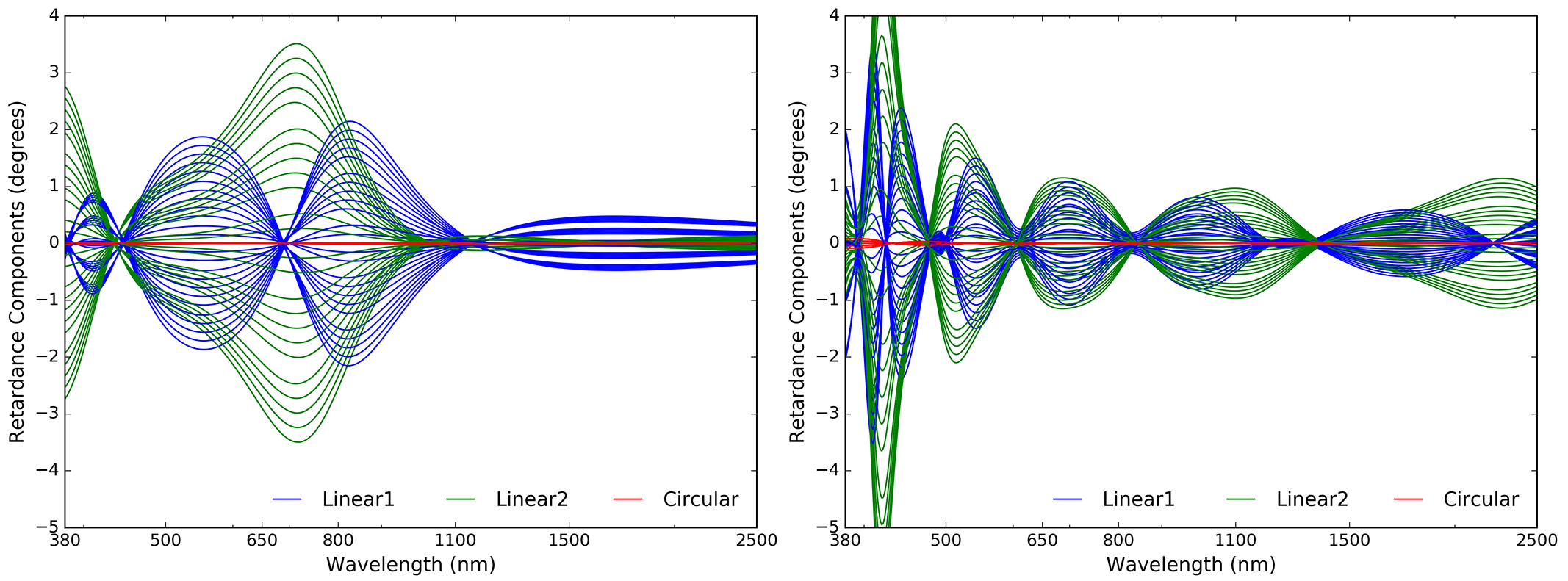}
}
\caption[] 
{\label{fig:thermal_errors_SARs} The spectral variation of elliptical retardance components in response to bulk temperature rise from 0$^\circ$ to 40$^\circ$ as well as thermal changes with depth of up to $\pm$4$^\circ$ linearly through the optic. The thermal models of all calibration are shown as they will see the thermal load and summit environmental temperatures.  DL-NIRSP calibration retarder is at left and the CryoNIRSP calibration retarders is at right. Note that the CryoNIRSP SAR is crystal MgF$_2$ and will see negligible thermal loading from the beam, but significant environmental changes in bulk temperature on the summit.  The simulation is included to show design sensitivity.}
\vspace{-6mm}
\end{center}
\end{figure}

We modeled the impact of thermal variations to the retarders for all combinations of depth gradient and bulk temperature rise.   Figure \ref{fig:thermal_errors_SARs} shows the DL-NIRSP retarder on the left and the Cryo-NIRSP retarder on the right.  We fit an axis-angle elliptical retarder model to the thermally perturbed Mueller matrix. Blue shows the first linear component of retardance (rotation about Q on the Poincar\'{e} sphere). Green shows the second component of linear retardance (rotation about U on the Poincar\'{e} sphere). Red shows the circular retardance component (rotation about V on the Poincar\'{e} sphere). The retardance variation is roughly a few degrees retardance per component. Updated thermal finite element models (FEMs) have been computed using revised (and directionally dependent) conductivity for crystal quartz, coating heat loads revised to reflect our as-measured coating absorptivity and revised optical models removing cover windows. Details of the various opto-mechanical models are in the Appendix. The new FEMs suggest the temperature gradients are a factor of three to five less with depth and radius compared to the window-covered optical models. The bulk material temperature rise is still at amplitudes of many degrees when the optic is used without the optical protection of an upstream polarizer, but the rise is also significantly slower due to the improved crystal conductivity and reduced loads. Models in the Appendix show operation from 0$^\circ$C to 40$^\circ$C from the baseline 20$^\circ$C along with depth gradients in the range of 0$^\circ$C to 4$^\circ$C. As in our above fringe thermal stability analysis, the retardance simulation uses physical expansion, thermo-optic coefficients and birefringent temperature sensitivities in the same amplitude ranges.

With this thermal perturbation analysis we are able to assess the temperature stability requirements for these retarder optics from both fringe and elliptical retardance stability perspectives. The thermal perturbation analysis was combined with finite element models to derive requirements and performance estimates for DKIST optics in response to optical absorption, cooling, mount conduction and other factors.  For DKIST, the thermal instabilities combined with polarization fringes will likely be one of the major limitations of the delivered data products.

\clearpage
\section{Thermal Models of Heated DKIST Retarders}
\label{sec:appendix_thermal_FEM}

We have detailed thermal finite element models for each crystal quartz and MgF$_2$ retarder that reflect the varying environmental temperatures as well as heat loads from a diverse set of use cases. We have performed detailed thermal finite element models to show the behavior of our retarders from 0$^\circ$ to 40$^\circ$ in the presence of depth dependent heating that changes substantially with configuration of the upstream optics. We also have assessed absorptivity of anti-reflection coatings and the index-matching oil between the coated crystals. A detailed presentation of all DKIST thermal models is beyond the scope of this document, but we outline here an example model and some highlights. The expected thermal behavior and corresponding stability of the DKIST calibration optics depend significantly on the assumed conductivity of the materials, cooling rates and input heat loads. Thermal gradients across the optic clear aperture and with depth through the part do cause more significant departures from the nominal retarder design. The spectral fringe dependence on thermal behavior is also an important contribution to system temporal stability. We also must compare the impact of fringes to the retardance stability in response to thermal changes. As the optics change temperature, the individual crystal plates have changing birefringence. The change in each crystal is somewhat compensated by the design as the pairs of plates subtract retardance from each other in the standard A-B-A Pancharatnam design. 

In addition, calibration can be performed with this optic combined with one of a few polarizers mounted upstream of the retarder. This polarizer aluminum wires reflect roughly half the light and additionally absorb roughly 10\% of the light through imperfect aluminum reflectivity. The fused silica substrate of the polarizer also absorbs wavelengths longer than about 5$\mu$m.  This polarizer thus reduces the heat absorbed by the quartz retarder by a factor of roughly 3x and substantially changes the depth dependent temperature distribution.  For the crystal MgF$_2$ retarder, the heat load is entirely removed by the polarizer.  Detailed consideration of thermal impacts of the various calibration use cases is required for DKIST.

\subsection{Thermal Finite Element Models For DKIST Retarders}

A thermal finite element model (FEM) was created for the calibration retarder by Hofstadter Analytical Services LLC. Initially we modeled four different heating scenarios at several durations of exposure to sunlight. In order to model the polarimetric effects of the thermal load, cumulative power absorbed through the depth of an optic and coating absorption at each coated interface were incorporated into the thermal FEM. 

\begin{wrapfigure}{r}{0.55\textwidth}
\vspace{-3mm}
\centering
\hbox{
\hspace{0em}
\includegraphics[width=0.55\textwidth]{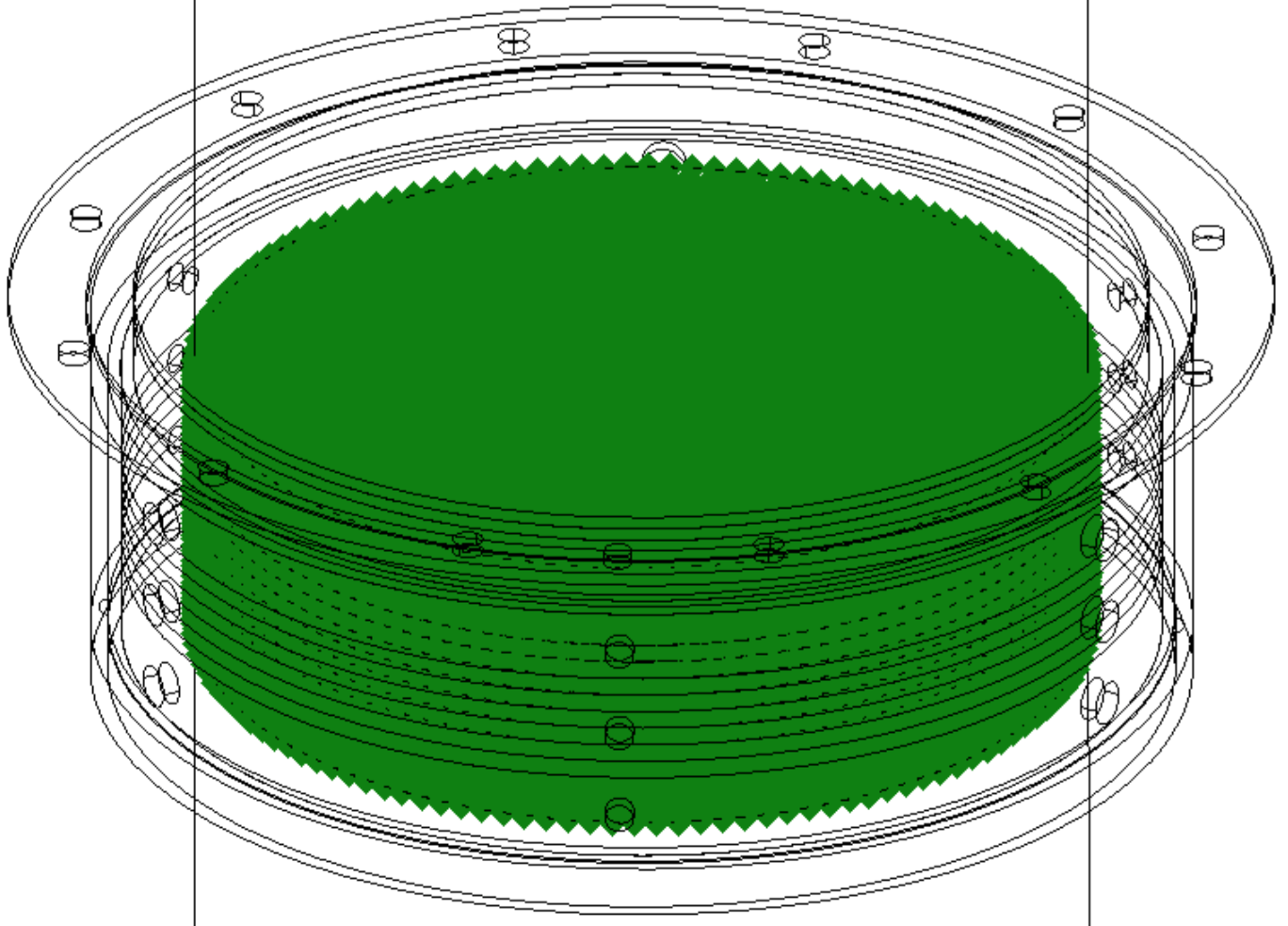}
}
\caption{\label{fig:femmesh} Thermal FEM model side view. Green points show the 17 layers that make up the crystals, interfaces and cover window bulk material. See text for details.}
\vspace{-3mm}
\end{wrapfigure}

As mentioned in H17\cite{Harrington:2017jh}, we recently made a very significant design change to remove the 10mm thick cover windows.  We include here the thermal analysis of those cover windows as this analysis, in addition to the fringe simulations of H17\cite{Harrington:2017jh} were important drivers of this change.  Often, high aspect ratio retarders use cover windows as a method of guaranteeing better transmitted wavefront error, beam deflection and durability.  But the thermal and fringe impacts must be considered against these possible performance improvements. 

Although thermal effects on both the calibration polarizer and retarder are of concern, this paper focuses on the heating of the retarder because it creates numerous polarimetric errors due to the six crystalline retarder stack up design. 

DKIST provided profiles of depth dependent bulk absorption and coating absorption to Hofstadter Analytical LLC to use in the thermal FEMs. The models spanned the full diameter and depth of the retarder and included the mounting structures (rotary stage, bearings and cell mount). Along the optical axis of the parts, there were nodes every 2 mm spanning the 10 mm thick substrates and six 2 mm thick crystals in the center. The parts were mounted in an aluminum cell with RTV between the part and the cell. 

The thermal Finite Element Model (FEM) shown in Figure \ref{fig:femmesh} illustrates the DKIST retarder component in the aluminum mounting cell. Each material layer has different transmission and absorption properties that depend on wavelength and thickness. The depth of absorption and heating depends on the input spectrum and the optical constants of the optic. Using the input power spectra along with the optical constants for Infrasil 302 fused silica, the heat budget and the flux absorbed with depth was calculated.

In the thermal FEM, the optics were modeled as 17 independent layers sampling the two cover windows and six crystals. In the associated stress FEM there were 4 to 8 stress model elements near the location of each thermal model node.  We used the temperature nodal data and the associated stress element data provided by Hofstadter Analytical LLC to interpolate stress elements on to the temperature node structure and extracted statistical information about the associated temperatures and stresses. For the thermal gradient data presented here, the temperature nodal data was interpolated to find the temperature of the center of each crystal plate. The nodal structure was centered on the coating locations to deposit coating absorbed heat at the correct depth. The crystal plate temperatures are the average of the top and bottom temperature nodes bracketing the plate location.

\subsection{Coating Absorptivity: Heating Impacts \& Photo-Thermal Measurements}

Absorption of anti-reflection coatings can be a very significant heating term when considering all 16 surfaces in a six-crystal plus two cover-window design. Initially, our first coating run with an initial vendor included coatings that absorbed over 30\% of wavelengths shorter than 400 nm. Subsequently these highly absorptive coatings were stripped and a new process developed to ensure low absorption.  We performed a thorough characterization of the coating absorption for every coating shot used on the DKIST retarders. We used Stanford Photo-Thermal Solutions (S-PTS) to verify coating absorptivity for this new process at six wavelengths throughout visible and near-infrared wavelengths (405 nm, 532 nm, 690 nm, 785 nm, 830 nm and 1064 nm) using their interferrometric technique\cite{2009SPIE.7193E..0DA}.

\begin{wrapfigure}{l}{0.60\textwidth}
\vspace{-4mm}
\centering
\hbox{
\hspace{-0.2em}
\includegraphics[width=0.58\textwidth]{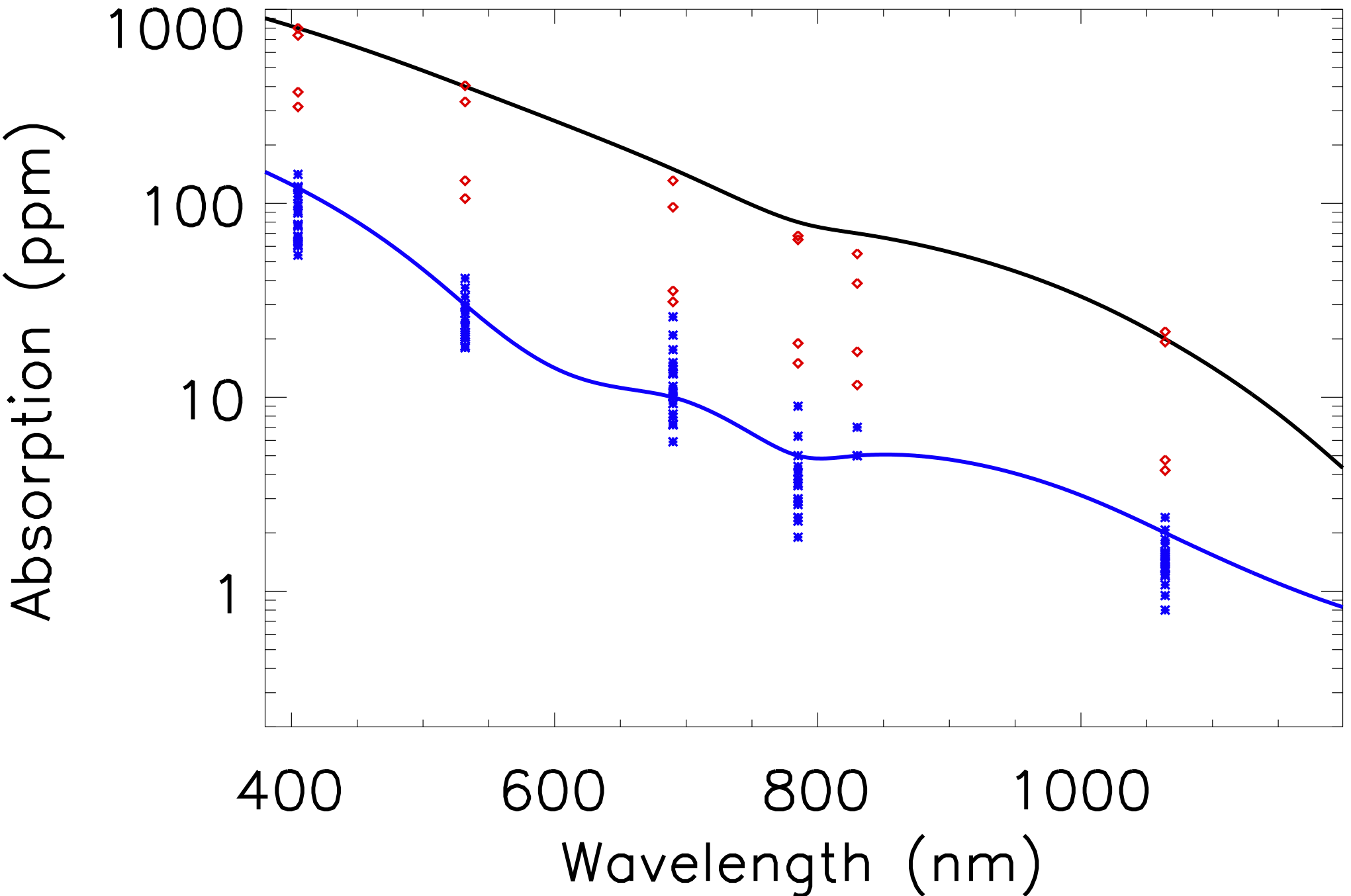}
}
\vspace{1mm}
\caption{\label{fig:coating_heat}  The isotropic MgF$_2$ anti-reflection coating absorptivity measured at S-PTS in parts per million. The black curve shows typical results from witness samples for a standard coating run. Blue shows typical values from our low-absorption process.}
\vspace{-3mm}
\end{wrapfigure}

Figure \ref{fig:coating_heat} shows the S-PTS measured coating absorption for all shots used to coat the DKIST retarder crystals.  The black curve showed a typical, non-contaminated coating shot.   The blue curve shows the new low-absorption coating process with integrated heat at roughly 1/10$^{th}$ the nominal levels.  This new process never resulted in a contaminated coating that required stripping. We have tested witness samples from all our coating shots and the blue symbols in Figure \ref{fig:coating_heat} show all data. The blue curve is typical of our low-absorption coatings per S-PTS\cite{2009SPIE.7193E..0DA}.  

We compute the coating heat as the cumulative sum over all wavelengths incident on the coating multiplied by the smooth coating absorption curves of Figure \ref{fig:coating_heat}.  The coating absorption is dominated by short wavelengths with a fairly smooth spectral dependence. the cumulative distribution is dominated by wavelengths in the 400 nm to 800 nm wavelength range where the solar sepctrum contains most of the incident power. The difference between coating absorption is roughly 55 milliwatts for the black curve and roughly 5 milliwatts for the blue curve. The nominal 55 mW absorption created nearly a watt of heating when considering 12 absorbing coatings on the six crystal retarders and the other four coatings on the now-removed two cover windows. This coating heat is a significant fraction of the total heat budget.

\subsection{Bulk Material Absorptivity: Crystal \& Window Transparancy}

The materials in the retarder have strongly varying spectral absorption. Crystal quartz and Infrasil absorb significantly at wavelengths longer than roughly 3000 nm. Crystal MgF$_2$ retarders absorb wavelengths longer than 6000 nm and were initially designed with CaF$_2$ cover windows that also absorb wavelengths than 7000 nm. 

As part of this study, we also used Stanford Photo-Thermal Solutions (S-PTS) to verify the crystal bulk material absorptivity from our material providers. Often, material data sheets show absorption at levels typical of spectrophotometric limits around 0.05\%.  Materials catalogs will also quote transmission for various materials as 99.95\% in typical curves when the actual material is orders of magnitude more transparent. For our 300 Watt incident load this unrealistic 0.05\% absorption value incorrectly becomes the dominant term in the heating budget. We sent samples of our crystal quartz, MgF$_2$, Infrasil and CaF$_2$ to S-PTS for verification and we did indeed find that absorption was less than 10 ppm for the samples in the middle of the expected transmission band.

\begin{wrapfigure}{l}{0.63\textwidth}
\vspace{-4mm}
\centering
\hbox{
\hspace{0em}
\includegraphics[width=0.61\textwidth]{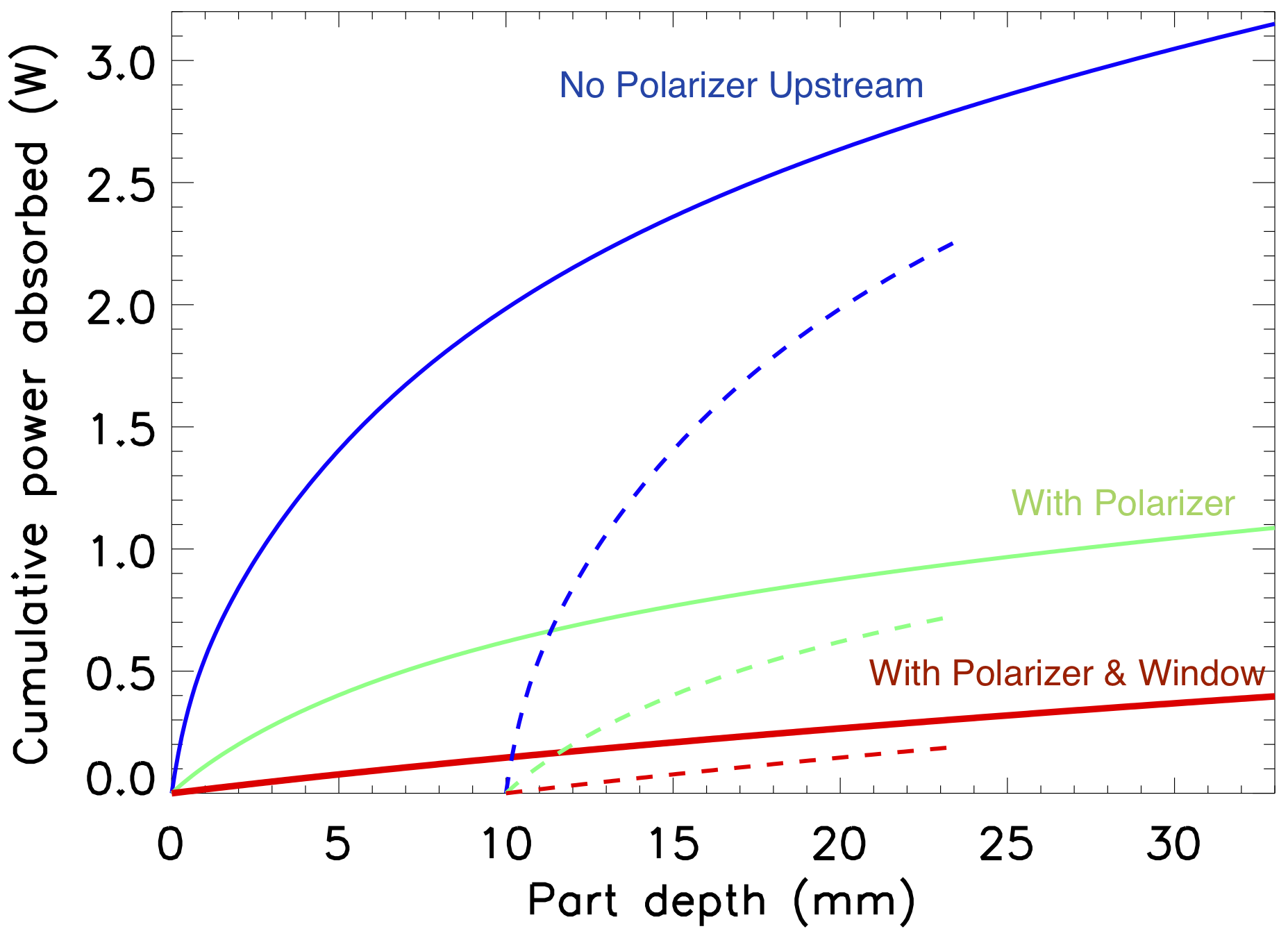}
}
\caption{\label{fig:heat_absorption_rates_quartz} The absorbed power distribution with depth through the quartz retarder. The nominal design had cover windows giving 33 mm total thickness. The crystals alone account for $\sim$12 mm of optical thickness in the middle of the part. Solid lines show the design with cover windows while dashed lines show the design without windows. The dashed lines start at a nominal depth of 10 mm for clear comparison with the thicker window-covered crystals. The blue curves show the cumulative heat load with depth when the optic is unprotected in the 300 W beam.  We show heat calculations with a polarizer (CalPol1) mounted above as green lines. Heat loads when the polarizer is used with another 25 mm thick Infrasil window (CalPol2) as red lines. }
\vspace{-3mm}
\end{wrapfigure}

To compute the heat load with depth through an optic, we use Beers-law for nominal absorption in a material along with the actual incident solar spectrum from far UV to thermal NIR. We used sequential layers of 0.1 mm thickness to recompute the absorption as functions of depth as well as to modify the spectral flux incident from one layer on the subsequent material layer.  By following this iterative process, we can correctly absorb the spectral flux at the proper depth and distribute the heat load correctly as the beam is sequentially absorbed in propagation through the optic.  

We also follow the same procedure for computing the spectral power removed from the incident beam by optics mounted upstream of the retarder. During DKIST calibration, we use either one of two polarizers or no upstream optic. One polarizer is a wire grid protected by a coating on a 1 mm fused silica substrate (CalPol1). The second polarizer is the same wire grid but with an additional 25 mm thick Infrasil window mounted downstream of the polarzer (CalPol2). This second window absorbs significantly more NIR wavelengths and removes load from the crystal retarder.  We assess the polarimetric impact of this window in other sections, but we note that this second window + polarizer assembly effectively removes more than 90\% of the thermal load on the quartz retarder.  The polarizer alone removes all thermal load from the MgF$_2$ retarder. 

Figure \ref{fig:heat_absorption_rates_quartz} shows the cumulative distribution for the optical power absorbed by the bulk material as a function of depth for the various crystal quartz retarder designs and use cases. Solid lines show heat loads for a retarder that includes the 10 mm thick cover windows.  Dashed lines show the heat loads for the quartz crystal stack without cover windows.  The blue curves show the quartz retarder without any optic mounted upstream, fully illuminated by the 300 Watt beam.  The solid line shows 3.1 Watts is absorbed in the nominal covered design while the dashed blue line shows that roughly 2.3 Watts is absorbed in the optic when no cover windows are used. For quartz, absorption of NIR wavelengths dominates the heat distribution. Removing the cover windows reduces the thermal load and it also does change the depth dependence as crystal is substantially more conductive than glass. 

The green curves show heat loads when the polarizer (CalPol1) is mounted upstream of the quartz retarder. The power absorbed by the optic without cover windows is roughly 0.7 Watts compared to 2.3 Watts when this no-cover-window optic is used alone in the beam without the protection of the polarizer. The red curves show the heat loads when the quartz is used with the combined wire grid polarizer and 25 mm thick Infrasil window (CalPol2) mounted above the quartz retarder. In this case, most of the NIR wavelengths are removed from the beam before the retarder. Both with-windows and without-windows cases see greatly reduced absorption. The no-cover-window optic sees 0.2 Watts, which is significantly smaller than the coating absorption loads described above. We recently removed the cover windows from the optics and changed the DKIST design, in part due to this thermal analysis.   

\begin{wrapfigure}{r}{0.63\textwidth}
\vspace{-2mm}
\centering
\hbox{
\hspace{0em}
\includegraphics[width=0.61\textwidth]{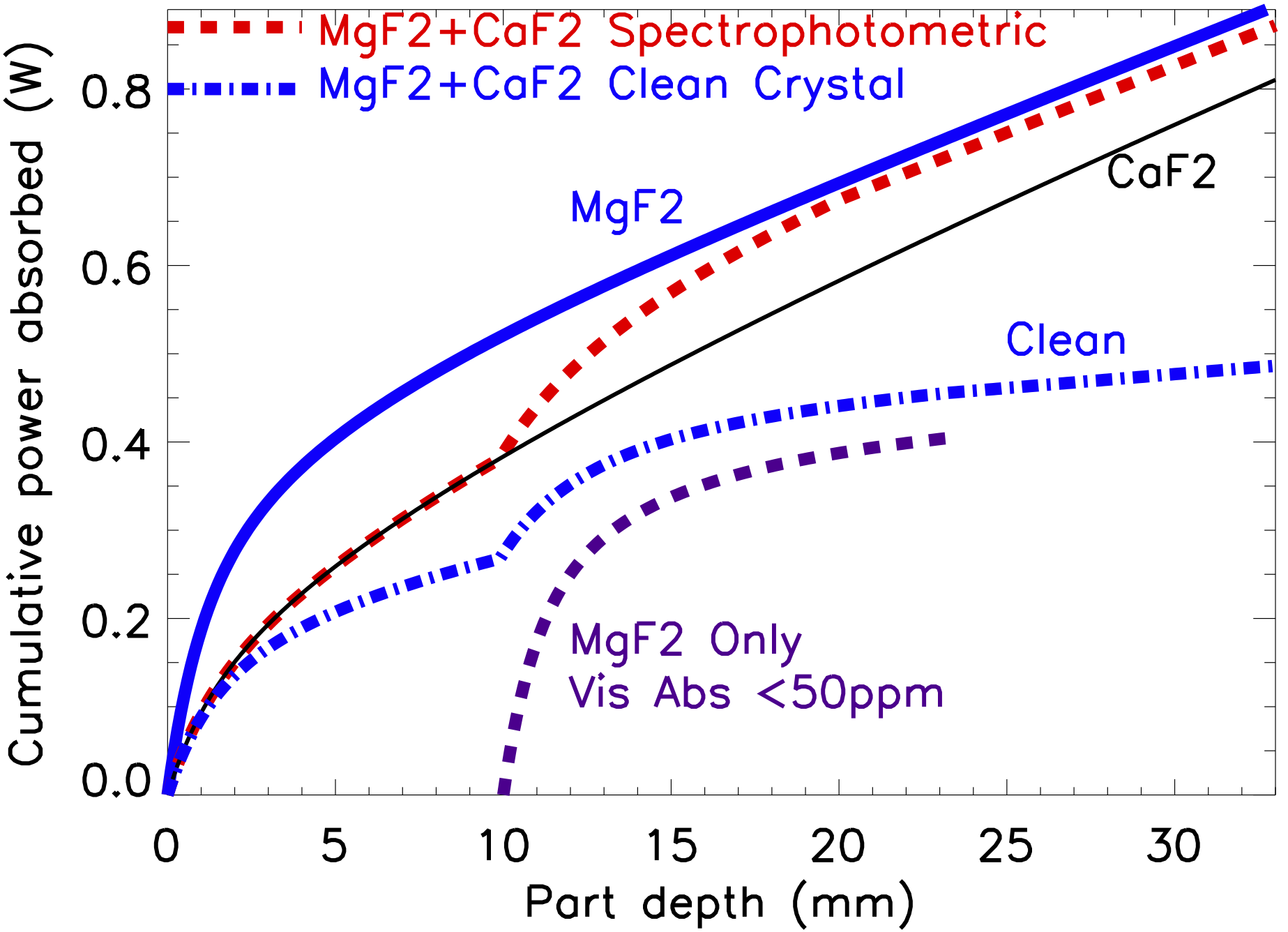}
}
\caption{\label{fig:heat_absorption_rates_mgf2} The heating distribution with depth through the optic for the MgF$_2$ retarders. The nominal design with cover windows has each optic over 33 mm thick. The curves originating at 0 mm part depth correspond to models with the CaF$_2$ cover windows. The purple dashed line starting at 10 mm part depth corresponds to the retarder model with only MgF$_2$ crystals reaching 0.4 Watts heating. The solid blue and black curves shows the heat load when assuming the 0.05\% spectrophotometric limit assumed on most material data sheets. The dashed red line shows the net heat load on the optic reaching 0.85 Watts with cover windows when both CaF$_2$ and MgF$_2$ materials are assumed to have these limits. The dot-dashed blue line shows the {\it clean} materials with photo-thermal absorption limits in the parts per million range where the optic heat load only reaches 0.5 Watts.}
\vspace{-3mm}
\end{wrapfigure}

Figure \ref{fig:heat_absorption_rates_mgf2} shows the crystal MgF$_2$ retarder with and without the now-removed CaF$_2$ cover windows. We only show heat loads without any polarizers mounted above the retarder. Mounting the polarizer above the MgF$_2$ retarder removes all heat load as there are no coatings and the wire-grid polarizer absorbs all wavelengths longer than 6000 nm. For the retarders with cover windows, the total optical thickness was over 33 mm. The blue curve shows MgF$_2$ absorption alone while the black curve shows the CaF$_2$ window absorption alone.  The dashed red curve shows the heat load when cover windows are used. The step at 10 mm optical depth represents the slightly shorter wavelength transmission band of MgF$_2$ absorbing around 6000 nm wavelengths after the CaF$_2$ cover window has removed the rest of the flux in the IR bandpass.  When the two windows at 10mm thickness each are removed, only the inner 12 mm of crystal MgF$_2$ optical path remains to absorb heat. We show the dashed purple curve where we use crystal-type absorption limits of less than 50 ppm at visible wavelengths following our measurements as opposed to spectrophotometric limits of 0.05\%  (500 ppm). In this no-window scenario, the heat is deposited strongly in the first 4 mm of the optic given the sharp transition from transparent to absorbing at IR wavelengths.  

Another minor consideration is the refractive index matching oil used between all layers.  This oil could possibly cause small absorption and possible degradation with time.  We have spectrophotometric measurements from 300 nm to 6000 nm wavelength for a 1 cm thick sample without any detectable absorption.  We have also done extensive testing for UV damage to this oil, including multiple years worth of effective exposure to 325 nm and 360 nm wavelengths.  No significant spectral absorption was detected after these irradiation tests.  Thus we do not include a heating term for the oil.

The spectral dependence of the bulk material heating for the DKIST retarders is dominated by near infrared wavelengths. The cumulative distribution functions show that nearly no signifcant power is absorbed by the quartz for wavelengths shorter than 2800nm. But between 3000 nm and 5000 nm wavelength, almost all the heat variation is seen. The wire grid polarizer effectively absorbs 5500 nm and longer wavengths with only 20\% transmission at 4500 nm wavelength. When using an Infrasil window in combination with a polarizer, the bulk heat load on the quartz retarders goes to nearly zero. A polarizer alone will remove the bulk heat load from the MgF$_2$ crystal retarders.  The crystals, coatings and oils are all very transparent at visible wavelengths requiring spectral propagation for accurate calculations of the thermal loads with depth through the optic considering varying optical configurations during DKIST calibration and operation.

\subsection{Six Heating Scenarios: With \& Without Upstream Polarizer \& 3 Coating Loads}

We consider thermal models of the quartz retarder to demonstrate the polarimetric impact of temporal, radial and depth dependence of the temperature distribution. We tested a range of coating scenarios for the isotropic MgF$_2$ anti-reflection coatings ranging from optimistic to pessimistic. We used coating heat values of 10 mW, 30 mW and 100 mW per coating when the polarizer is not mounted upstream. The coatings absorb stronger at short wavelengths, so the coating heat is reduced by roughly 2x when the polarizer is mounted, even though the bulk heating terms are changed significantly more than 3x. 

\begin{wraptable}{l}{0.57\textwidth}
\vspace{-2mm}
\caption{Thermal FEM Material Properties}
\label{table:materials_properties}
\centering
\begin{tabular}{l l l l l l l}
\hline
\hline
Material	& Modul.  	& Pois.	& CTE		& Cond.	& Spec.		& $\rho$		\\
Name	& Elast.  	& Ratio	& $\alpha$	& 		& Heat		& 			\\
\hline
Infrasil	& 70		 & 0.17	& 0.51		& 1.38	& 772		& 2.2 		\\
Quartz E	& 97.2	  & 0.56	& 7.1			& 10.7	& 710		& 2.65		\\
Quartz O	& 76.5	  & 0.22	& 13.2		& 6.2		& 710		& 2.65		\\
Al		& 68.9	  & 0.33	& 23.6		& 167	& 896		& 2.7			\\
RTV 		& 3.51	  & 0.40	& 270		& 0.21  	& 500		& 1.05		\\
Steel		& 193	  & 0.25	& 17.2		& 16.2	& 500		& 8.0			\\
CaF$_2$	& 75.8	  & 0.26	& 18.7 		& 9.71	& 853		& 3.18		\\
MgF$_2$	& 138.5	  & 0.27	& 13.7		& 11.6  	& 955		& 3.15		\\
\hline
\hline
\end{tabular}
\vspace{-3mm}
\end{wraptable}

We also use two optical configurations with and without the calibration wire grid polarizer to show the impact of different incident power absorbed with depth curves. The resulting four heating models are similar in behavior, but different in gradients and temperature rise. There are 14 coatings in the interior of the part (not exposed to air).  For these simulations, we also use the models for optics with cover windows. As glass is a poor conductor, this internal heat source exacerbates internal depth and radial temperature gradients. These models were a large part of the motivation to remove the cover-windows.  In addition, the crystal conductivity is a factor of roughy five more than glass.  

Crystal-only simulations have greatly reduced thermal gradients both with depth and across the clear aperture.  When using crystal-only models with polarizers and windows mounted above, the heat loads are significantly smaller.  These thermal models then become more strongly coupled to assumptions about heat transfer through the bonding RTV, the temperature conduction through rotation stage bearings, forced air cooling assumptions and several other model-specific variables.  For simplicity, we show the cover window scenarios in detail and use them to motivate subsequent removal of cover windows from the as-built retarders.

Table \ref{table:materials_properties} shows the materials properties assumed in the thermal and stress FEMs.  The modulus of elasticity is in units of Gigapascals (GPa) in the second column.  The CTE ($\alpha$) is in parts per million per $^\circ$C in the third column. Poissons ratio is unitless in the fourth column.  Conductivity is in Watts per meter per $^\circ$C in the fifth column.  Specific heat is in Joules per kg per $^\circ$C is in the sixth column.  Density ($\rho$) is in 10$^3$kg per cubic meter in the last column. We use Aluminum 6061, RTV 118, and 303 Stainless steel.  

The crystal quartz is assumed to be an orthotropic material, consistent with the ordinary index being aligned to Z in a uni-axial A-plane cut crystal retarder and having ordinary and extra-ordinary axes rotated about the optical axis per the achromatic design.  Rotations of these crystal axes are set by the achromatic retarder design, for example, [0$^\circ$, 90$^\circ$, 65$^\circ$, 155$^\circ$, 0$^\circ$, 90$^\circ$] for the DL NIRSP modulator.  

At Gregorian focus with a $\sim$300 Watt optical load, the coatings provide a wide range of heating variation. For the worst heating scenario using {\it no-polarizer} at high flux levels as well as the more pessimistic coating absorptivity, the 14 coatings can absorb at 333 ppm giving 1.4 Watts total (100 mW per coating). With a more optimistic coating absorption, the 14 internal coatings absorb at 100 ppm giving 0.4 Watts total load (30 mW per coating). When similar absorption rates are used with the polarizer mounted above the retarder, the heat load from the coatings drops to 0.7 Watts and 0.2 Watts for the better or worse respectively. As the coatings absorbed 10x more light at 532 nm than 1064 nm in the photo-thermal testing, we assume the changing infrared flux levels absorbed by the bulk material with varying configuration does not significantly change the assumed coating absorption terms. 

\begin{wrapfigure}{r}{0.52\textwidth}
\vspace{-4mm}
\centering
\hbox{
\hspace{0em}
\includegraphics[width=0.52\textwidth]{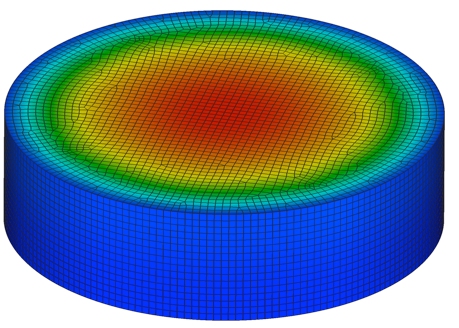}
}
\caption{\label{fig:therm3d} The temperature distribution in the model with cover windows at 33 mm total thickness. We assumed the high bulk heat {\it no polarizer} configuration with {\it better} coating absorption at 100 ppm per coating. The radial temperature gradient of roughly 7$^\circ$C is seen from red to blue colors. See text for details.}
\vspace{-4mm}
\end{wrapfigure}

The two different optical configurations have surprisingly different heat loads when considering bulk absorption.  Without the polarizer, there heat loads are 2.00 Watts bulk absorption in top Infrasil window, 0.67 Watts bulk absorption in crystal quartz layers and 0.33 Watts bulk absorption in bottom Infrasil window. With the polarizer, the heat loads are 0.62 watts bulk absorption in top Infrasil window, 0.26 Watts bulk absorption in the crystal quartz, 0.17 Watts bulk absorption in bottom Infrasil window.

An example thermal FEM output is shown in Figure \ref{fig:therm3d}.  The color scale varies linearly from blue at 33$^\circ$C to red at 39.66$^\circ$C, covering roughly a 7$^\circ$C range. The center of the optic is significantly hotter than the edge which conducts heat through the bonding RTV to the rotation stage that is fixed at ambient temperature.  Most thermal model outputs show similar behavior to Figure \ref{fig:therm3d} with a hot center, cooler edges and some depth dependence to all temperature gradients.

\begin{figure}[htbp]
\begin{center}
\vspace{-1mm}
\hbox{
\hspace{-0.0em}
\includegraphics[height=15.5cm, angle=0]{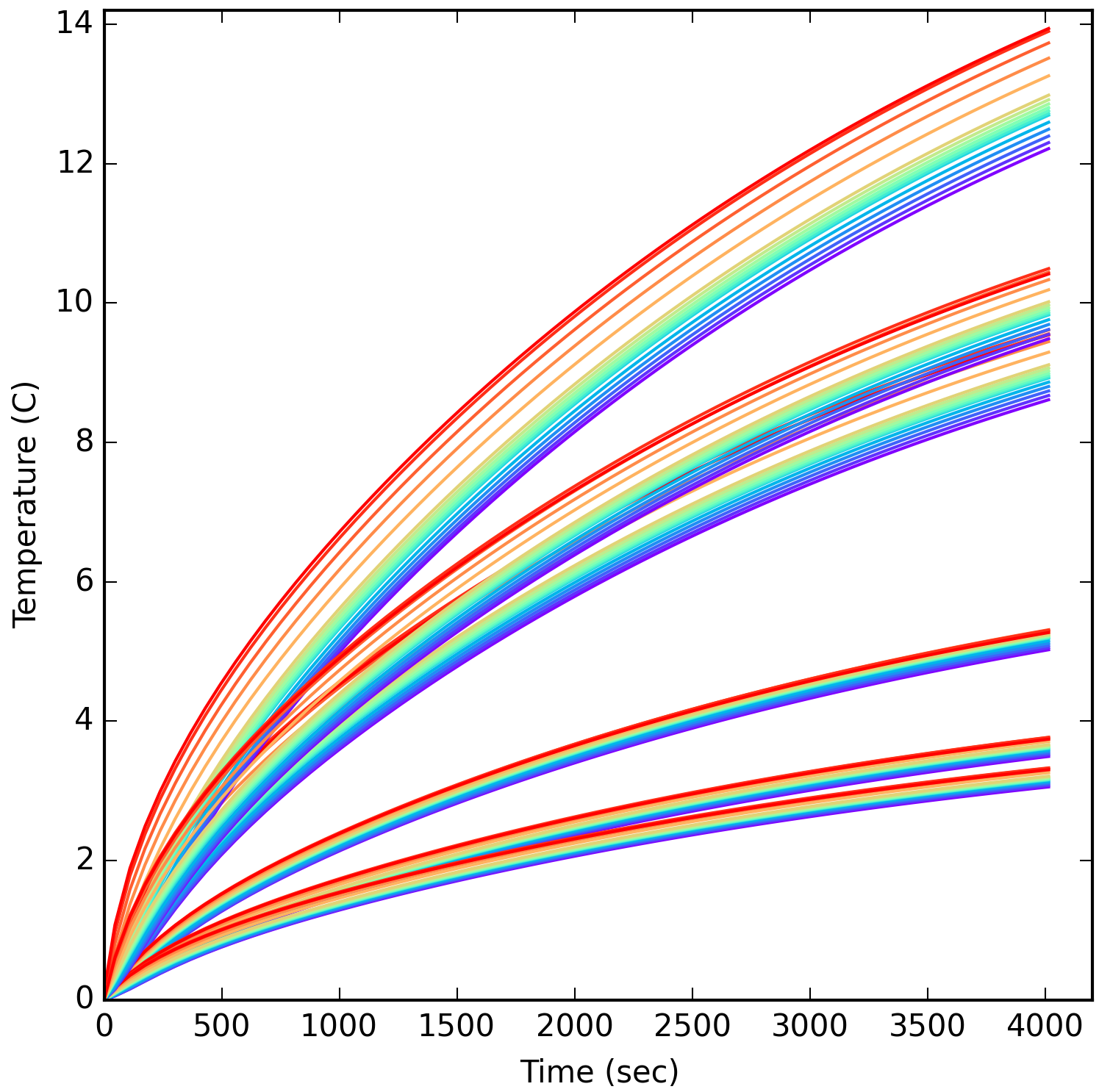}
}
\caption{\label{fig:quartz_retarder_heat_with_time_depth} Results of thermal finite element models (FEMs) for quartz calibration retarders in the 300 W beam. There are 6 curve families that each represent one model of retarder temperature versus time. For each family of curves, we show temperatures for 17 layers (depths along the optical axis) within the optic.  The red curve shows the top of the optic facing the incident beam. The blue curve shows the bottom of the part at beam exit. The color progression from red to green to blue shows the behavior with depth through the part. The further spread the colors are, the bigger the thermal gradients within the optic. The red and orange colors correspond to the top Infrasil window.  The green and yellow colors correspond to the 6 quartz crystals.  The blue colors correspond to the bottom Infrasil window.  The top three families correspond to scenarios where the calibration polarizer is not inserted in the beam and the full 300W reaches the quartz retarder.  We used coating heat values of 10 mW, 30 mW and 100 mW per coating. The bottom 3 families of curves correspond to scenarios where the calibration polarizer is inserted above the quartz retarder.  This polarizer reflects roughly half the light, absorbs roughly 10\% of the light (aluminum reflectivity) and the fused silica substrate also absorbs wavelengths longer than about 5$\mu$m.  This reduces the heat absorbed by the quartz retarder crystal and glass by a factor of roughly 3x. We used coating heat values of 5mW, 10mW and 50mW per coating for these three lower curve families. We ran the simulations for 8 hours to reach steady state temperatures roughly near the asymptotic values seen above here. The major limiation is the low conductivity of Infrasil as a glass insulator. }
\vspace{-2mm}
\end{center}
\end{figure}

Figure \ref{fig:quartz_retarder_heat_with_time_depth} shows families of heating curves corresponding to the three coating absorption levels and the two optical configurations for calibration (with / without upstream polarizer). The bulk temperatures rises roughly 3 to 5 times faster upon initial illumination when the polarizer is not mounted upstream of the quartz retarder. The temperature gradient with depth assumes a value nearly matching the steady-state value within less than 3 minutes. This gradient with depth is relatively constant through the quartz retarder with time over hours. The thermal gradient amplitudes are case specific but are roughly 0.2$^\circ$ when the polarizer is mounted and roughly 0.8$^\circ$ when the polarizer is not used. The different coating absorption values do not seem to change the thermal gradients significantly but do increase the heat load and hence drive temperatures higher faster. 

The bulk temperature of the part rises more than 2$^\circ$ in the first 20 minutes but the behavior of the different cases is quite varied. The temperature dependence of all 17 layers for each of the scenarios is shown in Figure \ref{fig:quartz_retarder_heat_with_time_depth}. Each nodal depth layer is a different color with red for the top layers, green for the middle layers and blue for the bottom layers. The scenarios without the polarizer have the highest temperature increases reaching 14$^\circ$C above ambient for the top of the optic absorbing the NIR flux and 12$^\circ$C above ambient for the bottom of the optic.  The top most family of curves corresponds to {\it worst} coating absorption values (330 ppm) and the higher power heat load without the polarizer mounted in the beam.  Reducing the coating heat from 100 mW per coating to the lower absorption values typical of our new process does reduce the heating curves by roughly 25\%.  However, these curves all show rapid temperature rise.

When the polarizer is mounted above the retarder, the temperature rises are much slower. Steady-state is achieved at significantly lower absolute temperatures. The polarizer-protected quartz retarder rises roughly 2$^\circ$ to 3$^\circ$ in the first 20 minutes of heating. If the polarizer is not in the beam, the retarder heats up 7$^\circ$ to 9$^\circ$ in the same time period.

\clearpage
\subsection{Thermal Gradients: Distribution Over the Clear Aperture With Depth And Time}

The thermal spatial gradients in the window-covered designs are largely independent of any convection or external air cooling as Infrasil is a good insulator. As the DKIST retarders are mounted near focal planes, the beam footprints on the optic sample varying spatial regions across the clear aperture. Thus, a calibration must assume some amplitude of field variation in the presence of temporal instability. The design challenge is to create a retarder that does not vary spatially to levels of significant impact. 

\begin{figure}[htbp]
\begin{center}
\vspace{-1mm}
\hbox{
\hspace{-0.3em}
\includegraphics[height=8.3cm, angle=0]{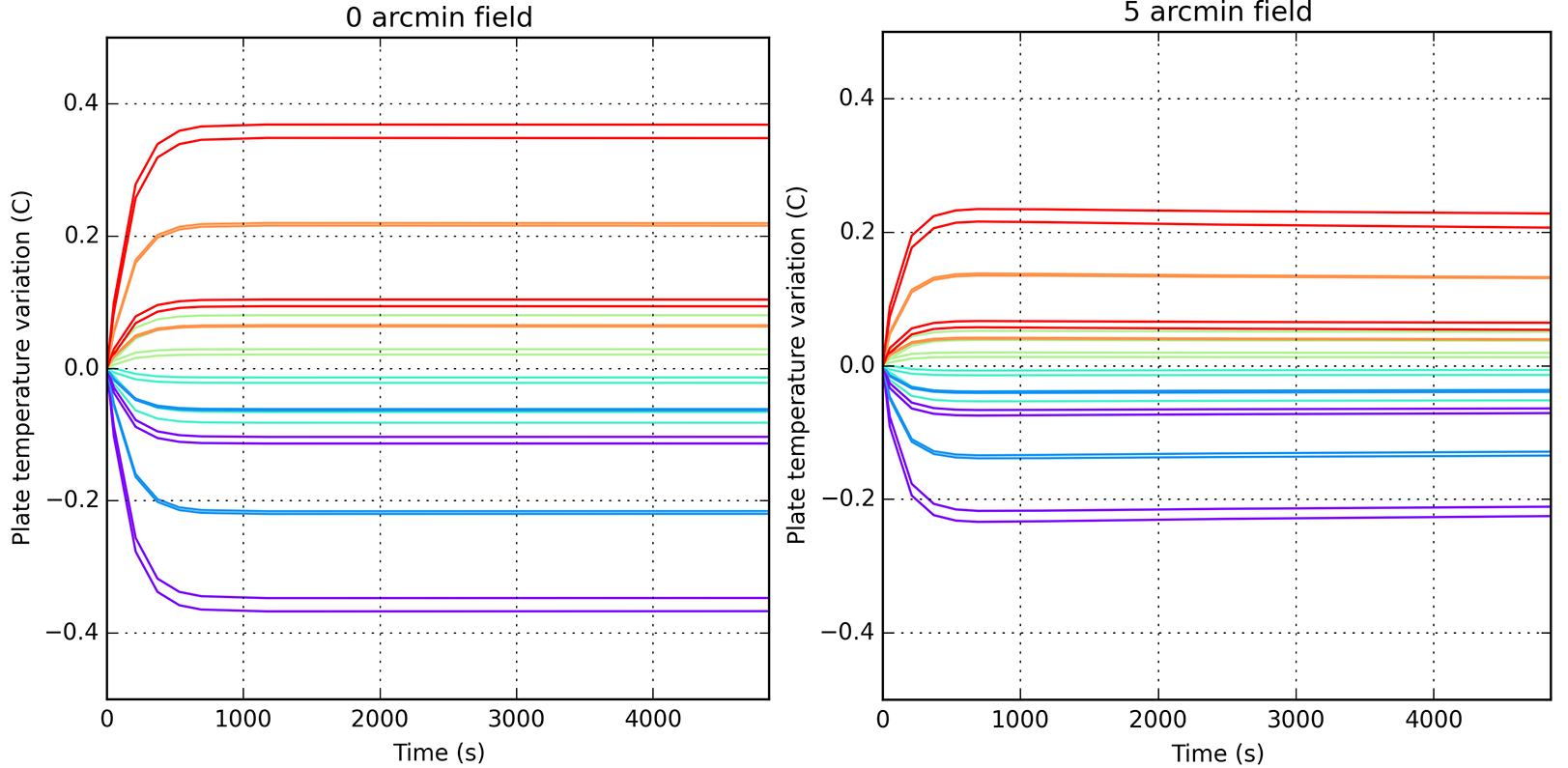}
}
\caption{\label{fig:tempgrad} The temperature gradients across the 6 plates of the quartz retarders as a function of time for all 4 heating scenarios. The temperature difference are shown for the optic center (left) and for the 5 arc minute field edge (right) at a distance of 49 mm from part center. The temperature gradient is established in $<$400 seconds of heating. The highest heat load scenario without the polarizer and assuming {\it worse} coating absorption of 330 ppm per surface are the outer most colored curves. The 100 ppm coating heat model gives almost identical results. Another 2 red curves show the lower heat scenario with the polarizer mounted in the beam for 330 ppm and 100 ppm coating absorption values respectively. }
\end{center}
\end{figure}

 Figure \ref{fig:tempgrad} highlights the radial and depth dependence of the temperature distribution. The top crystals get the hottest and are shown in red for all four scenarios. The bottom crystals are the coolest and are shown in blue.  The spread in temperatures between red and blue curves is the temperature depth gradient. The average temperature was subtracted from the temperature of each crystal plate for every time step modeled. The gradient at the part edge is roughly half the amplitude than at the center of the optic as seen by the difference in left and right graphics in Figure \ref{fig:tempgrad}. As there are two red curves very close to each other, we conclude that changing the coating absorption from 100 ppm to 330 ppm does not significantly change the thermal gradient. 

There are significant changes in this thermal gradient with radius from the center of the part out to the edge of the optic where the glass contacts the RTV and the cell mount. The thermal gradient is roughly double the amplitude at the center of the optic than near the edge of the illuminated region. This radial dependence will change the behavior of the Mueller matrix as a function of field since the optic is near a focal plane. For the retarder optics near Gregorian focus, the footprint for the 2.8 arc minute field requires a 66.3 mm clear aperture and the full 5 arc minute field requires a 98.1 mm clear aperture.

The gradient across the 6 crystal plates is established quite quickly. The gradient reaches $>$80\% of it's steady state amplitude within $<$300 seconds. Figure \ref{fig:tempgrad} shows the difference between the average plate temperature and the 6 individual plate temperatures for the 4 scenarios. Red shows the top plate, purple shows the bottom plate.

\begin{wrapfigure}{r}{0.54\textwidth}
\vspace{-2mm}
\centering
\hbox{
\hspace{0em}
\includegraphics[width=0.53\textwidth]{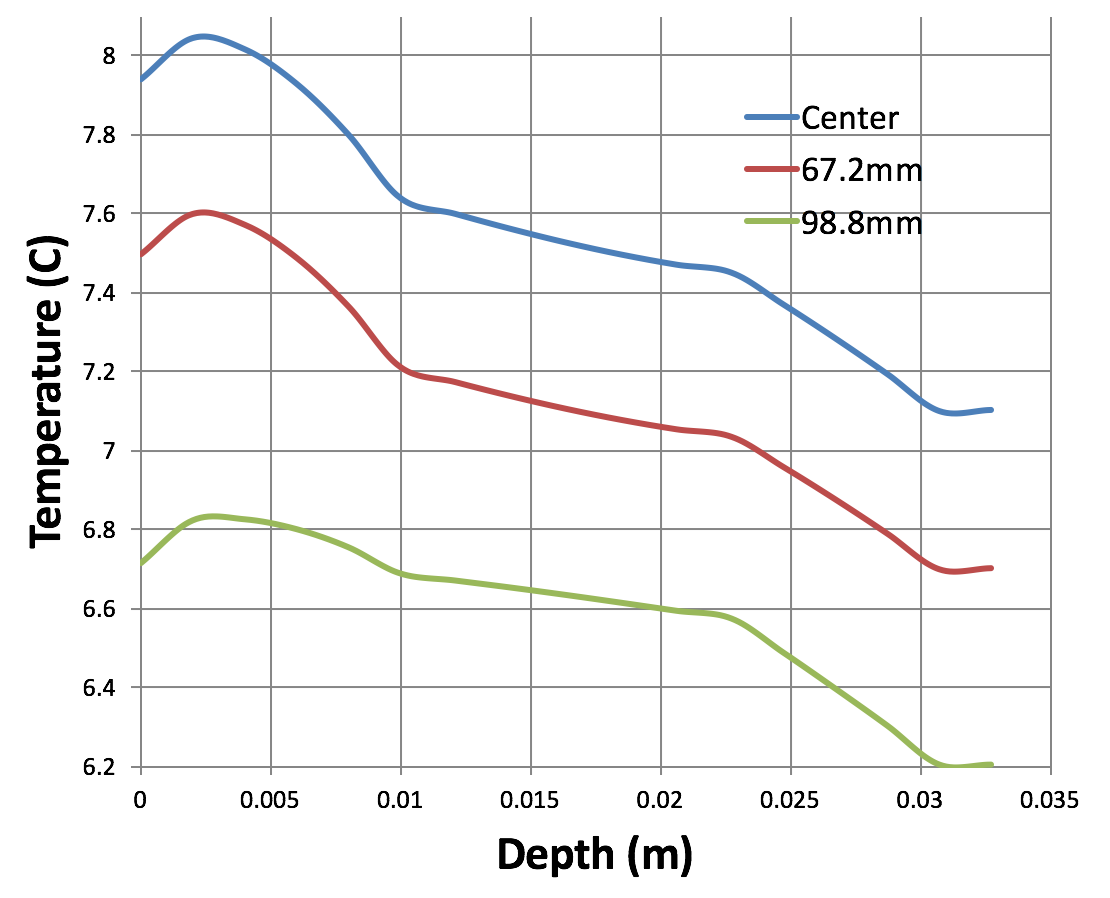}
}
\caption{\label{fig:therm_DL_temp_vs_depth_with_radius} The temperature versus depth through the 33 mm thick optic at three select clear aperture locations for a quartz six-crystal retarder with windows on each side.}
\vspace{-4mm}
\end{wrapfigure}

Temperature gradients of $\pm$0.4$^\circ$C steady state are seen in the high heat scenario without the polarizer mounted in the beam. The 5 arc-minute field edge has a gradient with roughly half the amplitude. For both positions on the optic, the thermal gradient is near the steady-state amplitude in $<$400 seconds even though the time to steady state is about 8 hours in these simulations.

Figure \ref{fig:therm_DL_temp_vs_depth_with_radius} shows the depth dependence of temperature at three select clear aperture locations for a thermal model beginning at -5$^\circ$C. After 7200 seconds of heating, the optic is roughly 12$^\circ$C above the ambient -5$^\circ$C. The center of the optic is roughly 0.9$^\circ$C warmer than the edge of the clear aperture at 98.8 mm diameter.  The top Infrasil cover window has a somewhat parabolic shaped temperature profile as the thermal model includes forced-air cooling on the exterior surface.  The cooling however is quite ineffective given the low conductivity and long thermal time constants.  The six crystal stack occupies depths from 10 mm to 22.6 mm and the increased crystal conductivity flattens the thermal gradient with depth in this region.  The center of the optic has roughly a 0.2$^\circ$C gradient with the clear aperture edge seeing roughly half this gradient.  The bottom window is cooler than the top window and sees a more linear depth gradient.

\subsection{Thermal Impact of Removing Cover Windows: Reduced Gradients \& Loads}

Removing the cover windows drastically reduces the thermal gradients with depth through the six crystal retarder optic as well as radially across the clear aperture of the optic. Here we show revised thermal models for {\it no-cover-window} retarders under three typical calibration configurations.

\begin{wrapfigure}{l}{0.55\textwidth}
\vspace{-3mm}
\centering
\hbox{
\hspace{0em}
\includegraphics[width=0.53\textwidth]{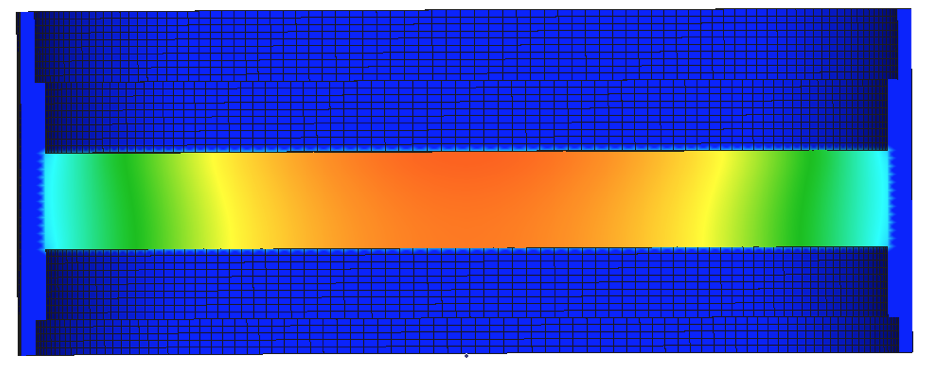}
}
\caption{\label{fig:therm_Cryo_removed_windows_temp_distribution} A slice through the 3D temperature model for the CryoNIRSP SAR without windows. Blue colors show the cooler temperature of the metal cell. The temperature scales from 20.5$^\circ$C to 20.9$^\circ$C from blue to red. There is a 0.4$^\circ$C gradient from the center of the clear aperture towards the edge of the optic bonded to the cell.  There is a very small gradient with depth seen as the slight change in color between top and bottom of the optical exterior interfaces.}
\vspace{-2mm}
\end{wrapfigure}

The spatial gradient behavior of the quartz calibration optics is essentially the same as Figure \ref{fig:therm_Cryo_removed_windows_temp_distribution}, but the models include conduction through the rotation stage bearings into the mount. The time to the formal steady-state solution is still several hours in these quartz models as they include slow conductivity through the RTV bonding material raising the temperature of a much larger thermal mass. But the heating rates are greatly reduced and as such, the temporal changes are quite slow.

\begin{wrapfigure}{r}{0.55\textwidth}
\vspace{-1mm}
\centering
\hbox{
\hspace{-0.5em}
\includegraphics[width=0.55\textwidth]{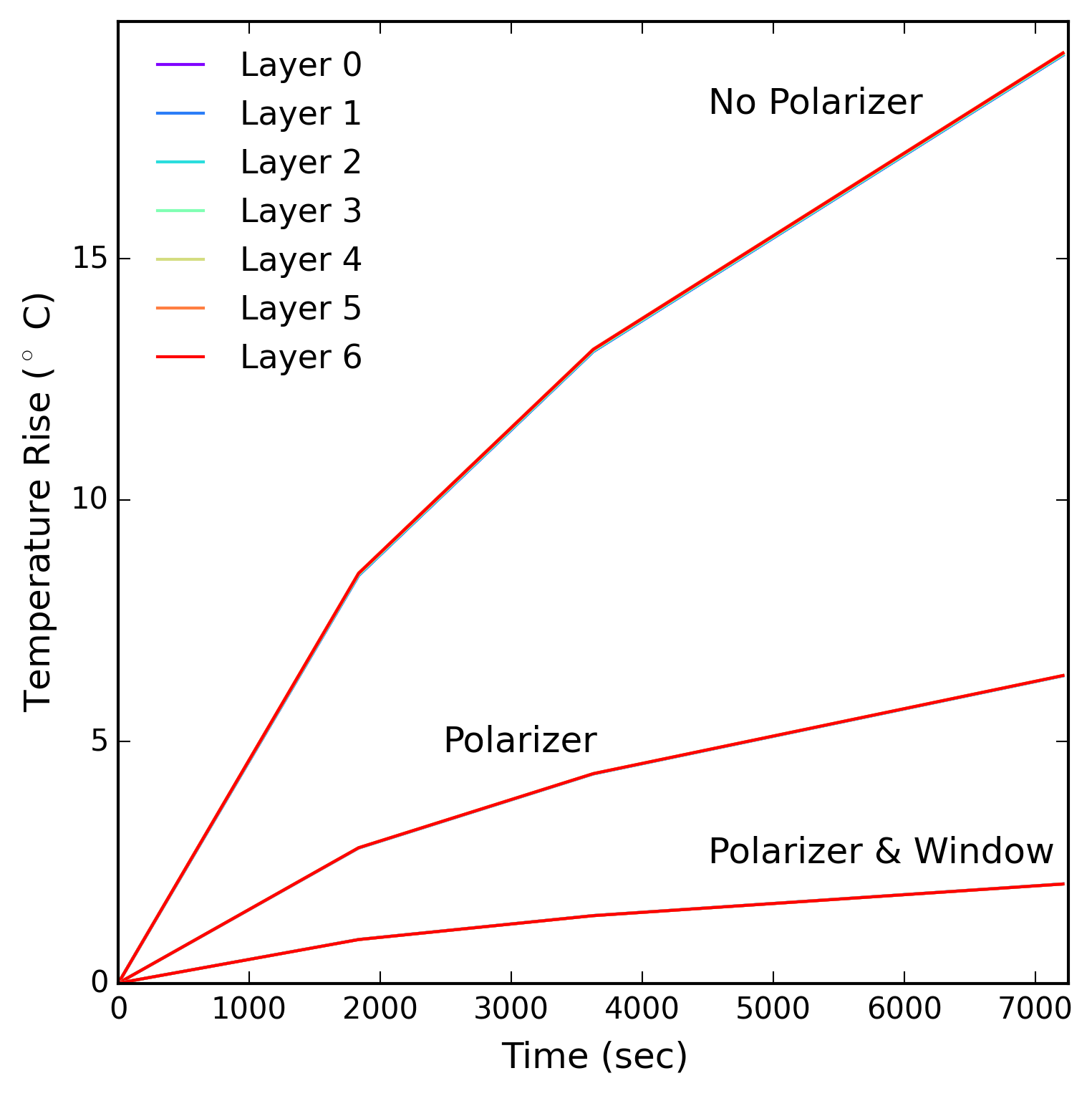}
}
\caption{\label{fig:therm_DL_removed_windows_temp_vs_time} The temperature versus time for three scenarios of the DL SAR six crystal retarder but now without Infrasil cover windows. Each family of curves represents three scenarios.  Like Figure \ref{fig:quartz_retarder_heat_with_time_depth} there are 7 different colors representing the temperature at the top and bottom of each crystal. However, with the revised thermal models, there is almost no change in temperature with depth, so there is hardly any distinguishing of the colors in each curve, unlike the clear gradients seen in Figure \ref{fig:quartz_retarder_heat_with_time_depth}. See text for details.}
\vspace{-4mm}
\end{wrapfigure}

Figure \ref{fig:therm_DL_removed_windows_temp_vs_time} shows revised models for temperature varying with time. The highest heat load would be seen when the SAR is used alone in the beam without protection from the calibration polarizer.  The load is 2.25 Watts with a depth dependence as above strongly concentrated towards the top of the optic. For this model, we assumed 10 mW per coating as an additional heat load.  The temperature rises 19.2$^\circ$C in 7200 seconds, equivalent to 2 hours.  Note that in Figure \ref{fig:therm_DL_removed_windows_temp_vs_time}, we do plot all seven thermal model layers corresponding to top and bottom interfaces for all six crystals.  The thermal gradient is roughly 0.03$^\circ$C and is essentially invisible on this graphic.  This effectively removes thermal gradients from the list of retarder Mueller matrix errors. 

The second scenario in Figure \ref{fig:therm_DL_removed_windows_temp_vs_time} is where the quartz crystal retarder is used with the calibration polarizer mounted ahead in the beam.  The polarizer reflects more than half the incident flux after accounting for the absorption of the aluminum wires. The 1 mm thick fused silica polarizer substrate absorbs all wavelengths longer than roughly 5500 nm, further reducing the load on the retarder.  These factors combined reduce the load to 0.73 Watts and we also assume 5 mW per coating. In Figure \ref{fig:therm_DL_removed_windows_temp_vs_time}, this configuration results in roughly 6.3$^\circ$C heating in two hours of retarder use. The final scenario of Figure \ref{fig:therm_DL_removed_windows_temp_vs_time} is where the quartz retarder is used with a combination of a polarizer and additional 25 mm thick Infrasil window mounted above.  This additional 25 mm of Infrasil reduces the heat load to 0.19 Watts but leaves the coating heat unchanged at 5 mW per coating. In this configuration, the optic heats 2.1$^\circ$C in two hours of use.  

Similar improvement in thermal behavior is seen in the MgF$_2$ calibration retarder, the Cryo-NIRSP SAR.  For this optic when used without a polarizer, the bulk thermal load is 0.40 Watts distributed with depth as above. When this optic is used with the polarizer mounted above, there is no heat load. 

Thermal variation across the clear aperture is still present without the cover windows, but at greatly reduced magnitudes. The significantly higher conductivity of the crystals combined with the lack of thick insulating layers reduces these gradients by a factor of roughly five. Figure \ref{fig:therm_Cryo_removed_windows_temp_distribution} shows a a model for the Cryo-NIRSP calibration retarder used without any polarizer and the 0.48 Watt load.  The thermal variation across the clear aperture of the optic is roughly 0.3$^\circ$.  Depth gradients are nearly negligible.  For this model, the time to steady-state is only 3600 seconds at a temperature only 0.8$^\circ$C above ambient, but these models do not include conduction to the rotary stage and simply fix the cell at a constant ambient temperature.

\subsection{Stress Birefringence Spatial Distribution: Clear Aperture Variation}

Given the strong thermal changes and gradients, the potential for stress birefringence is a concern for the project. The stress optic coefficient for Fused Silica is roughly $\sim$4 nm of phase per mm of thickness per MPa of pressure.

An order-of-magnitude estimate shows that this effect could be the a significant source of error, but only for the window-covered designs under strong thermal loading.  A 25mm thick part at 1MPa pressure can introduce 100nm of phase retardance error. This spatial variation creates aperture dependence (birefringence) and bulk changes (stress-retardance) across the part that impacts our ability to calibrate the telescope. Some of our worst-case models showed stresses approaching a few hundredths of a wave stress values under various time and absorptivity scenarios. These initial results partially motivated this new study.

The thermal FEM was coupled to a stress analysis. The stress model includes many types of stress computations for each node throughout the optical elements and the mounting structure (rotary stages). We include the bonding RTV, expansion of the various mounting and rotation-stage elements.

\begin{figure}[htbp]
\begin{center}
\vspace{-1mm}
\hbox{
\hspace{-0.3em}
\includegraphics[height=8.3cm, angle=0]{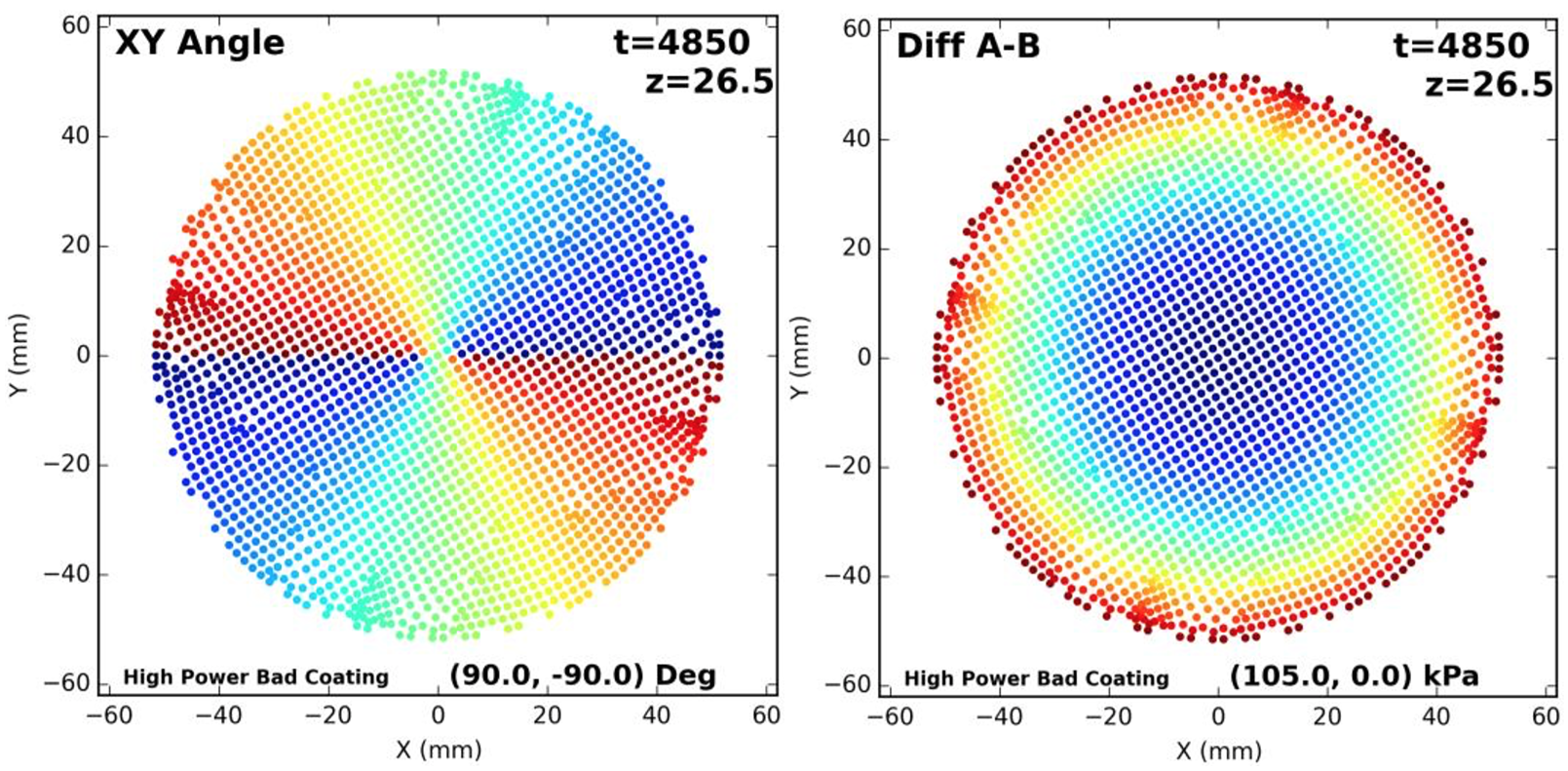}
}
\caption{\label{fig:stressdistribution} The stress analysis computed at a time of 4850 seconds for a layer at 26.5mm depth near the top of the crystal stack.  The higher flux scenario without the polarizer mounted in the beam was used with {\it bad} coating absorption of 330ppm.  Both panels show the spatial distribution of the stresses across the 103mm clear aperture of the part.  Each point shows the stresses computed at a node of the thermal FEM.  The left hand panel shows the angle of the stress in the XY plane. The color scale is linear from red to blue as the angle is changed from -90$^\circ$ to +90$^\circ$.  The right hand panel shows the difference in magnitude between the principal stresses. The color scale is linear from blue to red as stress increases from 0kPa to 105kPa. Note that the dark blue values at the center of the right panel indicate that the stress is in uniform compression (both X and Y magnitudes are equal) such that the stress birefringence is zero. }
\end{center}
\vspace{-4mm}
\end{figure}

We can treat the XY plane stresses as roughly normal to the optical propagation through the optic. We can then estimate the stress birefringence seen by a beam propagating vertically through the optic. This assumption is reasonable for an F/ 13 converging beam with incidence angles mostly below 5$^\circ$.  There will be some angle of incidence and field-of-view effects, but the dominant stress effect is caused by XY stress imbalances.  

The essential result is that the stress birefringence is a smooth radial function driven by heating of the interior of the optic. Infrasil, like all glasses, is an insulator.  Heat deposited by bulk and coating absorption heats the middle of the part.  The glass begins to expand and the part center experiences compression.  The part edges are cooler than the center and thus expand less. This expanding interior drives the outside of the part in to azimuthal tension (positive stresses).  The result is stress birefringence with an azimuthal structure with an amplitude that is a smooth function of radius.

The principal stresses are computed in the model which are translated to the angle of the stress birefringence and magnitude. Figure \ref{fig:stressdistribution} shows both azimuthal angle and magnitude of the stresses as an example spatial distribution of the stress. The model is computed after 4850 seconds of illumination (heating), at a depth (layer) of 26.5mm near the top of the retarder crystal stack using the higher flux no-polarizer heating scenario and assumed {\it worse} coating absorption of 330 ppm per coating.

To assess the impact of some structural model conditions imposed by the boundary conditions of the model, tests were run on models that allowed the retarder crystal plates to slide freely while the default models here retain structural rigidity. There are some shear forces that couple the vertical (z) dimension to the radial XY forces.  However, these forces are small and can be neglected for the purposes of estimating stress birefringence.  The ficticius stress values are $\sim$2 kPa amplitude compared to the principal in-plane stresses of 59 kPa and 109 kPa. The impact of stress birefringence can be estimated at the field edges. The required clear aperture at the calibration retarder is 66.2 mm for the 2.8 arc minute field.  The radius is 33.1mm and this is mostly contained inside the region of uniform compression shown in Figure \ref{fig:stressdistribution}.

It should be noted at this point that the inner 66.2mm shows stress difference values that are substantially below the peak values.  The center of the optic is largely in uniform compression.  Stress birefringence does not seem to be a large effect given these models.    

With a stress-optic coefficient of 4 nm per mm per MPa and stress amplitudes of $\sim$50 kPa through a 30 mm part, we get 6nm of phase retardation.  This is 0.01 waves of retardance at 600 nm wavelength and was similar in magnitude to requirements imposed for polishing errors.  Since the scaling of retardance with stress is linear, stress values below 10 kPa will have no practical impact on the calibration procedure.

\subsection{Thermal Summary: Temporal Stability for Fringes And Design}

By creating a detailed thermal model and including measurements for several types of heat sources, we have a reasonable expectation of thermal performance for the DKIST retarders under the 300 Watt optical load. By knowing the temporal, radial and depth dependence of the temperature distribution, we can model the instabilites of polarization fringes as well as the net change in elliptical retardance. We examined in detail how window-covered retarder designs exacerbate temperature effects and create significant temperature gradients. Not only do these temporal instabilities change the polarization fringe pattern, but they create ellipcital retardance variation across the clear aperture of the part varying with time. 

In response to these simulations, and the basic polarization fringe amplitude simulations of H17\cite{Harrington:2017jh}, we removed the cover windows from the retarders.  When using crystal-only designs, the greatly increased thermal conductivity reduces thermal gradients both with depth and across the clear aperture. 

The MgF$_2$ crystal retarders don't see significant heating when used with the calibration polarizer mounted in the beam ahead of the retarder.  When CaF$_2$ cover windows were used with this optic, the limiting heating is from the absorption in the coatings on the windows. When the CaF$_2$ windows are removed, the heat load is dominated by absorption at wavelengths longer than 6000nm. This 0.5 W heat load does cause the MgF$_2$ retarder optic to rise 0.8 C in 3000 seconds to reach steady state. When the polarizer is used upstream of the MgF$_2$ retarder, no IR flux reaches the optic and the heating is negligible. 

The quartz retarders see significant heat load. When cover windows are used are 16 anti-reflection coatings, and more than double the absorbed heat from the quartz alone at wavelengths longer than roughly 3500nm. The heat load was over 3.1 Watts when using the nominal design without protection from the polarizer. Even without the Infrasil cover-windows, the load is 2.7 Watts without the polarizer.  However, the Infrasil cover-windows are insulators and trapped the heat in the optic, greatly increasing the thermal time constant and exacerbating all thermal issues. The time to steady-state is roughy 8 hours.

Given that temporal stability is a requirement for calibration, these thermal simulations strongly influenced decision making.  When removing the Infrasil cover windows and using the quartz retarder with a polarizer plus window in the calibration proces, the laod is reduced to less than 0.2 Watts.  Given the crystal conductivity, the quartz retarder steady state temperature is spatially uniform to better than 0.5$^\circ$C and the steady state temperature is within 1$^\circ$C of the environment.  When used without protection of an upstream optic in the 300 Watt beam, the improved crystal conductivity without insulating windows greatly improves the temporal stability and reduces gradients. Presenting detailed thermal results is beyond the scope of this article, but the fringe sensitivity to temperature couples tightly to these thermal performance parameters. Design of solar retarders must account for temporal drift of polarization fringes and several types of heat sources to assess impact of the design stability in a calibration process.

\clearpage
\section{Meadowlark SPEX Laboratory Setup Details}
\label{sec:appendix_spex}

\begin{wrapfigure}{r}{0.62\textwidth}
\centering
\vspace{-3mm}
\begin{tabular}{c} 
\hbox{
\hspace{-1.0em}
\includegraphics[height=6.6cm, angle=0]{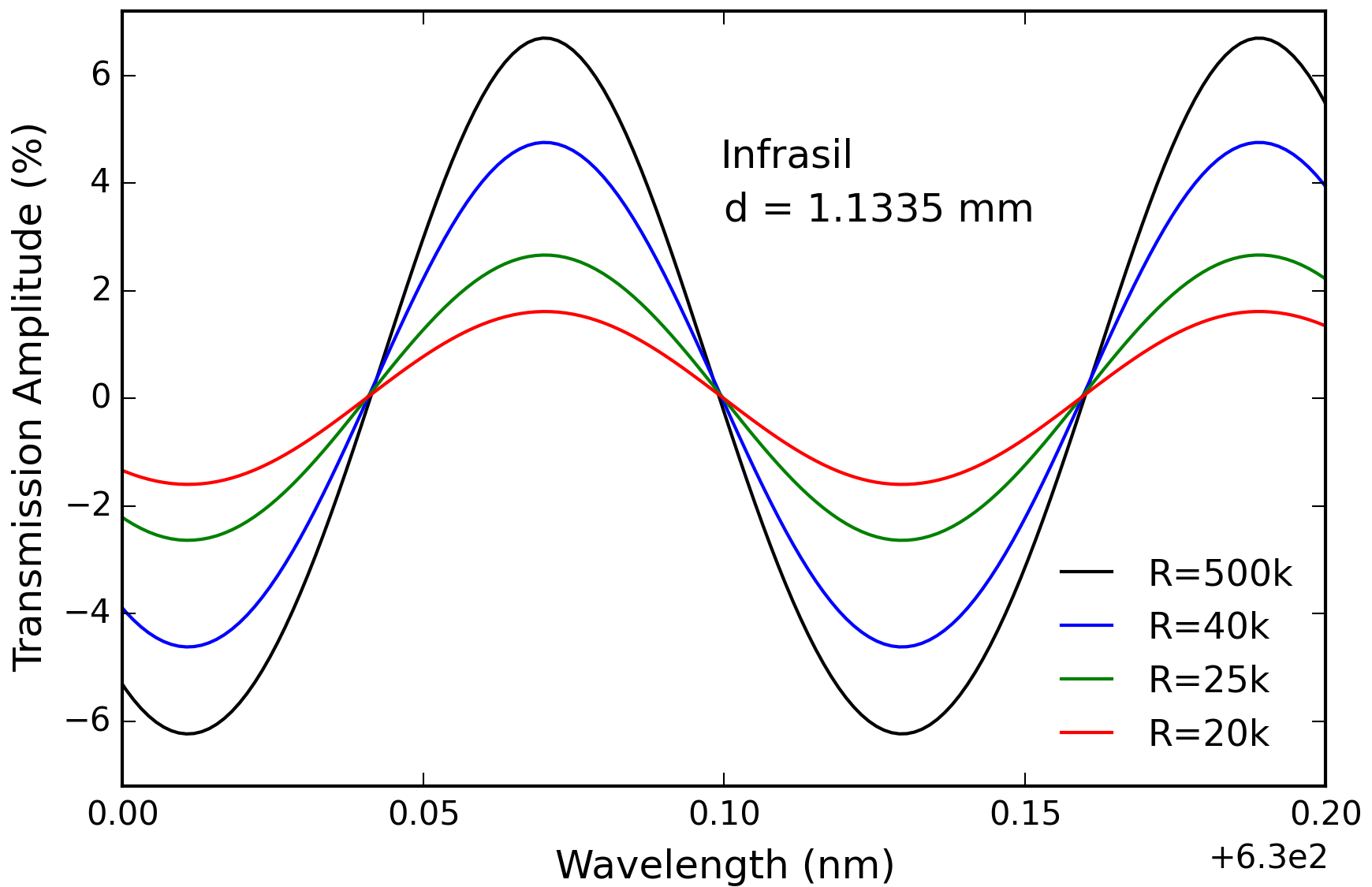}
}
\end{tabular}
\caption[Convolution & Fringe Amplitude] 
{ \label{fig:convolve_infrasil}  The fringe amplitude predicted in the Berreman code after convolution with Gaussian profiles corresponding to varying spectrograph resolving power. See text for details.   }
\vspace{-4mm}
 \end{wrapfigure}

In this Appendix, we outline some details of the experimental setup. The instrument profile of 0.016 nm full-width half-maximum was measured with a neon discharge lamp at 653 nm. The profile has Gaussian shape giving a resolving power of 40,800.  Other spectral lines measured at 585 nm, 609 nm,  633 nm and 725 nm gave resolving powers in the range of 32,000 to 49,000.  Over this wavelength range, the resolving power should not change much, possibly pointing to mild internal optical mis-alignments.  In Figure \ref{fig:convolve_infrasil} we show the impact of the instrument profile on the detected fringe amplitude. For our 1.1335 mm thick Infrasil window, we expect the fringe amplitude to be degraded at or below the blue curve in Figure \ref{fig:convolve_infrasil}.

Meadowlark staff estimate the window was square to the incoming beam to better than 1$^\circ$ for the collimated measurement and better than 5$^\circ$ for the F/ 8 measurement due to mechanical space constraints. In several experiments we conducted, the fringe amplitude was not significantly impacted by the manual alignment procedure. Repeated measurements of fringes showed amplitudes detected were within a small fraction of a percent.

We were suspicious that the optical alignment and other light source issues with the Spex system were causing some sensitivity and fringe amplitude reduction. The original fringe measurements by Meadowlark Optics presented in our previous H17 reference and in earlier sections above only achieved roughly half the predicted fringe amplitude, even after accounting for possible resolution degradation. In addition, the measured spectral resolving power of the Spex system was significantly less than theoretical, suggesting alignment issues. The optics collimating and directing the beam into the Spex instrument was rebuilt with an iris and new optics for measuring the fringe amplitude as a function of system F/ number. The fiber collimator was changed from an OAP assembly to a kinematic-mounted lens tube assembly. The fiber was mounted inside a 1 inch diameter lens tube along with a Thor Labs AC-254 50 mm focal length achromatic doublet. A laser cut circular aperture mask was mounted in the tube immediately after the collimating lens with a 10.0 mm diameter. The fiber was mounted to the input end of another tube and collimation achieved by threading tubes to the proper separation. This assembly was then threaded in a kinematic mount. We also put a second laser cut mask and iris roughly 20 cm of optical path later to allow for control of the collimated beam diameter.  This mount and iris allowed us to assess the impact of optical alignment as well as control the beam F/ number to measure impact on fringe amplitudes. 

With this optical change to a 50 mm collimator, the fiber core is now 1:1 reimaged onto the slit by the 50 mm focal length lens. This optical change also reduced the incidence angle variation from $\pm$0.38$^\circ$ to $\pm$0.11$^\circ$ with the 200 $\mu$m diameter core fiber. The entrance aperture and new iris both vignettes more area of the beam, reducing the signal level, even though the fiber core image is smaller on the slit, providing greater throughput linearly.

\subsection{MgF2 Crystal Retarder Lab Data \& Models}

We also tested a smaller MgF$_2$ crystal retarder in the SPEX setup to verify fringe amplitude and period predictions. The clear aperture of this crystal retarder is only 6.4 mm. The F/ 8 beam stop on the collimating mirror corresponds to a 6.4 mm footprint on the retarder in the collimated beam, critically filling the aperture. The MgF$_2$crystal retarder fast and slow axes were oriented 45$^\circ$ with respect to the grating rulings and mirror fold axes. The MgF$_2$ crystal thickness is measured to be 927.0 $\mu$m $\pm$0.5$\mu$m. The transmitted wavefront error (TWE) is measured at 0.044 waves at 632.8 nm peak-to-peak over an aperture of 6 mm diameter. Beam deviation was measured to be 1.6 arc-seconds. The beam footprint was reduced from 6 mm for the collimated beam to about 3 mm for the F/ 8 beam. A Fourier analysis of the data found the fringe period to be at 0.155 nm as predicted. We could not detect the difference between the theoretical periods of 0.1541 nm for the extraordinary beam and 0.1555 nm for the ordinary beam.  The fringe had a minimum near 628.5 nm wavelength with the amplitude rising to about 3\% at 634 nm.  

Calculations with our Berreman code showed similar behavior to the Quartz retarder presented in H17\cite{Harrington:2017jh}. The fringe amplitude maximum was theoretically 9\% as expected for a 2.6\% surface reflection with transmission ranging from 99.9\% to 89.5\%. The refractive index of 1.389 for the extraordinary beam and 1.377 for the ordinary beam produce an amplitude modulation with a period of about 15.5 nm. We only detected fringe amplitudes of 4\% peak to peak but we had used a wavelength range near one of the amplitude minima where extraordinary and ordinary fringes destructively interfere.  We did not pursue this sample further as the behavior was as expected.

\clearpage

\section{Measured Fringes In A Six-crystal Retarder Using an F/ 13 Beam In the Summit Environment with Keck \& LRISp}
\label{sec:appendix_keck}

In this section we show on-telescope measurements of fringe periods and amplitudes for a six-crystal superachromatic retarder used in an F/ 13 beam in an astronomical spectropolarimeter mounted on the Keck 10 meter diameter telescope located near the summit of Maunakea, Hawaii. This six-crystal retarder uses nearly the same design strategy as DKIST and provides an excellent on-sky demonstration of fringe amplitude reduction in the F/ 13 beam. We also can use this prior work to show an on-sky demonstration of fringe thermal stability in an on-summit environment as this retarder is inside an instrument at Cassegrain focus exposed to environmental temperature fluctuations at night. 

The Keck telescope has a low resolution imaging spectrograph with a polarimetric unit (LRISp) \cite{Keck:2014uc,Keck:2012ub,Goodrich:2003kv,Keck:2007wf,Harrington:2015cq,Keck:2011vc,1998SPIE.3355...81M,McLean:2003co,1995PASP..107..375O,Rockosi:2010ez,1998SPIE.3355...81M,Adkins:2010dn}. The 10 meter diameter primary mirror combined with this Cassegrain-mounted spectropolarimeter leads to high sensitivity on faint targets such as galaxies, stars or comets. We outline some of the initial design choices for the LRISp retarders including considerations of crystal thickness. We use simple analytic calculations to show retardance predictions for the design using the same process as for the DKIST calibration retarders. The two main observations relevant to this work is that the fringe amplitudes measured for this retarder are consistent with our predictions for an F/ 13 converging beam. We also use this optic to verify the thermal fringe instabilities are consistent with the thickness of the crystals and the use of this retarder in a thermally uncontrolled summit environment. We present a design and LRISp data for a Pancharatnam style retarder \cite{Pancharatnam:1955iw} that uses 0.40 mm quartz crystals, 0.34 mm MgF$_2$ crystals with the angle of 59$^\circ$ between crystal pairs. We show some analytical solutions, basic design tolerances, fringe predictions and measurements for such a design as applied to a night-time astronomical spectropolarimeter.

\begin{wrapfigure}{r}{0.45\textwidth}
\centering
\vspace{-6mm}
\begin{equation}
\label{pan_stack1}
\cos \frac{\Delta}{2}  =  \cos \frac{\delta_B}{2}  \cos \delta_A  - \sin \frac{\delta_B}{2}  \sin \delta_A  \cos 2\theta  
\end{equation}
\begin{equation}
\cot 2\Theta =  \frac{ \sin \delta_A  \cot \frac{\delta_B}{2}  +  \cos \delta_A \cos  2\theta }   {  \sin 2\theta }
\label{pan_stack2}
\end{equation}
\vspace{-8mm}
\end{wrapfigure}

A common retarder design tool was introduced by Pancharatnam \cite{Pancharatnam:1955iw} to make a super-achromatic retarder as a combination of three bi-crystalline achromats.  By using three bi-crystalline achromats together, many designs could greatly increase the wavelength range for achromatic linear retardance of various specification. There are many degrees of freedom if one chooses different materials, retardance values and orientations for all six crystals. 

The Pancharatnam designs are usually simplified by choosing just 2 materials and making the outer two bi-crystalline retarders identical. This simple design uses an A-B-A type alignment where the two outer bicrystallline pairs are mounted with their fast axes aligned. Provided the bi-crystalline pairs are treated as perfect linear retarders, there is a simple theoretical formula for the linear retardance of such an A-B-A design. If we take the retardance of the A crystals as $\delta_A$ and the B crystals as $\delta_B$, and the relative orientation between the A and B crystal pairs as $\theta$, we can write the formula for the resulting superachromatic optic retardance ($\Delta$) and fast axis orientation ($\Theta$) as in Equations \ref{pan_stack1} and \ref{pan_stack2} \cite{Pancharatnam:1955iw}.

Often, a further constraint is to make all three crystal pairs identical for manufacturing simplicity.  There is still an orientation offset between the inner B pair and the outer A pairs. This way, a simple Pancharatnam design would only use two materials (such as Quartz and MgF$_2$ crystal) and a manufacturer would only polish each material to one specific thickness. This way, the retarder has three identical bi-crystalline achromats with an orientation of [0$^\circ$,  X$^\circ$, 0$^\circ$] and only two thicknesses to vary for a three-variable optimization problem.  

The polarization optics in LRISp consist of a quarter-wave and a half-wave superachromatic Pancharatnam \cite{Pancharatnam:1955iw} retarder mounted in a two wheels just ahead of the spectrograph entrance slit. The modulation strategy coded in to the LRISp software is the standard {\it Stokes definition} scheme where a half wave plate is rotated in increments of 22.5$^\circ$ to create exposures that can be subtracted to directly measure one component of the Stokes vector. 

Note that the two LRISp retarders were manufactured by Halle \cite{1991PASP..103.1314G,Goodrich:2003kv,Goodrich:1995fg}. Per Goodrich \cite{1991PASP..103.1314G,Goodrich:2003kv,Goodrich:1995fg}, Halle had initially tried a {\it subtraction} method similar to DKIST with $\sim$2 mm thick plates and a nominal thickness difference to specify the retardance.  Halle had difficulty aligning (clocking / rotating) the retarders and the assembled parts had {\it unacceptable ripples} as occurs with imperfect subtraction between thick crystal plates. The sensitivity to polarimetric artifacts is amplified by the crystal thickness, as also found for DKIST designs \cite{2014SPIE.9147E..0FE,Sueoka:2014cm,Sueoka:2016vo}. This difficulty caused the Halle team to switch to a thin crystal design \cite{1991PASP..103.1314G,Goodrich:2003kv,Goodrich:1995fg}. The nominal thickness for the LRISp half-wave part is 404.33 $\mu$m for each quartz crystal and 339.82 for the MgF$_2$ crystal \cite{1991PASP..103.1314G,Goodrich:2003kv,Goodrich:1995fg}.  

\begin{figure}[htbp]
\begin{center}
\vspace{-1mm}
\includegraphics[width=0.99\linewidth, angle=0]{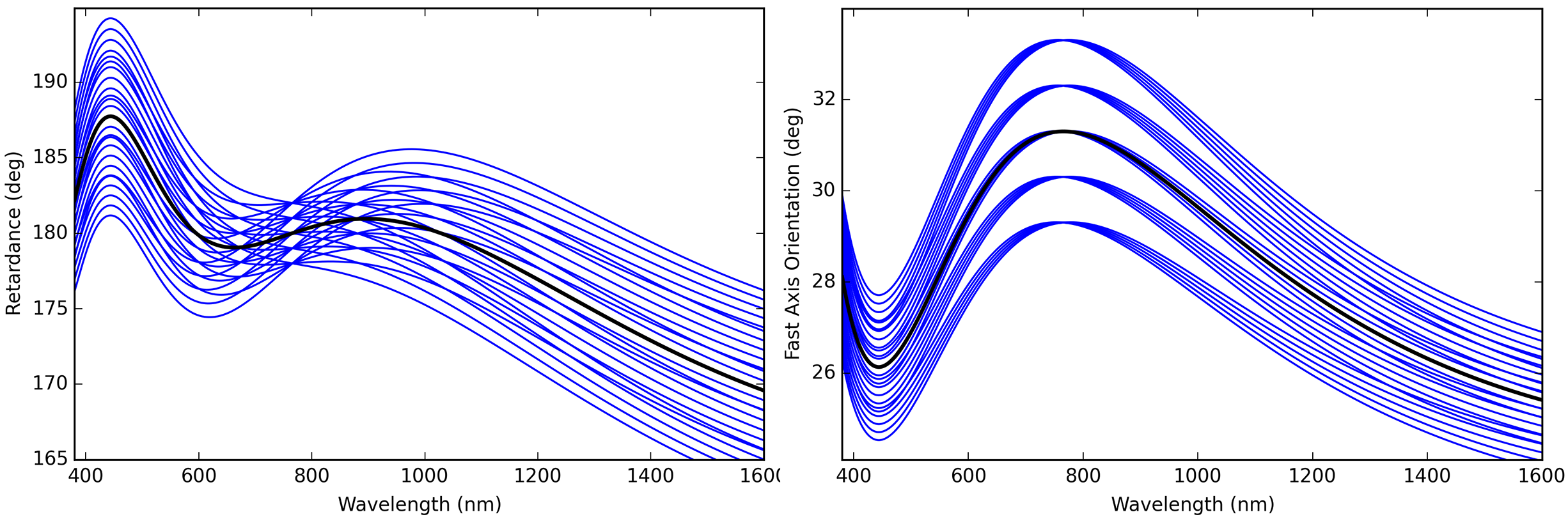}
\caption{The linear retardance and fast axis orientation for the LRISp super-achromatic half-wave retarder design (black) and some simple design perturbations of crystal thickness (blue). We show here the impact on linear retardance of $\pm$1$^\circ$ and $\pm$2$^\circ$ variations in the linear retardance and orientation of only the middle retarder. Note that many other manufacturing imperfections can be simulated but lead to non-zero circular retardance aong with relatively high spectral frequency oscillations in retardance. See text for details.}
\label{fig:lrisp_hwp_design_tolerance}
\vspace{-3mm}
\end{center}
\end{figure}

We show a simple design perturbation analysis for the LRISp half-wave plate design in Figure \ref{fig:lrisp_hwp_design_tolerance}. We take the nominal bi-crystalline parameters and change the middle part retardance and fast axis by $\pm$1$^\circ$ and $\pm$2$^\circ$. The variations in the linear retardance of just this one crystal pair cause design variations of roughly 10$^\circ$ in linear retardance and a few degrees in fast-axis orientation. Material between the crystals is a concern in modeling fringes in a many-crystal optic. In Goodrich et al. 1991 \cite{1991PASP..103.1314G}, there is mention that several manufacturers {\it assemble and glue} the crystals together. The Halle company specifications for their current super-achromatic retarders states their optics currently are {\it cemented}. There likely will be a material between the crystals with an unknown but small thickness and an unknown non-zero mismatch in refractive index with wavelength between the crystals.

\begin{wrapfigure}{l}{0.53\textwidth}
\centering
\vspace{-4mm}
\hbox{
\hspace{-0.8em}
\includegraphics[height=5.8cm, angle=0]{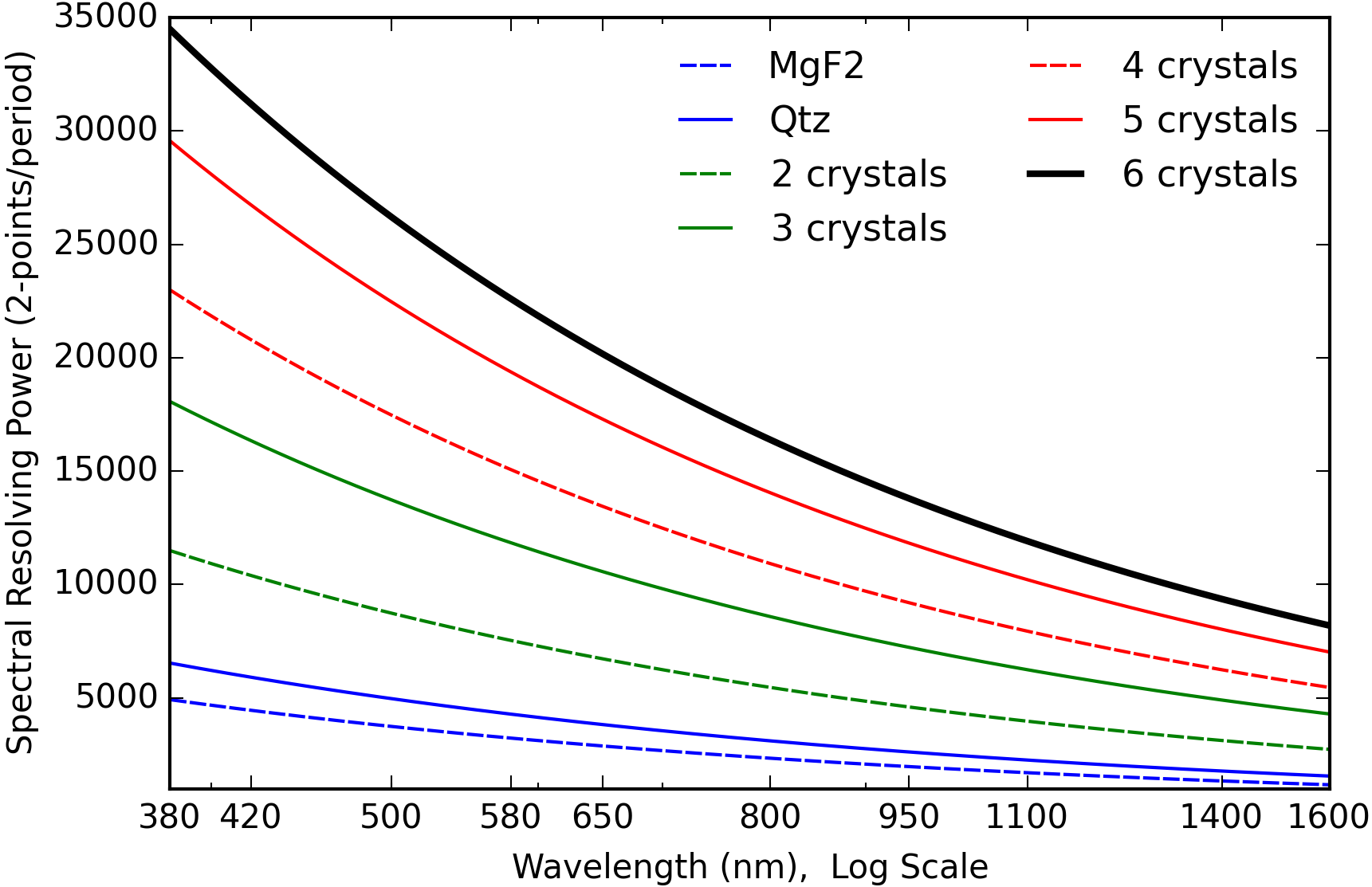}
}
\caption[LRISp HWP Fringes] 
{ \label{fig:LRISp_HWP_Fringe_Resolution} Spectral resolving power required to detect fringes for LRISp. The two blue curves correspond to a single quartz or MgF$_2$ crystal. The next lines show predicted fringe periods corresponding to successive propagation through the SiO$_2$-MgF$_2$-SiO$_2$-MgF$_2$-SiO$_2$-MgF$_2$ interfaces in the retarder. The series ends with the black curve as the period for the entire six crystal stack.}
\vspace{-5mm}
\end{wrapfigure}

Figure \ref{fig:LRISp_HWP_Fringe_Resolution} shows the spectral spectral resolving power required to measure all components of the fringes. The individual MgF$_2$ and SiO$_2$ crystals are shown in blue with thicknesses 0.40 mm and 0.34 mm respectively. The curve requires 2 points sampled per period at a spectral resolving power of R$\sim$5,000 and 6,000 respectively at 400 nm wavelength. The higher curves show how the spectral fringe period gets smaller as the back-reflected wave sees an ever thicker optical path. The solid green curve shows the fringe caused by the wave interfering through two MgF$_2$ and SiO$_2$ crystals. The highest curve would correspond to the entire stack of crystals, requiring a resolving power of 30,000. Given that we only achieved R$\sim$3,000 we are only detecting fringes of the single crystals and we could be subject to errors comparing fringe amplitudes to models due to under-sampling.

\begin{figure}[htbp]
\begin{center}
\vspace{-0mm}
\hbox{
\hspace{-1.0em}
\includegraphics[width=0.99\linewidth, angle=0]{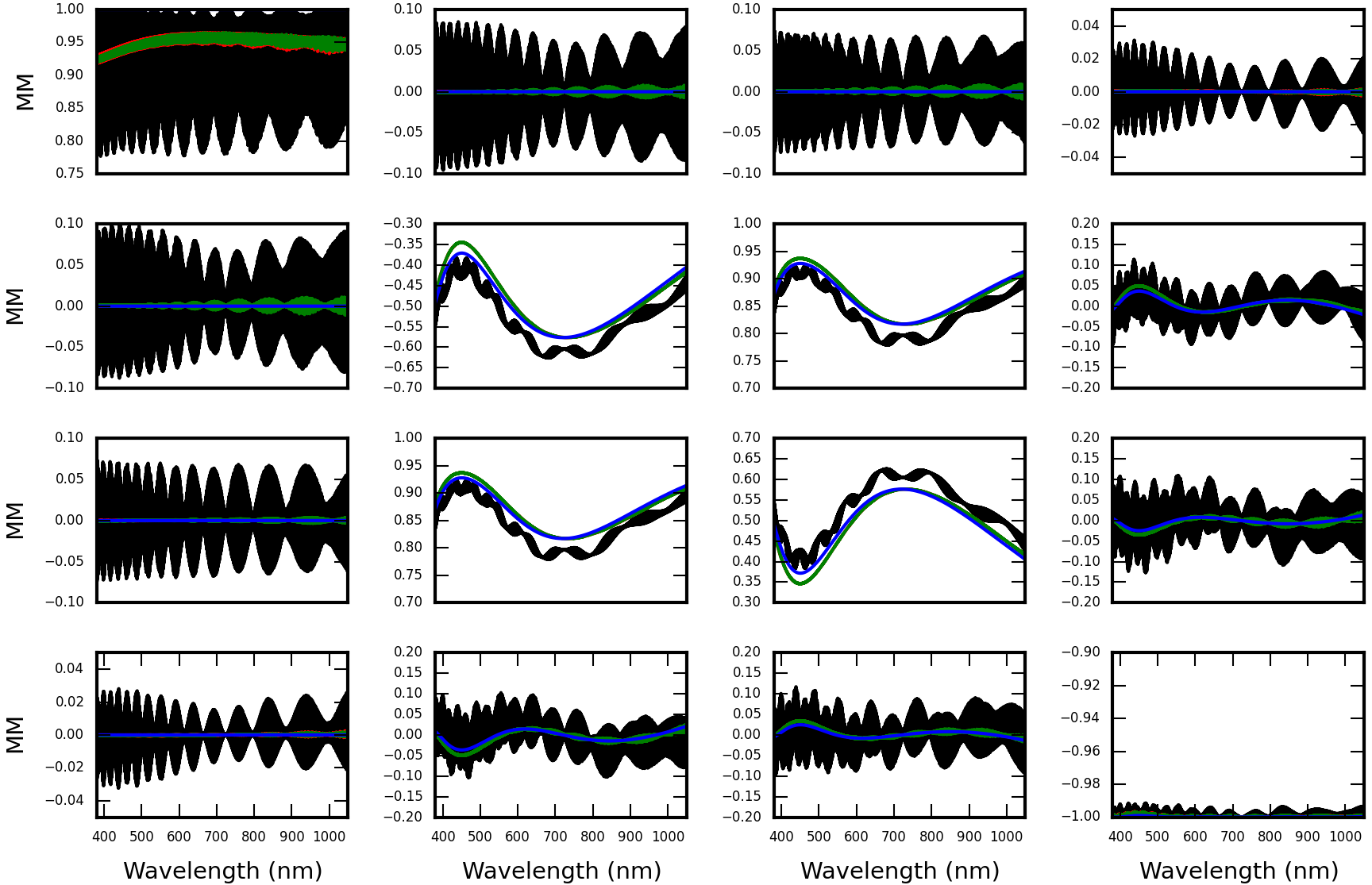}
}
\vspace{2mm}
\caption{\label{fig:LRISp_HWP_Berreman_MMt} The Berreman calculus Mueller matrix in transmission for the LRISp half-wave plate in various spectral resolving powers and optical configurations.  Black shows an optically contacted model with only a single AR coating on the quartz to air interface at R=500,000 as a simple but highly fringed optic. Crystal clocking errors (rotational mis-alignments) of 0.5$^\circ$ were introduced to the SiO$_2$ plate in the B bi-crystalline retarder to demonstrate ripples in the spectra from subtraction plate mis-alignments. In addition, the second A bi-crystalline retarder was misaligned with the first by 0.5$^\circ$ to show net ellipticity resulting from A-B-A misalignments. Green shows a model with n=1.46 cement between crystals at a physical thickness of 150$\mu$m with the same single AR coating on the quartz to air interface at the observed R=2,500. We convolved the high resolution data set with a Gaussian profile to match the measured LRISp spectral instrument profile. Significant amplitude reduction is seen. The red curve shows a model with 5 mm thick fused silica cover windows, anti-reflection coatings on both windows and a cement layer at n=1.46 with a physical thickness of 75$\mu$m. The red curve is barely visible in the $II$ element and is nearly identical to the green curve showing fringes are dominated by the single crystals at these resolving powers. The blue curve shows the theoretical retarder Mueller matrix derived for the unperturbed design using a stack of three ideal bi-crystalline achromats. In ideal retarder models, no diattenuation or transmission fringes are present. See text for details.}
\vspace{-7mm}
\end{center}
\end{figure}

An example of the Berreman theoretical transmitted Mueller matrix is shown in Figure \ref{fig:LRISp_HWP_Berreman_MMt} with simple optical contact of all crystals, no epoxy, no anti-reflection coatings on any surfaces. We adopt a standard astronomical convention for displaying Mueller matrices.  We normalize every element by the $II$ element to remove the influence of transmission on the other matrix elements as seen in Equation \ref{eqn:MM_IntensNorm}.

The Mueller matrix of Figure \ref{fig:LRISp_HWP_Berreman_MMt} does show transmission fringes at amplitudes up to 20\%, diattenuation terms up to 10\% and significant oscillation in the retardance, similar to those measured in the lab \cite{Harrington:2017jh}. The black curve shows the collimated beam prediction at infinite spectral resolving power.  Blue shows the theoretical Mueller matrix derived from a stack of ideal linear retarders including the perturbation analysis outlined in the text.  Green shows the Berreman prediction but at degraded spectral resolving power by convolution with the appropriate Gaussian instrument profile.  The LRISp retarder is in an F/ 13 beam but the marginal ray only sees less than half a wave of path difference compared to the chief ray after reflection inside a single crystal.  This reduces fringes when averaging over the aperture by a small factor, but not below detection limits.

The DKIST project had funded more accurate and modern measurements of crystal birefringence over a wider wavelength range \cite{Sueoka:2016vo}. Other studies such as Mahler et al \cite{2011ApOpt..50..755M} similarly point out variation among studies and vendor-reported models. For this paper, we are using the DKIST revised formulas for the refractive indices and birefringence  \cite{Sueoka:2016vo}.  Our models may vary slightly from other studies.  For reference, we needed to change the design MgF$_2$ crystal thickness by about 12 $\mu$m for our design to match the theoretical curves shown in Goodrich \cite{1991PASP..103.1314G}. Likely some slight mis-match in the designs presented here will be caused by different refractive index formulas. However, this is of minimal significance to the fringe predictions as the 12 $\mu$m of crystal thickness difference corresponds to $<$0.4\% fringe period change.  

In Table \ref{table:LRIS_Ret_Design} we show a possible layout of optical interfaces for the LRISp retarder design. The Halle manufacturers website for super-achromatic retarders also shows the use of cover windows for their standard visible wavelength design 380 nm to 1100 nm wavelength. However, cover windows are not used for their standard infrared design 600 nm to 2700 nm wavelength. They state that {\it cement} is used but without specifying thickness or refractive index. In addition, they state that a standard quarter-wave anti-reflection coating is applied as a single layer of MgF$_2$. Given these options, we assume cover windows and a cement are possibilities for the LRISp optic. In Table \ref{table:LRIS_Ret_Design}, we list the cement as Epx and give a nominal thickness of 75 $\mu$m.

\begin{wraptable}{r}{0.40\textwidth}
	\vspace{-3mm}
	\caption{LRISp Retarder Design}
	\label{table:LRIS_Ret_Design}
	\centering
	\begin{tabular}{l l l l}
		\hline
		\hline
		Material	& Thickness	& $\theta$		& Note		\\
		Name	& $\mu$m		&  deg.		&		\\
		\hline
		\hline
		AR 		& 0.1223			& -			& CWL?	\\
		FS		& 5000			& -			& ?		\\
		Epx		& 75				& -			& n=1.46?\\
		Qtz		& 403.1036		& 0			&		\\
		Epx		& 75				& -			& n=1.46?\\
		MgF		& 339.820			& 90		&		\\
		Epx		& 75				& -			& n=1.46?\\
		Qtz		& 403.1036		& 58.7		& +0.5$^\circ$		\\
		Epx		& 75				& -			& n=1.46?\\
		MgF		& 339.820			& 148.7		& 	\\
		Epx		& 75				& -			& n=1.46?\\
		Qtz		& 403.1036		& 0			& +0.5$^\circ$ 		\\
		Epx		& 75				& -			& n=1.46?		\\
		MgF		& 339.820			& 90			& +0.5$^\circ$ 		\\
		Epx		& 75				& -			& n=1.46?\\
		FS		& 5000			& -			& ?		\\
		AR 		& 0.1223			& -			& CWL?	\\
		\hline
		\hline
	\end{tabular}
\vspace{-4mm}
\end{wraptable}

We additionally make the optimistic assumption that the refractive index is an average between crystal quartz and crystal MgF$_2$ at n=1.46. This index would likely be a design goal for minimization of fringes. We list a fused silica cover window as FS and use a nominal 5 mm physical thickness but also have models at 2 mm thickness. We do not know the central wavelength of the AR coating and thus chose 675 nm and 500 nm for models covering a range of possibilities. Given the uncertainties, we compute several different Berreman models with or without cover windows, with AR coatings and with cement layers of varying thickness and index. We also compare this to optically contacted models. We also solve analytically for the physical thickness of the crystal quartz plate using our refractive index equations to ensure an exact retardance at the design wavelength using the Berreman calculus, denoted at Qtz in Table \ref{table:LRIS_Ret_Design}. We also note application of rotational errors of 0.5$^\circ$  to crystals 3, 5 and 6 for later comparison on the impact of manufacturing tolerances.

As seen in H17\cite{Harrington:2017jh} and above, thin spaces between crystals filled with air, oil or cement can change fringe amplitudes over broad wavelength ranges. The gaps introduce a fringe period at $\lambda^2$/2dn which can have a large spectral period, much larger than from the millimeter-thickness crystals. In DKIST laboratory optics, cement layers are measured in the 30 $\mu$m to 100 $\mu$m range. The refractive index matching oil layers between crystal optics are measured to be in the range 5 $\mu$m to 15 $\mu$m H17\cite{Harrington:2017jh}. Harrington et al. 2015, hereafter called H15, \cite{Harrington:2015cq} outlined a data reduction pipeline to process dual-beam spectropolarimetric data with this instrument. A collaboration has been using this instrument for high precision spectropolarimetry where fringes must be very well separated from stellar signals. \cite{2017ApJ...847...61B,2017ApJ...847...60K}  As part of using this instrument in 2014, we performed a range of additional calibrations to characterize the internal cross-talk using the daytime sky \cite{Harrington:2015dl,Harrington:2017eja,Harrington:2010km,Harrington:2011fz} as well as many internal calibrations to establish orientations of the retarders. We found a spectral resolving power of R$\sim$ 2500 at 800nm wavelength rising to R$\sim$ 3500 at 1000 nm wavelength. The spectral sampling was high, rising from 56 pm to 59 pm over the same wavelength range. This sampling gives roughly 5 detector pixels per full width half max of the monochromatic slit image derived from gas arc discharge lamp spectra. The resolving power is only about 0.3 nm (the optical full-width-half-max of a monochromatic input).

\begin{figure}[htbp]
\begin{center}
\vspace{-1mm}
\includegraphics[width=0.99\linewidth, angle=0]{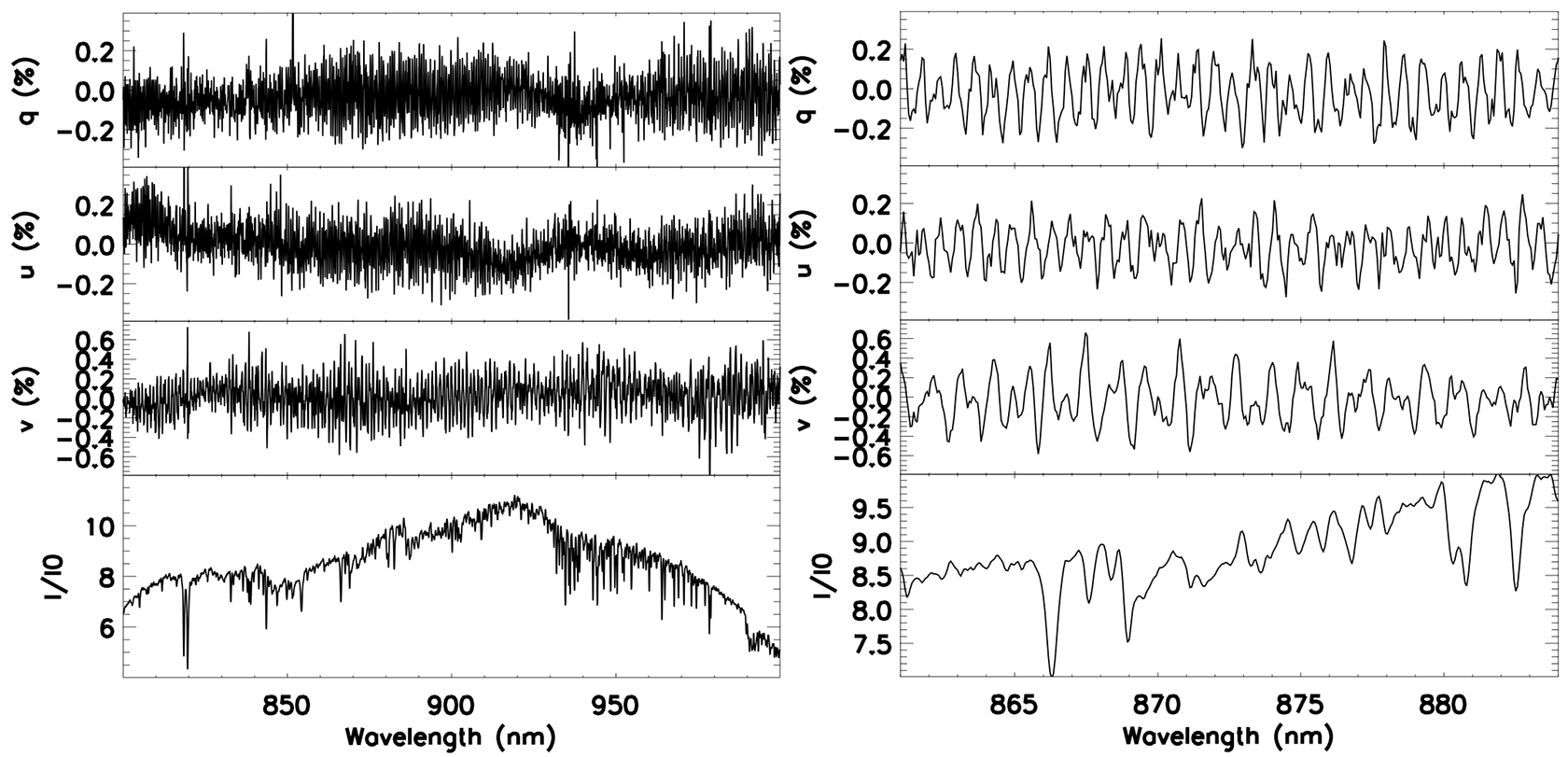}
\caption{\label{fig:evlac_spectrum} The $I$ spectrum and normalized $quv$ spectrum derived with LRISp for the magnetic star EV Lac.  The fractional $quv$ spectra as well as the detected normalized intensity are shown here covering the full wavelength range on the left and covering a narrow 850 nm to 890 nm bandpass on the right. Strong polarization fringes are seen in all $quv$ data. The statistical signal to noise ratio is below 0.03\% with fringes at $\pm$0.2\% to $\pm$0.6\% amplitudes, ten times the statistical noise levels. The $q$ and $u$ measurements use only the rotating half-wave retarder. The $v$ measurements use a second quarter-wave retarder inserted at fixed orientation in the beam ahead of the half-wave modulator. \cite{Harrington:2015cq} See text for details.}
\vspace{-4mm}
\end{center}
\end{figure}

An example full Stokes observation is shown in Figure \ref{fig:evlac_spectrum}. This star (EV Lac) was observed over a few tens of minutes and was well exposed in each image. There is very little continuum polarization in this target and the $quv$ spectra are dominated by polarization fringes.  

\begin{wraptable}{l}{0.48\textwidth}
\vspace{-2mm}
\caption{LRISp 864 nm Beam Properties}
\label{table:lrisp_fringe_beam_properties}
\centering
\begin{tabular}{l l l l}
\hline\hline
- 		& 0.34mm		& 0.40mm	 & 0.74mm		\\
-		& MgF$_2$		& SiO$_2$	 & Both			\\
\hline
\hline
E-Index		& 1.386			& 1.546		&				\\
O-Index		& 1.375			& 1.537		&				\\
E- Fringe	& 0.760 nm		& 0.574 nm	& 0.327 nm		\\
O- Fringe	& 0.770 nm		& 0.578 nm	& 0.329 nm		\\
\hline
Sample		& 13			& 10		& Pixels		\\
Resolve		& 2.3			& 1.7		& FWHMs			\\
\hline
Chief OP	& 1091.0 		& 1431.5 	& 2522.5		\\
\hline
Marg. F/ 13	& 0.42			& 0.44		& 0.86			\\
\hline
\hline
\end{tabular}
\vspace{-3mm}
\end{wraptable}

As this source is essentially unpolarized and the LRISp instrument is known to have very small induced polarization ($<0.2\%$)\cite{1991PASP..103.1314G,Goodrich:2003kv,Goodrich:1995fg,1995AJ....110.2597T}, all the artifacts in the $quv$ spectra are due to diattenuation of the plates. To illustrate the robustness of the fringes, we attempted to extract the spectra from the images using a wide range of settings for the various filters and algorithms in our analysis software. This demodulation scheme either requires further calibration or assumes no cross-talk or other polarization imperfections and does require six exposures (at least). The half-wave plate is closer to the focal plane and is always in the beam. To accomplish measurement of circular polarization, the fixed quarter-wave linear retarder is rotated into the beam ahead of the half-wave linear retarder. The alignment of the fast axes in the mount as well as chromatic variation thus limit the validity of the assumptions behind a simple {\it Stokes definition} demodulation by just subtracting image pairs. We followed this standard sequence but then observed polarized standard stars as well as the daytime sky to assess the cross-talk in the system\cite{Harrington:2015cq}.

The predicted fringe period for just a single crystal is barely within the resolving power of LRISp \cite{Harrington:2015cq}. At 846 nm wavelength, the measured full-width-half-max of a monochromatic input is about 0.33 nm well sampled with five spectral pixels. At this wavelength, quartz has refractive indices of n=1.54 while MgF$_2$ has indices of n=1.38. Using the thicknesses of 0.40 mm and 0.34 mm for each crystal, we see that the spectral fringe periods are roughly 0.58 nm for quartz and 0.77 nm for the MgF$_2$. This puts the predicted fringe period at roughly two times the instrument profile optical FWHM, being dispersed over roughly 10 detector pixels. 

Table \ref{table:lrisp_fringe_beam_properties} shows properties of the LRISp modulator crystals in an F/ 13 beam. Each column corresponds to increasing thickness of crystal from the single MgF$_2$ crystal at 0.34 mm to a single SiO$_2$ crystal at 0.40 mm to the combined bi-crystalline achromat MgF$_2$ and SiO$_2$ at 0.74 mm total thickness. For the resolution and sampling calculations, we use a spectral resolving power of 2500 giving a 0.338 nm FWHM and spectral sampling of 56 pm per pixel. We show the extra-ordinary and ordinary beam refractive indices in the first two rows.  The spectral fringe period for each crystal extraordinary and ordinary beam is shown in the third and fourth rows. We then compute the spectral sampling in pixels for the fifth row for the average of the extraordinary and ordinary beams. The sixth row shows how well LRISp resolves the fringes in terms of optical FWHM per spectral fringe period.  Only roughly 2 optical FWHMs separate the fringe peaks showing very poorly resolved fringes and a degraded fringe amplitude (which we simulate below).

\begin{figure}[htbp]
\begin{center}
\vspace{-1mm}
\includegraphics[width=0.99\linewidth, angle=0]{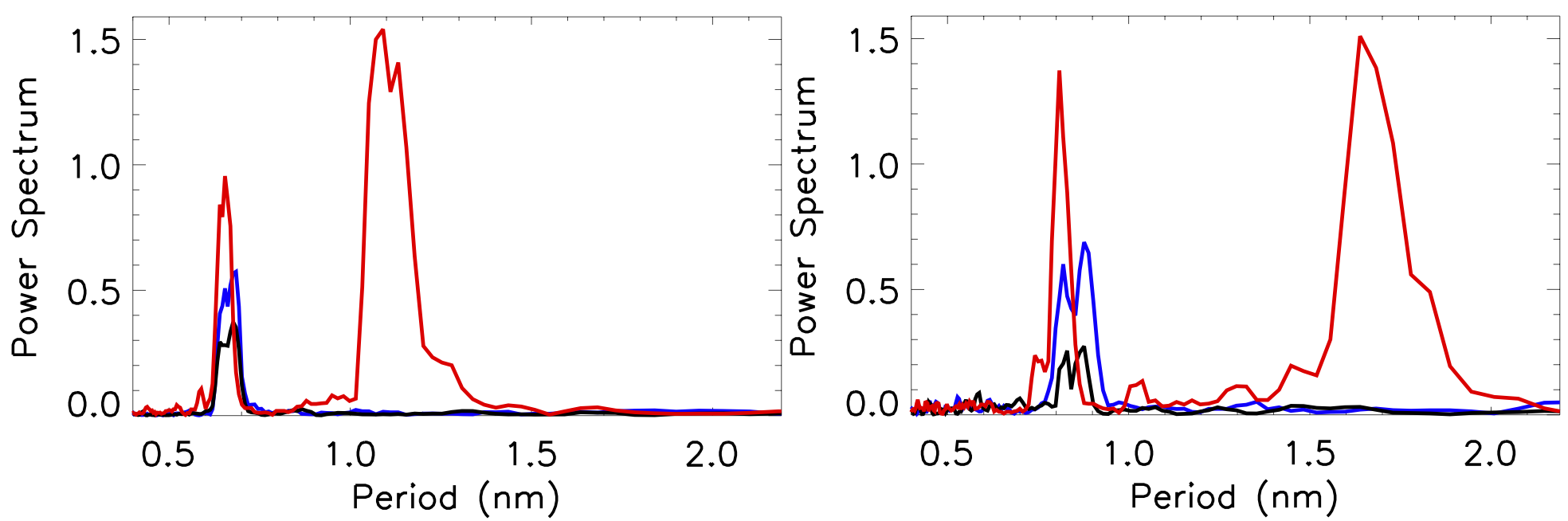}
\caption{\label{fig:lrisp_unpol_fft_power} The fringe power of the $QUV$ spectrum for LRISp observations of EV Lac taken on August 22nd, 2014. The left panel shows the FFT of a 20nm bandpass centered at 846nm. The right shows 964nm central wavelength. Red shows Stokes $v$ when both quarter- and half- wave retarders are in the F/ 13 beam. Blue corresponds to Stokes $q$ and yellow is Stokes $u$. The power is computed as $FFT^2$ and plotted as a function of spectral period in nm. Note that the predicted fringe periods are $\sim$0.770nm for the MgF$_2$ crystal at 846nm and 1.00nm at 964nm. The quartz fringe is 0.58nm at 846nm wavelength increasing to 0.754nm at 964nm wavelength. The observations using half-wave plate only (blue, black) show a single peak of fringe power at the expected period for the single 0.4 mm crystal quartz plate. The red curve showing $v$ has an additional peak that is only present when both quarter- and half- wave retarders are in the beam. }
\vspace{-5mm}
\end{center}
\end{figure}

Row seven lists the chief ray optical path through the crystal.  Row eight lists the marginal ray path difference between chief and marginal rays for an F/ 13 beam. At 864 nm wavelength, the back-reflected chief ray sees 1091 waves of optical path when propagating through a single 0.34 mm thick MgF$_2$ crystal while the marginal ray for an F/ 13 beam sees an additional 0.42 waves of optical path.

As pointed out in H15\cite{Harrington:2015cq}, we found the Fourier spectrum power had peaks very similar to Figure \ref{fig:lrisp_unpol_fft_power} for the various targets observed on three separate campaigns. The fringe power spectra are shown for $q$ in blue, $u$ in black and $v$ in red. As our stellar sources are effectively unpolarized in the continuum as is with the Cassegrain focus of the Keck telescope, fringes are dominated by diattenuation terms in the retarder Mueller matrix. As measurements of Stokes $v$ require both the quarter- and the half- wave retarders, there are possible interference effects between both retarders. The observations with both retarders in the beam (quarter-wave in front of half-wave) should and do have the same peaks as the blue and black curves. But the $v$ measurements show additional power in broad peaks at higher frequencies. All curves share power at fringe periods below 1 nm. Only the $v$ spectra show additional power at longer periods when two optics are in the beam. 

It is encouraging that the 846 nm and 964 nm wvelength observations show substantial fringe power where LRISp has spectral resolution at the predicted periods. At 846 nm wavelength this is the $\sim$0.770 nm fringe from MgF$_2$ which increases to 1.00 nm at a wavelength of 964 nm. We also expect a contribution to the fringe from the spectral period corresponding to a wave propagating through both MgF$_2$ and SiO$_2$ crystals. We consider the agreement between the predictions and the observations of Figure \ref{fig:lrisp_unpol_fft_power} to be quite good given that as-built crystals can have significantly different thicknesses.

With the Berreman calculus, we model the entire system with both retarders as well as any possible bonding material between crystals. H15 modeled this as interference between sum and difference terms but did not model the full Mueller matrix.  We created several Berreman models for the LRISp six-crystal retarder. Given the refractive index mis-match between the MgF$_2$ and SiO$_2$ crystals, an optical contact would create a significant reflection at all interfaces. The Mueller matrix computed in the collimated beam of the Berreman formalism does show transmission fringes at amplitudes up to 20\%, diattenuation terms up to 10\% and significant oscillation in the retardance, similar to those measured in the lab \cite{Harrington:2017jh}. However, the F/ 13 beam will have significant impact on the actual measured fringes.  We also note that in the H15 LRISp data set, we also had complete spectral coverage from 380 nm to over 700 nm using the blue arm of the instrument fed by a dichroic beam splitter reflection. The resolving power was only 450 to 790 from shortest to longest wavelengths, severely under-resolving the spectral fringes. As seen in H15, there were no fringes detected in the blue data sets largely because of the significantly shorter spectral fringe period and several times lower spectral resolving power of that configuration. 

\begin{wrapfigure}{l}{0.63\textwidth}
\centering
\vspace{-2mm}
\begin{tabular}{c} 
\hbox{
\hspace{-1.0em}
\includegraphics[height=7.0cm, angle=0]{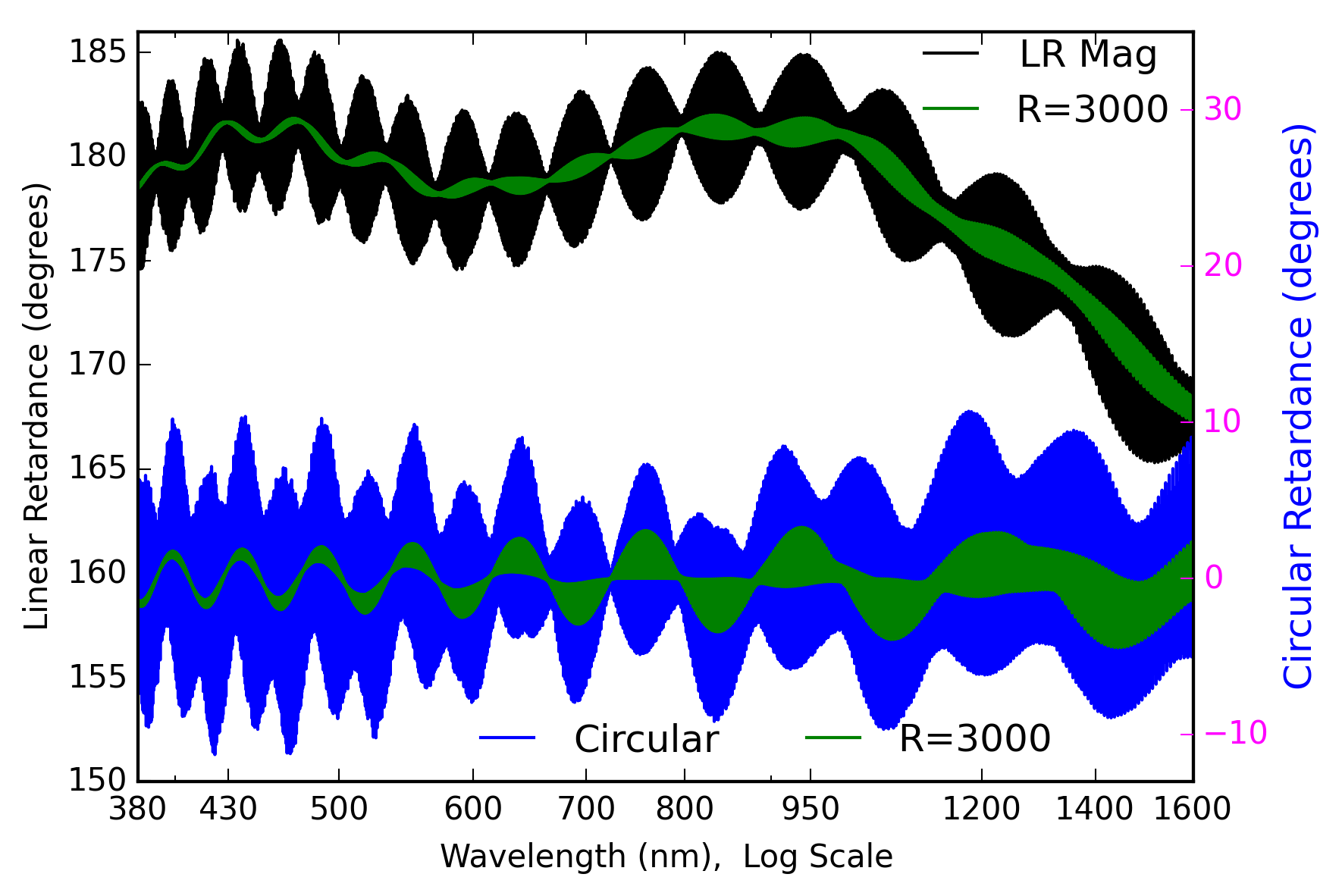}
}
\end{tabular}
\caption[LRISp HWP ER] 
{ \label{fig:LRISp_HWP_Berreman_EllipticalRetardance} Elliptical retarder fit parameters to the Berreman-derived transmission Mueller matrix for the LRISp half-wave plate. The clocking errors (rotational mis-alignments) were used for demonstration. Black shows the magnitude of linear retardance. See text for details.}
\vspace{-3mm}
\end{wrapfigure}

In Figure \ref{fig:LRISp_HWP_Berreman_EllipticalRetardance}, we show elliptical retarder model fits to the Berreman Mueller matrix.  The black curve shows the nominal linear retardance magnitude on the left hand Y-axis.  The Berreman model matches the nominal theoretical 180$^\circ$ retardance over the entire 380 nm to 1000 nm wavelength bandpass. The circular retardance is shown in blue using the right-hand Y-axis. The elliptical retardance fringes oscillate at the expected spectral period with circular retardance fringe amplitudes up to $\pm$10$^\circ$. The slight non-zero average in circular retardance comes from the retarder orientation mis-alignments simulated following typical manufacturing tolerances we applied to crystals 3, 5 and 6 as shown in Table \ref{table:LRIS_Ret_Design}.  We note that we did reproduce the Goodrich 1991 Figure 5 retardance predictions for a few different refractive index formulas. Slight changes in the refractive index formula to have minimal impact on the conclusions derived here. 

\begin{figure}[htbp]
\begin{center}
\vspace{-1mm}
\includegraphics[width=0.99\linewidth, angle=0]{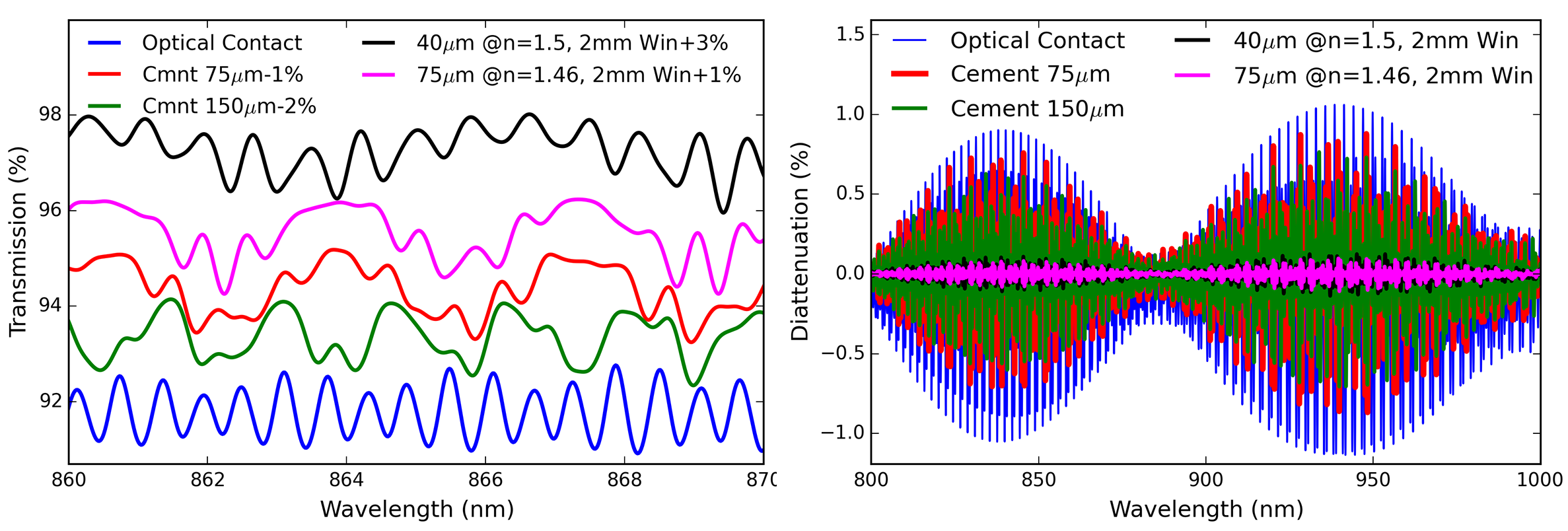}
\caption{\label{fig:lrisp_transmission_diattenuation_fringe} The transmission and $IQ$ diattenuation fringes for a range of Berreman models. We degraded the model resolution to match the R=2,500 LRISp instrument profile. Blue shows an optically contacted model.  Red shows a model with cement at layer thickness of 75 $\mu$m and a refractive index of n=1.46. Green shows a model with the cement thickness increased to 150 $\mu$m. Black shows a model with 2 mm thick fused silica cover windows and cement between all optics at a thickness of 40 $\mu$m with an index of n=1.50.  Magenta shows this window model with a cement at 75 $\mu$m thickness and index of n=1.46.  }
\vspace{-5mm}
\end{center}
\end{figure}

The amplitude of the predicted transmission and diattenuation fringes depends strongly on the system resolving power. By convolving all models with Gaussian instrument profiles of the equivalent resolving power of R=2,500, fringe amplitudes are reduced from over 10\% to less than 1\%. Cover windows also severely impact the predicted diattenuation amplitudes reducing the fringes further. We also no not have manufacturers data on the cement presence, thickness or refractive index. We present a range of models to demonstrate the variation caused by optical changes in Figure \ref{fig:lrisp_transmission_diattenuation_fringe}. The left plot shows the transmission with vertical offsets applied. The optically contacted model uses crystal-crystal interfaces only with a single layer of isotropic MgF$_2$ applied to the quartz to air interface. The cemented models use thicknesses between 40 $\mu$m and 150 $\mu$m. We also show the impact of slight refractive index differences in the cement using 1.46 and 1.50 as possible intermediate values between the MgF$_2$ crystal at n=1.38 and SiO$_2$ crystal at n=1.55. These models demonstrate significant impact of all design possibilities on the transmission and diattenuation fringe amplitudes.

Assessing the measured fringe amplitude against design possibilities also requires accounting for the F/ number reducing fringe amplitudes. At a wavelength of 846 nm, the chief ray propagating through a single 0.34 mm thick MgF$_2$ crystal would see 1100 waves of optical path. The single 0.40 mm SiO$_2$ crystal chief ray back-reflected path is 1500 waves.  At F/ 13, these individual crystals would produce roughly half a wave of optical path variation from beam center to the marginal ray edge. The back-reflection causing interference at the spectral period of any single crystal thankfully is mitigated by the smaller refractive index mis-match for the internal interfaces. 

The SiO$_2$ and MgF$_2$ interfaces see an index difference of 1.546 to 1.386 giving an internal reflectivity of only 0.3\% assuming optical contact. With a cement of intermediate index and fraction of a wave thickness, reflection could be further reduced. A single SiO$_2$ crystal to air reflection is 4.6\% while MgF$_2$ crystal to air is 2.6\% at these wavelengths. It is unknown whether any anti-reflection coatings or cover windows were applied to the retarder as none are mentioned in the various LRISp document packages. \cite{Keck:2007wf,Goodrich:2003kv,1995AJ....110.2597T,Goodrich:1995fg,1991PASP..103.1314G} The Berreman model of Figure \ref{fig:LRISp_HWP_Berreman_MMt} predicts transmission fringes up to 20\% and diattenuation of $\pm$10\% for an optically contacted, uncoated full resolution model. This is roughly two orders of magnitude larger than observed. There is an order of magnitude reduction in fringe amplitude from low spectral resolving power shown in Figure \ref{fig:lrisp_transmission_diattenuation_fringe}. We expect fringes in the range of $\pm$1\% for the likely retarder configuration of no cover window with index matching cement and AR coatings. If cover windows are used, the fringes are further reduced by the presumed AR coatings and increase of the spectral fringe period. Further reduction in fringe amplitude will be seen from the beam F/ number. 

As shown in Table \ref{table:lrisp_fringe_beam_properties}, the single-crystal marginal ray sees roughly 0.4 waves of OPD compared to the chief ray. This gives a further fringe amplitude reduction but of a factor 2 or less. For the entire six-crystal stack, the beam traverses over 6,000 waves of optical path. The marginal ray path is over 2.5 waves longer than the chief ray back-reflection. With several waves of aperture average, we would expect an order of magnitude reduction in polarization spectral fringe for the fastest spectral periods. Given that these fringes are then severely unresolved, their presence is expected to be negligible in the data set. The collimated Berreman predictions suggest diattenuation values of 10\% but we detected 0.2\% magnitudes.  After accounting for the low resolving power via instrument profile convolution, we see a reduction to well below 1\%.  After assessing variables for the cement layer thickness and index along with the aperture average from the converging F/ 13 beam, we achieve model fringe magnitudes in the range of 0.2\% similar to those detected.

\subsection{Summary of the Keck LIRSp Fringe Analysis in an F/13 Beam}

We showed in this section that we can use the Berreman calculus and considerations of the F/ 13 beam to reproduce the general characteristics of detected fringes for an on-summit spectropolarimeter. We can predict the fringe amplitude and the temporal instability of the fringes in response to the instruments uncontrolled thermal environment. 

Given the individual crystal thickness is about five times thinner than the DKIST retarders, the thermal sensitivity would be less than one fifth wave phase per $^\circ$C temperature change. The Maunakea summit environment is typically temperature-stable to better than 1$^\circ$C after sundown. With such small, thin crystals, the thermal timescale for adjustment to exterior environmental changes is much faster and we can assume the retarder tracks ambient temperature far faster than the 80-minute timescale we modeled for the DKIST retarders. However, with all-night operation and possible temperature change at $^\circ$C magnitudes, even this thin retarder will have unstable fringes. This is consistent with the data reduction algorithms for fringe removal required in H15 with slow drifts in fringes and irreproducibility of the fringe pattern between nights as well as seasons. We conclude that this six-crystal retarder confirms our on-summit expectations for fringe amplitudes as functions of beam F/ number and additionally confirms the fringe thermal instabilities. 

Our fringe amplitude predictions are limited by the low resolving power of LRISp.  However, this new F/ 13 approximation suggests that the thin crystals do not see more than a factor of few reduction of the slowest fringe periods.  We are also limited by the lack of knowledge of if / what the bonding between crystals may be.  Significant changes to the fringe properties occur if there is a refractive-index matched epoxy between the quartz and MgF$_2$.  However, we still detect the slowest fringe period components at exactly the predicted period and with about the correct amplitude after consideration of the low spectral resolving power and slight reduction from a fraction of a spatial fringe across the clear aperture. This observational data shows that all frequency components will be present in many-crystal retarders. Fringes are not removed by averaging over many of the fastest fringe periods as the LRISp low resolving power was not sufficient to completely smooth fringes from the detected spectra.

\clearpage

\bibliography{ms_ver02} 			
\bibliographystyle{spiebib}		

\end{document}